\documentclass[preprint,8pt]{elsarticle}

\usepackage{amssymb}
\usepackage{graphicx}
\usepackage{amssymb}
\usepackage{graphicx}
\usepackage{amsfonts}
\usepackage{geometry}
\usepackage{epsfig}
\usepackage{CJK}
\usepackage{amsmath}
\usepackage{epstopdf}
\usepackage{mathrsfs}
\usepackage{bbding}
\usepackage{cases}
\usepackage{subfigure}
\usepackage{float}
\usepackage{lineno,hyperref}
\usepackage{lineno,hyperref}
\usepackage{lipsum}
\usepackage{amssymb}
\usepackage{graphicx}
\usepackage{amsfonts}
\usepackage{geometry}
\usepackage{epsfig}
\usepackage{CJK}
\usepackage{amsmath}
\usepackage{graphicx}
\usepackage{epstopdf}
\usepackage{mathrsfs}
\usepackage{bbding}
\usepackage{cases}
\usepackage{subfigure}
\usepackage{float}
\usepackage{lineno,hyperref}
\modulolinenumbers[5]

\journal{}

\begin{document}

\begin{frontmatter}



\title{Mixed interactions of localized waves in the three-component coupled derivative nonlinear Schr\"{o}dinger equations}

\author[]{Tao Xu$^{1,2}$}
\author[]{Yong Chen$^{1, 2,3}$\corref{mycorrespondingauthor}}
\cortext[mycorrespondingauthor]{Corresponding author.}
\ead{ychen@sei.ecnu.edu.cn}
\address{1 Shanghai Key Laboratory of Trustworthy Computing, East China Normal University, Shanghai, 200062, China\\
2 MOE International Joint Lab of Trustworthy Software, East China Normal University, Shanghai, 200062, China\\
3 Department of Physics, Zhejiang Normal University, Jinhua 321004, China}

\begin{abstract}
The Darboux transformation of the three-component coupled derivative nonlinear Schr\"{o}dinger equations is constructed, based on the special vector solution elaborately generated  from the corresponding Lax pair,  various  interactions of localized waves are derived. Here, we focus on the higher-order interactional solutions among higher-order rogue waves (RWs), multi-soliton and multi-breather.  Instead of considering various arrangements among the three components $q_1$, $q_2$ and $q_3$,  we define the same combination as the same type solution. Based on our method, these interactional solutions are completely classified into six types among these three components $q_1$, $q_2$ and $q_3$. In these six types interactional solutions, there are four mixed interactions of localized waves in three different components. In particular, the free parameters $\alpha$ and $\beta$ paly an important role in dynamics structures of the interactional solutions, for example, different nonlinear localized waves merge with each other by increasing the absolute values of $\alpha$ and $\beta$.
\end{abstract}


\end{frontmatter}

\section{Introduction}
In the past several decades, there have been a variety of  researches on  nonlinear localized waves including  bright \cite {6-1,6-2}  or dark solitons \cite{6-3,6-4}, breathers \cite{6-5,6-6} and rogue waves \cite{6-7,6-8,6-9,6-10} in the field of nonlinear science. When self-focusing (self-defocusing) effect interplays with dispersion effect, soliton  including bright or dark soliton  may be formed. Besides, these waves keep their amplitudes and speeds unchanged during propagating. Owing to the instability of small amplitude perturbations, breathers localized in time or (and) time can be generated  and own some periodic properties, which are fundamentally different from soliton solutions. Up to the present, there  mainly exists two special kinds of breathers such as  Akhmediev breathers (ABs) \cite {6-11,6-12} and Kuznetsov-Ma breathers (KMBs) \cite{6-13}.  ABs propagate periodically in space and localize in time and KMBs propagate periodically in time and localize in space. As the rogue wave prototype, a simple rational solution-the Peregrine soliton was first derived by Peregrine in 1983 \cite{6-14} and it can be seen as the limiting case of ABs or KMBs. Localizing in both space and time, rogue wave (RW) always appears from nowhere and disappears without a trace,  whose peak amplitude is usually more than twice of the background wave height \cite{6-15}. There have been many experimental observations \cite{6-16,6-17} and theoretical researches \cite{6-18,6-19,6-20} about rogue wave in different nonlinear models.

In recent years, different interactions of localized waves have been reported in many single-component systems. The hybrid solutions between RWs and conidal periodic waves were constructed in the focusing nonlinear Schr\"{o}dinger (NLS) equation through Darboux transformation (DT) method \cite{6-21}. Through  the consistent Riccati expansion or consistent tanh expansion  method, interaction solutions between solitons and some other types of nonlinear wave can be obtained directly in many nonlinear systems \cite{6-22,6-23}. Hybrid solutions including RWs interacting  with solitons and breathers at the same time was derived in the Boussinesq equation by Hirota bilinear method \cite{6-24}. Besides, some novel semi-rational solutions were constructed in several nonlocal nonlinear integrable models \cite{6-25,6-26}, which included lump solitons interacted with RWs, solitons and breathers, respectively, lump solitons interacted with breathers and periodic line waves at the same time, etc. Utilizing the Hirota bilinear method, the authors constructed the novel rogue wave triggered by the interaction between lump soliton and a pair of resonance kink stripe solitons \cite{6-26-1,6-26-2}.

Additionally, the studies for interactions of localized waves have been extended to multi-component coupled systems. It has been reported that dark-dark, bright-dark soliton existed in various coupled nonlinear systems by KP reduction technique \cite{6-3,6-27,6-28}. Bright-dark-rogue solutions were constructed in both two-component NLS \cite {6-29} equations and Hirota \cite{6-30} equations by DT. Interactional solutions including breather and dark soliton, breather and anti-dark soliton, dark soliton and anti-dark soliton were all constructed in  multi-component coupled nonlocal NLS equations \cite{6-31,6-32}. These results indicate that more novel and  abundant nonlinear localized waves may be obtained in the coupled systems than the ones in single-component systems. In this paper, we focus on constructing some novel interactional solutions in the following  three-component coupled derivative nonlinear Schr\"{o}dinger (DNLS) equations \cite{6-33,6-34}
 \begin{equation}\label{xt-6-1}
\begin{cases}
iq_{1t}+q_{1xx}-\frac{2i}{3}\epsilon[(|q_1|^2+|q_2|^2+|q_3|^2)q_1]_x=0,\\
iq_{2t}+q_{2xx}-\frac{2i}{3}\epsilon[(|q_1|^2+|q_2|^2+|q_3|^2)q_2]_x=0,\\
iq_{3t}+q_{3xx}-\frac{2i}{3}\epsilon[(|q_1|^2+|q_2|^2+|q_3|^2)q_3]_x=0.
\end{cases}
\end{equation}
Here, $q_1$, $q_2$ and $q_3$ are the complex envelops of three fields along the coordinate $x$ and $t$ is the time, each non-numeric subscripted variable denotes for partial differentiation and $\epsilon=\pm1$. The coupled system is relevant to the polarized Alfv$\acute{e}$n waves  in plasma physics and the ultra-short pulse field. The DNLS equations (\ref{xt-6-1}) govern the evolution equation of an Alfv$\acute{e}$n wave propagating along the magnetic field with weak nonlinearity and dispersion. In the ultra-short pulses field,  the width of optical pulse is in the order of femtosecond ($10^{-15}$ s) and the spectrum of these ultrashort pulses is approximately of the order $10^{15}s^{-1}$,  the NLS equation is less accurate. Here, the DNLS system can describe the propagation of the ultra-short pulses.

The RWs of the single-component DNLS equation were constructed by DT \cite{6-35,6-35-1,6-36}.  The higher-order semi-rational solutions of the single-component DNLS equation was obtained in \cite{6-36-1}, such as higher-order RWs interacting with higher-order breathers. There have been may other results about single-component DNLS equation, such as soliton solutions \cite{6-36-2}, stationary solutions \cite{6-36-3} and breather solutions \cite{6-36-4}, etc. Besides, the two-component coupled DNLS system was derived by Morris and Dodd \cite{6-33}. N-soliton solutions of the two-component case of the coupled system (\ref{xt-6-1}) was constructed by DT \cite{6-34}. In \cite{6-37}, Baronio et al. constructed the first-order interactional solutions of the two-component coupled NLS equations, which included first-order RW, first-order RW interacting with one-bright (dark) soliton and one-breather, respectively. Compared to first-order RW,  the higher-order RWs can describe RWs of significantly higher ratio of
peak to the background amplitude \cite{6-37-1}.  It is necessary to investigate the interactional solutions between higher-order RWs and other nonlinear localized waves. Combined a special vector solution of the Lax pair and generalized DT, we successfully constructed higher-order interactions of localized waves in some  multi-component coupled systems \cite{6-38,6-39,6-40,6-41,6-42}. Actually, we generalized Baronio's results in \cite{6-37} to higher-order cases in the same two-component NLS system \cite{6-38} and some other multi-component coupled systems \cite{6-39,6-40,6-41,6-42}. Compared to two-component systems \cite{6-38,6-39,6-42}, there can exist some novel and  interesting mixed interactions of localized waves among three different components in three-component ones \cite{6-40,6-41}. Here, we extend the two-component coupled DNLS equations in \cite{6-33} and \cite{6-34} to three-component case \cite{6-1}, and construct the corresponding Lax pair. Based on the fact that the DNLS system is important in plasma physics and the ultra-short pulse field, it is very necessary to  investigate the mixed interactions of localized waves for the three-component system (\ref{xt-6-1}).

In this paper,  hybrid (interactional) solution is defined by RW interacting with other two states (soliton or breather) in each component. Besides, it is named mixed hybrid (interactional) solution that different hybrid (interactional) solutions combine together in the three components $q_j~(j=1,2,3)$ at the same time. Choosing the appropriate periodic seed solutions of the system (\ref{xt-6-1}), a special vector solution of the corresponding Lax pair can be elaborately constructed. Based on this kind of special vector solution, some novel and interesting interactional solutions are derived in three-component coupled DNLS equations (\ref{xt-6-1}) through DT technique.  Various higher-order interactions of localized waves are constructed, among them the first- and second-order hybrid solutions are demonstrated in detail. Through defining the same combination as the same type solution among the three components $q_1$, $q_2$ and $q_3$, the first- and second-order interactional solutions are classified in six cases: (1) the interactional solutions degenerate to rational ones and these three components are all rogue waves; (2) two components are hybrid solutions between rogue wave and breather (RW$+$breather), and one component is hybrid solutions between RW and amplitude-varying soliton (RW$+$amplitude-varying soliton); (3) two components are RW$+$amplitude-varying soliton, and one component is RW$+$bright soliton; (4) two components are RW$+$breather, and one component is RW$+$bright soliton; (5) two components are RW$+$bright soliton, and one component is RW$+$amplitude-varying soliton; (6) three components are all RW$+$breather. We can find that the remaining four cases are all mixed interactional solutions in the coupled system (\ref{xt-6-1}) except for cases (1) and (6).

Furthermore, the disturbing parameters $\alpha$ and $\beta$ play an important role in controlling the dynamics of interacting process in different nonlinear localized waves. It can be found that  higher-order RWs merge with either multi-bright  soliton, multi-amplitude-varying soliton or multi-breather significantly by increasing the absolute values of   $\alpha$ and $\beta$.  These results received in our paper paovide evidence of some attractive interactions between higher-order RWs and multi-soliton or multi-breather.  These behaviors can also be interpreted as a
mechanism of generation of higher-order RWs out of multi slowly moving boomeronic solitons. Besides,  Baronio et al. gave the experimental conditions for observing this first-order interactional solutions including RW and one-bright (dark) soliton, RW and one-breather in two-component coupled NLS equations \cite{6-37}. We expect the higher-order localized waves presented in our work will be verified in
physical experiments in the future.  Additionally, we can draw a conclusion that these kinds of mixed interactions of localized waves may only be obtained by DT in the nonlinear systems,  whose components are more than 3 with the corresponding Lax pair including the matrices larger than $3\times3$.

This article is organized as follows. In Section 2, the DT of the three-component coupled DNLS equations is constructed. In Section 3, higher-order mixed interactional solutions are obtained. Especially, the first- and second-order hybrid solutions are discussed in detail. Besides, different nonlinear localized waves merge with each other by increasing the absolute values of $\alpha$ and $\beta$. The last section contains several conclusions and discussions.

\section{ Darboux transformation for the three-component DNLS equations}
For convenience, the parameter $\epsilon$ in the coupled system (\ref{xt-6-1}) is chosen as $\epsilon=-1$ in the following content. We extend two-component coupled DNLS equations \cite{6-33,6-34} to three-component case (\ref{xt-6-1}) and construct the corresponding Lax pair as follows
\begin{eqnarray}
&&\Phi_{x}=U\Phi=\lambda^2U_2+\lambda U_1,\label{xt-6-2}\\
&& \Phi_{t}=V\Phi=\lambda^4V_4+\lambda^3V_3+\lambda^2V_2+\lambda V_1,\label{xt-6-3}
\end{eqnarray}
with
\begin{eqnarray*}
&&U_2=\begin{pmatrix}-2I& &&\\&I&& \\&&I&\\&&&I\end{pmatrix},\quad U_1=\begin{pmatrix} 0&q_1&q_2&q_3\\-q_1^{*}&0&0&0\\-q_2^{*}&0&0&0\\ -q_2^{*}&0&0&0 \end{pmatrix}, \quad V_4=\begin{pmatrix}-9I& &&\\&0&& \\&&0&\\&&&0\end{pmatrix},\\
&& V_3=3U_1,\quad V_2=\begin{pmatrix}i(|q_1|^2+|q_2|^2+|q_3|^2)&0&0&0\\0&-i|q_1|^2&-iq_1^{*}q_2&-iq_1^{*}q_3\\0&-iq_2^{*}q_1&-i|q_2|^ 2&-iq_2^{*}q_3\\0&-iq_3^{*}q_1&-iq_3^{*}q_2&-i|q_3|^2 \end{pmatrix},\\
&&V_1=\begin{pmatrix}\begin{smallmatrix}0&iq_{1x}-\frac{2}{3}(|q_1|^2+|q_2|^2+|q_3|^2)q_1&iq_{2x}-\frac{2}{3}(|q_1|^2+|q_2|^2+|q_3|^2)q_2&iq_{3x}-\frac{2}{3}(|q_1|^2+|q_2|^2+|q_3|^2)q_3\\
                                      iq_{1x}^{*}+\frac{2}{3}(|q_1|^2+|q_2|^2+|q_3|^2)q_1^{*} &0&0&0\\
                                      iq_{2x}^{*}+\frac{2}{3}(|q_1|^2+|q_2|^2+|q_3|^2)q_2^{*} &0&0&0\\
                                       iq_{3x}^{*}+\frac{2}{3}(|q_1|^2+|q_2|^2+|q_3|^2)q_3^{*} &0&0&0 \end{smallmatrix} \end{pmatrix}.
\end{eqnarray*}
where $\Phi=(\phi(x,t),\varphi(x,t),\chi(x,t),\psi(x,t))^T$,~$T$ denotes the transpose of the vector and $\lambda$ is the spectral parameter. Additionally,  the three-component
coupled DNLS system (\ref{xt-6-1}) can be straightforwardly derived from the following compatibility condition $U_t-V_x+[U,V]=0$.

In \cite{6-34}, the DT of the two-component coupled DNLS equations was constructed and the corresponding DT was also generalized to the multi-component case. Based on the DT constructed in \cite{6-34}, the elementary DT of the coupled system (\ref{xt-6-1}) can be expressed as
\begin{eqnarray}
&&T=\frac{\lambda^2-\lambda_1^{*}}{\lambda_1^{*2}}I+\frac{\lambda^{*2}-\lambda_1^{2}}{\lambda_1^{*2}}N(\lambda)
\begin{pmatrix}\lambda\phi_{10}^{*}&\lambda_1^{*}\varphi_{10}^{*}&\lambda_1^{*}\chi_{30}^{*} &\lambda_1^{*}\psi_{10}^{*}\\
                         \lambda_1^{*}\phi_{10}^{*}&\lambda\varphi_{10}^{*}&\lambda\chi_{30}^{*}&\lambda\psi_{40}^{*}\\
                            \lambda_1^{*}\phi_{10}^{*}&\lambda\varphi_{10}^{*}&\lambda\chi_{30}^{*}&\lambda\psi_{40}^{*}\\
                            \lambda_1^{*}\phi_{10}^{*}&\lambda\varphi_{10}^{*}&\lambda\chi_{30}^{*}&\lambda\psi_{40}^{*}\end{pmatrix},\label{xt-6-4}\\
 &&q_1[1]=q_1-\frac{\lambda_1^{*2}-\lambda_1^2}{|\lambda_1|^2}\left(\frac{\phi_{10}\varphi_{10}^{*}}{\lambda_1|\phi_{10}|^2+\lambda_1^{*}(|\varphi_{10}|^2+|\chi_{10}|^2+|\psi_{10}|^2)}\right)_x,\label{xt-6-5}\\
&&q_2[1]=q_2-\frac{\lambda_1^{*2}-\lambda_1^2}{|\lambda_1|^2}\left(\frac{\phi_{10}\chi_{10}^{*}}{\lambda_1|\phi_{10}|^2+\lambda_1^{*}(|\varphi_{10}|^2+|\chi_{10}|^2+|\psi_{10}|^2)}\right)_x,\label{xt-6-6}\\
&&q_3[1]=q_3-\frac{\lambda_1^{*2}-\lambda_1^2}{|\lambda_1|^2}\left(\frac{\phi_{10}\psi_{10}^{*}}{\lambda_1|\phi_{10}|^2+\lambda_1^{*}(|\varphi_{10}|^2+|\chi_{10}|^2+|\psi_{10}|^2)}\right)_x,\label{xt-6-7}
\end{eqnarray}
where
\begin{eqnarray*}
&&I=\begin{pmatrix}1&&&\\&1&&\\&&1&\\&&&1\end{pmatrix},\quad N(\lambda)=\begin{pmatrix}\frac{\lambda\phi_{10}}{\lambda_1D}&&&\\ &\frac{\lambda\varphi_{10}}{\lambda_1D^{*}}&&\\ &&\frac{\lambda\chi_{10}}{\lambda_1D^{*}}&\\ &&&\frac{\lambda\psi_{10}}{\lambda_1D^{*}} \end{pmatrix},\\
&&D=\lambda_1|\phi_{10}|^2+\lambda_1^{*}(|\varphi_{10}|^2+|\chi_{10}|^2+|\psi_{10}|^2),
\end{eqnarray*}
here,  the  subscript $x$ in (\ref{xt-6-5})-(\ref{xt-6-7}) denotes for partial differentiation  and $\Phi_{10}=(\phi_{10}(x,t),\varphi_{10}(x,t),\chi_{10}(x,t),\psi_{10}(x,t))^T$ is the solution of the Lax pair  (\ref{xt-6-2}-\ref{xt-6-3})  with $\lambda=\lambda_1$.

Setting $\Phi_1(\lambda_1+\delta)=(\phi_1(x,t),\varphi_1(x,t),\chi_1(x,t),\psi_1(x,t))^T$ be a special vector eigenfunction of the Lax pair (\ref{xt-6-2}-\ref{xt-6-3}) with seed solution of the three-component coupled DNLS system (\ref{xt-6-1}) being chosen as $q_1=q_1[0],q_2=q_2[0],q_3=q_3[0]$ and  $\lambda=\lambda_1+\delta$. Expanding the column vector $\Phi_1(\lambda_1+\delta)$ at $\delta=0$, we have
\begin{eqnarray*}
\Phi_1=\Phi_1^{[0]}+\Phi_1^{[1]}\delta+\Phi_1^{[2]}\delta^2+\cdots+\Phi_1^{[N]}\delta^N+\cdots,
\end{eqnarray*}
where
\begin{center}
$\Phi_1^{[l]}=(\phi_1^{[l]},\varphi_1^{[l]},\chi_1^{[l]},\psi_1^{[l]})^T$,\quad $\Phi_1^{[l]}=\dfrac{1}{l!}\dfrac{\partial^l \Phi_1}{\partial \delta^l}|_{\delta=0}\quad(l=0,1,2,3\cdots)$.
\end{center}
It can be directly found that $\Phi_1^{[0]}$ is the special solution of the Lax pair  (\ref{xt-6-2}-\ref{xt-6-3}) with $q_1=q_1[0],q_2=q_2[0],q_3=q_3[0]$ and $\lambda=\lambda_1$. Based on the above facts, the one-fold generalized DT can be written as follows
\begin{eqnarray}
&&\Phi_1=T[1]\Phi, \quad T[1]=M_0[1]\lambda^2+M_1[1]\lambda-I,\\
&&q_1[1]=q_1[0]-\frac{\lambda_1^{*2}-\lambda_1^2}{|\lambda_1|^2}(\frac{\phi_1[0]\varphi_1[0]}{D_1})_x,\label{xt-6-11}\\
&&q_2[1]=q_2[0]-\frac{\lambda_1^{*2}-\lambda_1^2}{|\lambda_1|^2}(\frac{\phi_1[0]\chi_1[0]}{D_1})_x,\\
&&q_3[1]=q_3[0]-\frac{\lambda_1^{*2}-\lambda_1^2}{|\lambda_1|^2}(\frac{\phi_1[0]\psi_1[0]}{D_1})_x,\label{xt-6-12}
\end{eqnarray}
where
\begin{eqnarray*}
&&\Phi_1^{[0]}=(\phi_1^{[0]},\varphi_1^{[0]},\chi_1^{[0]},\psi_1^{[0]})^T=(\phi_1[0],\varphi_1[0],\chi_1[0],\psi_1[0])^T,\\
&&D_1=\lambda_1|\phi_1[0]|^2+\lambda_1^{*}(|\varphi_1[0]|^2+|\chi_1[0]|^2+|\psi_1[0]|^2),\\
&&M_0[1]=\frac{1}{\lambda_1^{*2}}I+\frac{\lambda_1^{*2}-\lambda_1^2}{\lambda_1^{*2}\lambda_1}\begin{pmatrix}\dfrac{\phi_1[0]\phi_1[0]^{*}}{D_1}&0&0&0\\
0&\dfrac{\varphi_1[0]\varphi_1[0]^{*}}{D_1^{*}} & \dfrac{\varphi_1[0]\chi_1[0]^{*}}{D_1^{*}}&\dfrac{\varphi_1[0]\psi_1[0]^{*}}{D_1^{*}}\\
0&\dfrac{\chi_1[0]\varphi_1[0]^{*}}{D_1^{*}} & \dfrac{\chi_1[0]\chi_1[0]^{*}}{D_1^{*}}&\dfrac{\chi_1[0]\psi_1[0]^{*}}{D_1^{*}}\\
0&\dfrac{\psi_1[0]\varphi_1[0]^{*}}{D_1^{*}} & \dfrac{\psi_1[0]\chi_1[0]^{*}}{D_1^{*}}&\dfrac{\psi_1[0]\psi_1[0]^{*}}{D_1^{*}}
 \end{pmatrix},\\
 &&M_1[1]=\frac{\lambda_1^{*2}-\lambda_1^2}{\lambda_1\lambda_1^{*}}\begin{pmatrix} 0&\dfrac{\phi_1[0]\varphi_1[0]^{*}}{D_1}&\dfrac{\phi_1[0]\chi_1[0]^{*}}{D_1}&\dfrac{\phi_1[0]\psi_1[0]^{*}}{D_1} \\
 \dfrac{\varphi_2[0]\phi_1[0]^{*}}{D_1^{*}}&0&0&0\\
  \dfrac{\chi_2[0]\phi_1[0]^{*}}{D_1^{*}}&0&0&0\\
   \dfrac{\psi_2[0]\phi_1[0]^{*}}{D_1^{*}}&0&0&0 \end{pmatrix},
\end{eqnarray*}

Considering this kind of  limit
\begin{eqnarray*}
\lim\limits_{\delta\rightarrow{0}}\dfrac{T[1]|_{\lambda=\lambda_1+\delta}\Phi_1}{\delta}=\lim\limits_{\delta\rightarrow{0}}\frac{M_0[1]\lambda^2+M_1[1]\lambda-I}{\delta}
=T_1[1]\Phi_1^{[1]}+(2\lambda_1M_0[1]+M_1[1])\Phi_1^{[0]}\equiv\Phi_1[1],
\end{eqnarray*}
the following equality $T_1[1]\Phi_1^{[0]}{=}T[1]|_{\lambda{=}\lambda_1}\Phi_1^{[0]}{=}(M_0[1]\lambda_1^2{+}M_1[1]\lambda_1{-}I)\Phi_1^{[0]}{=}0$ has been utilized in the above process. Besides, we can find that $\Phi_1[1]$ is a solution of the Lax pair  (\ref{xt-6-2}-\ref{xt-6-3}) under the following condition $q_1=q_1[1],q_2=q_2[1],q_2=q_3[1]$ and $\lambda=\lambda_1$. Then the two-fold generalized DT can be expressed as follows
\begin{eqnarray}
&&\Phi_2=T[2]T[1]\Phi, \quad T[2]=M_0[2]\lambda^2+M_1[2]\lambda-I,\\
&&q_1[2]=q_1[1]-\frac{\lambda_1^{*2}-\lambda_1^2}{|\lambda_1|^2}(\frac{\phi_1[1]\varphi_1[1]}{D_2})_x,\label{xt-6-17}\\
&&q_2[2]=q_2[1]-\frac{\lambda_1^{*2}-\lambda_1^2}{|\lambda_1|^2}(\frac{\phi_1[1]\chi_1[1]}{D_2})_x,\\
&&q_3[2]=q_3[1]-\frac{\lambda_1^{*2}-\lambda_1^2}{|\lambda_1|^2}(\frac{\phi_1[1]\psi_1[1]}{D_2})_x,\label{xt-6-18}
\end{eqnarray}
where
\begin{eqnarray*}
&&\Phi_1[1]=(\phi_1[1],\varphi_1[1],\chi_1[1],\psi_1[1])^T, T_1[1]{=}T[1]|_{\lambda=\lambda_1}{=}M_0[1]\lambda_1^2+M_1[1]\lambda_1-I,\\
&&D_2=\lambda_1|\phi_1[1]|^2+\lambda_1^{*}(|\varphi_1[1]|^2+|\chi_1[1]|^2+|\psi_1[1]|^2),\\
&&M_0[2]=\frac{1}{\lambda_1^{*2}}I+\frac{\lambda_1^{*2}-\lambda_1^2}{\lambda_1^{*2}\lambda_1}\begin{pmatrix}\dfrac{\phi_1[1]\phi_1[1]^{*}}{D_2}&0&0&0\\
0&\dfrac{\varphi_1[1]\varphi_1[1]^{*}}{D_2^{*}} & \dfrac{\varphi_1[1]\chi_1[1]^{*}}{D_2^{*}}&\dfrac{\varphi_1[1]\psi_1[1]^{*}}{D_2^{*}}\\
0&\dfrac{\chi_1[1]\varphi_1[1]^{*}}{D_2^{*}} & \dfrac{\chi_1[1]\chi_1[1]^{*}}{D_2^{*}}&\dfrac{\chi_1[1]\psi_1[1]^{*}}{D_2^{*}}\\
0&\dfrac{\psi_1[1]\varphi_1[1]^{*}}{D_2^{*}} & \dfrac{\psi_1[1]\chi_1[1]^{*}}{D_2^{*}}&\dfrac{\psi_1[1]\psi_1[1]^{*}}{D_2^{*}}
 \end{pmatrix},\\
 &&M_1[2]=\frac{\lambda_1^{*2}-\lambda_1^2}{\lambda_1\lambda_1^{*}}\begin{pmatrix} 0&\dfrac{\phi_1[1]\varphi_1[1]^{*}}{D_2}&\dfrac{\phi_1[1]\chi_1[1]^{*}}{D_2}&\dfrac{\phi_1[1]\psi_1[1]^{*}}{D_2} \\
 \dfrac{\varphi_2[1]\phi_1[1]^{*}}{D_2^{*}}&0&0&0\\
  \dfrac{\chi_2[1]\phi_1[1]^{*}}{D_2^{*}}&0&0&0\\
   \dfrac{\psi_2[1]\phi_1[1]^{*}}{D_2^{*}}&0&0&0 \end{pmatrix}.
   \end{eqnarray*}

   Iterating the above procedures, the $N$-fold generalized DT of the three-component coupled DNLS equations (\ref{xt-6-1}) can be derived as follows
   \begin{eqnarray}
&&\Phi_N=T[N]T[N-1] \cdots T[2]T[1]\Phi, \quad T[j]=M_0[j]\lambda^2+M_1[j]\lambda-I~(j=1,2,\cdots,N),\\
&&q_1[N]=q_1[N-1]-\frac{\lambda_1^{*2}-\lambda_1^2}{|\lambda_1|^2}(\frac{\phi_1[N-1]\varphi_1[N-1]}{D_N})_x,\label{xt-6-9}\\
&&q_2[N]=q_2[N-1]-\frac{\lambda_1^{*2}-\lambda_1^2}{|\lambda_1|^2}(\frac{\phi_1[N-1]\chi_1[N-1]}{D_N})_x,\\
&&q_3[N]=q_3[N-1]-\frac{\lambda_1^{*2}-\lambda_1^2}{|\lambda_1|^2}(\frac{\phi_1[N-1]\psi_1[N-1]}{D_N})_x,\label{xt-6-10}
\end{eqnarray}
where
\begin{eqnarray}
 &&  \hspace{-0.7cm}T_1[j]=T[j]|_{\lambda=\lambda_1}=M_0[j]\lambda_1^2+M_1[j]\lambda_1-I,\\
&& \hspace{-0.7cm}\Phi_1[N-1]=(\phi_1[N-1],\varphi_1[N-1],\chi_1[N-1],\psi_1[N-1])^T=\lim\limits_{\delta\rightarrow{0}}\dfrac{\Phi_N}{\delta^{N}},\label{xt-6-8}\\
&&\hspace{-0.7cm}D_N=\lambda_1|\phi_1[N-1]|^2+\lambda_1^{*}(|\varphi_1[N-1]|^2+|\chi_1[N-1]|^2+|\psi_1[N-1]|^2),\\
&&\hspace{-0.7cm}M_0[j]{=}\frac{1}{\lambda_1^{*2}}I{+}\frac{\lambda_1^{*2}{-}\lambda_1^2}{\lambda_1^{*2}\lambda_1}\begin{pmatrix}\frac{\phi_1[j{-}1]\phi_1[j{-}1]^{*}}{D_j}&0&0&0\\
0&\frac{\varphi_1[j{-}1]\varphi_1[j{-}1]^{*}}{D_j^{*}} & \frac{\varphi_1[j{-}1]\chi_1[j{-}1]^{*}}{D_j^{*}}&\frac{\varphi_1[j{-}1]\psi_1[j{-}1]^{*}}{D_j^{*}}\\
0&\frac{\chi_1[j{-}1]\varphi_1[j{-}1]^{*}}{D_j^{*}} & \frac{\chi_1[j{-}1]\chi_1[j{-}1]^{*}}{D_j^{*}}&\frac{\chi_1[j{-}1]\psi_1[j{-}1]^{*}}{D_j^{*}}\\
0&\frac{\psi_1[j{-}1]\varphi_1[j{-}1]^{*}}{D_j^{*}} & \frac{\psi_1[j{-}1]\chi_1[j{-}1]^{*}}{D_j^{*}}&\frac{\psi_1[j{-}1]\psi_1[j{-}1]^{*}}{D_j^{*}}
 \end{pmatrix},\\
 &&\hspace{-0.7cm}M_1[j]=\frac{\lambda_1^{*2}-\lambda_1^2}{\lambda_1\lambda_1^{*}}\begin{pmatrix} 0&\frac{\phi_1[j-1]\varphi_1[j-1]^{*}}{D_j}&\frac{\phi_1[j-1]\chi_1[j-1]^{*}}{D_j}&\frac{\phi_1[j-1]\psi_1[j-1]^{*}}{D_j} \\
 \frac{\varphi_2[j-1]\phi_1[j-1]^{*}}{D_j^{*}}&0&0&0\\
  \frac{\chi_2[j-1]\phi_1[j-1]^{*}}{D_j^{*}}&0&0&0\\
   \frac{\psi_2[j-1]\phi_1[j-1]^{*}}{D_j^{*}}&0&0&0 \end{pmatrix}.
   \end{eqnarray}

   Owing to  complexity and irregularity of the vector solution $\Phi_1[N-1]$ in Eq. (\ref{xt-6-8}),  the compact iterative formula cannot be derived directly, but the arbitrary $\Phi_1[j]~(j=1,2,3,\cdots,N)$ can be calculated through the formula (\ref{xt-6-8}). Additionally, the $N$-order mixed interactions of localized waves of the three-component coupled DNLS system (\ref{xt-6-1}) can be generated via the formulae (\ref{xt-6-9})-(\ref{xt-6-10}). In order to avoid tedious calculation of determinant with high order matrix and utilize the symbolic computation, the iteration form of the DT for the coupled system (\ref{xt-6-1}) is chosen. Furthermore, the expressions of higher-order interactional solutions of the coupled system (\ref{xt-6-1}) are complicated and tedious, then the first- and second-order interactional solutions are discussed in detail.

   \section{Higher-order mixed  interactions of localized waves}  
In order to construct various interactional solutions of the three-component DNLS equations (\ref{xt-6-1}) ,  a general nontrivial seed solution  can be chosen as
\begin{equation}
q_1[0]=d_1e^{-\tfrac{2}{3}i\theta},\quad q_2[0]=d_2e^{-\tfrac{2}{3}i\theta},\quad q_3[0]=d_3e^{-\tfrac{2}{3}i\theta}, \label{xt-6-13}
\end{equation}
where $\theta=(d_1^2+d_2^2+d_3^2)x$, $d_j~(j=1,2,3)$ are all real constants. For convenience, the above seed solution is chosen periodically in space variable $x$ without depending on time variable $t$. Starting with the above seed solution $q_1=q_1[0],q_2=q_2[0],q_3=q_3[0]$ with the spectral parameter $\lambda$, then the special vector solution of the Lax pair (\ref{xt-6-2}-\ref{xt-6-3}) can be elaborately derived as follows
\begin{eqnarray}
\Phi_1=\begin{pmatrix} (l_1e^{M_1+M_2}-l_2e^{M_1-M_2})e^{-\tfrac{i}{3}\theta}\\
                                        \rho_1(l_1e^{M_1-M_2}-l_2e^{M_1+M_2})e^{\tfrac{i}{3}\theta}-(\alpha d_2+\beta d_3)e^{ix\lambda^2}\\
                                           \rho_2(l_1e^{M_1-M_2}-l_2e^{M_1+M_2})e^{\tfrac{i}{3}\theta}+\alpha d_1e^{ix\lambda^2}\\
                                           \rho_3(l_1e^{M_1-M_2}-l_2e^{M_1+M_2})e^{\tfrac{i}{3}\theta}+\beta d_1e^{ix\lambda^2}  \end{pmatrix},\label{xt-6-14}
\end{eqnarray}
where
\begin{eqnarray*}
&&l_1=\frac{i(9\lambda^2-2\tau-\sqrt{81\lambda^4+4\tau^2})^{\tfrac{1}{2}}}{\sqrt{81\lambda^4+4\tau^2}}, \quad l_2=\frac{i(9\lambda^2-2\tau+\sqrt{81\lambda^4+4\tau^2})^{\tfrac{1}{2}}}{\sqrt{81\lambda^4+4\tau^2}},\\
&&M_1=-\frac{i}{2}\lambda^2(x+9\lambda^2t),\quad M_2=\frac{i}{6}\sqrt{81\lambda^4+4\tau^2}(x+3\lambda^2t+\sum_{k=1}^{N}s_k\epsilon^{2k}),\\
&&\rho_1=\frac{d_1}{\sqrt{\tau}},\quad \rho_2=\frac{d_2}{\sqrt{\tau}},\quad \rho_3=\frac{d_3}{\sqrt{\tau}},\quad s_k=m_k+in_k, \quad \tau=d_1^2+d_2^2+d_3^2,
\end{eqnarray*}
the parameters $m_j,n_j~(j=1,2,\cdots,N), \alpha$ and $\beta$ are all real free constants.

For generating the special vector solution (\ref{xt-6-14}), we should convert the variable coefficient differential expressions in the corresponding Lax pair (\ref{xt-6-2})-(\ref{xt-6-3}) with the seed solution  (\ref{xt-6-13}) to constant coefficient ones by a gauge transformation. Setting the gauge transformation as $\Psi=N\Phi$, the Lax pair (\ref{xt-6-2})-(\ref{xt-6-3}) can be transformed to
\begin{eqnarray*}
&&\Psi_x=U_0\Psi=(N_xN^{-1}+NUN^{-1})\Psi,\\
&&\Psi_t=V_0\Psi=(N_tN^{-1}+NVN^{-1})\Psi,
\end{eqnarray*}
where $N$${=}$diag$(e^{\tfrac{i}{2}\theta},e^{-\tfrac{i}{6}\theta},e^{-\tfrac{i}{6}\theta},e^{-\tfrac{i}{6}\theta})$. It is easily found that the characteristic equation of $U_0$ is a quartic equation. In order to generate some novel interactional solutions of the coupled system (\ref{xt-6-1}), the case that  the  characteristic equation of $U_0$ owing two group double roots is considered. Furthermore, all solutions in the fundamental solution of $U_0$ and $V_0$ are included in Eq. (\ref{xt-6-14}). If one constructs the RWs of the coupled system (\ref{xt-6-1}), the following expressions $-(\alpha d_2+\beta d_3)e^{ix\lambda^2}$, $\alpha d_1e^{ix\lambda^2}$ and $\beta d_1e^{ix\lambda^2}$ are not needed. Here, the above three expressions are necessary to generate some interactional solutions. We call these three expressions disturbing terms of the special vector solution (\ref{xt-6-14}) and the parameters $\alpha$ and $\beta$ the corresponding  disturbing coefficients. Considering the seed solution to be vanished  ($d_j=0~(j=1,2,3)$) or non-vanished ($d_j\neq0$) background, and the disturbing coefficients $\alpha$ and $\beta$ to be zero (with  disturbing terms) or non-zero (without  disturbing terms), we expect to construct various mixed interactions of localized waves of the coupled system (\ref{xt-6-1}).

Here, we choose the spectral parameter $\lambda=\frac{\sqrt{\tau}}{3}(1+i+f^2)$ with arbitrary small parameter $f$, and expand the Taylor expansion of the special vector function $\Phi_1$ in (\ref{xt-6-14}) at $f=0$ as follows
\begin{equation*}
\Phi_1=\Phi_1^{[0]}+\Phi_1^{[1]}f^2+\Phi_1^{[2]}f^4+\cdots+\Phi_1^{[j]}f^{2j}+\cdots,
\end{equation*}
with
\begin{equation*}
\Phi_1^{[j]}=(\phi_1^{[j]},\varphi_1^{[j]},\chi_1^{[j]},\psi_1^{[j]})^T=\dfrac{1}{(2j)!}\dfrac{\partial^l \Phi_1}{\partial f^j}|_{f=0}\quad(j=0,1,2,3\cdots),
\end{equation*}
and
\begin{eqnarray*}
&&\phi_1^{[0]}=\frac{6\tau(1-i)x+4\tau^2(1+i)t-9i}{9\sqrt{(-2+2i)\tau}}e^{\tfrac{\tau}{9}[(1-3i)x+2i\tau t]},\\
&&\varphi_1^{[0]}=\left( -\alpha\,d_{{2}}-\beta\,d_{{3}} \right) {{\rm e}^{-\frac{2}{9}\,\tau\,
x}}+{\frac { \left( 1+i \right) d_{{1}} \left( 12\,i\tau\,x-8
\,{\tau}^{2}t-9-9\,i \right) {{\rm e}^{\tfrac{\tau}{9}\,[ \left( 1+3\,i \right) x
+2\,i\tau\,t]}}}{18\sqrt {\tau}\sqrt { \left( -2+2\,i \right) \tau}}},\\
&&\chi_1^{[0]}=\alpha\,d_{{1}}{{\rm e}^{-\tfrac{2}{9}\,\tau\,x}}+{\frac { \left( 1+i
 \right) d_{{2}} \left( 12\,i\tau\,x-8\,{\tau}^{2}t-9-9\,i \right) {
{\rm e}^{\tfrac{\tau}{9}[ \left( 1+3\,i \right) x+2\,i\tau\,t]}}}{18\sqrt {\tau}
\sqrt { \left( -2+2\,i \right) \tau}}},\\
&&\psi_1^{[0]}=\beta\,d_{{1}}{{\rm e}^{-\tfrac{2}{9}\tau\,x}}+{\frac { \left( 1+i
 \right) d_{{3}} \left( 12\,i\tau\,x-8\,{\tau}^{2}t-9-9\,i \right) {
{\rm e}^{\tfrac{\tau}{9}[ \left( 1+3\,i \right) x+2\,i\tau\,t]}}}{18\sqrt {\tau}
\sqrt { \left( -2+2\,i \right) \tau}}},\\
&&\phi_1^{[1]}={\frac {i}{17496\sqrt { \left( -2+2\,i \right) \tau}}}[{-}864\,{\tau}^{3}{x}^{3}{+}({-}1728\,it{\tau}^{4}{+}1944\,i{\tau}^{2}{-}
4536\,{\tau}^{2}) {x}^{2}{+}( 1152\,{\tau}^{5}{t}^{2}{-}14688
\,it{\tau}^{3}-2592\,{\tau}^{3}t{+}\\
&&\hspace{0.5cm}4860\,i\tau{-}7776\,\tau) x{+}256\,i{\tau}^{6}{t}^{3}{-}864\,i{t}^{2}{\tau}^{4}{+}7776\,{\tau}^{4}{t}^{2}{-}27216\,i{\tau}^{2}t{-}11664\,i\tau\,m_{{1}}{-}11664\,i\tau\,n_{{1}}{-}9720\,{\tau}^{2}t{-}\\
&&\hspace{0.5cm}11664\,\tau\,m_{{1}}{+}11664\,\tau\,n_{{1}}{+}2187{-}6561i]{{\rm e}^{\tfrac{1}{9}\,\tau\,[ \left( 1-3\,i \right) x+2\,i\tau\,t]}},\\
&&\varphi_1^{[1]}=\frac{2}{9}(1-i) \tau\,x( \alpha\,d_{{2}}+\beta\,d_{{3}}) {{\rm e}^{-\tfrac{2}{9}\,\tau\,x}}-{\frac {1}{17496}}\,{\frac {\Omega\,d_{{1}}{{\rm e}^{\tfrac{1}{9}\,\tau\,[
 \left( 1+3\,i \right) x+2\,it\tau]}}}{\sqrt {\tau}\sqrt { \left( -2+2
\,i \right) \tau}}},\\
&&\chi_1^{[1]}= \frac{2}{9}(-1+i) \tau\,\alpha\,xd_{{1}}{{\rm e}^{-\tfrac{2}{9}\,\tau
\,x}}-{\frac {1}{17496}}\,{\frac {d_{{2}}\Omega\,{{\rm e}^{\tfrac{1}{9}\,\tau\,
[ \left( 1+3\,i \right) x+2\,it\tau]}}}{\sqrt {\tau}\sqrt { \left( -2+
2\,i \right) \tau}}},\\
&&\psi_1^{[1]}=\frac{2}{9}(-1+i)\tau\,\beta\,xd_{{1}}{{\rm e}^{-\tfrac{2}{9}\,\tau
\,x}}-{\frac {1}{17496}}\,{\frac {d_{{3}}\Omega\,{{\rm e}^{\tfrac{1}{9}\,\tau\,
[ \left( 1+3\,i \right) x+2\,it\tau]}}}{\sqrt {\tau}\sqrt { \left( -2+
2\,i \right) \tau}}},\\
&&\Omega={-}864\,i{\tau}^{3}{x}^{3}{+}(1728\,{\tau}^{4}t{-}648\,i{\tau}^{2}{+}1944\,{\tau}^{2}) {x}^{2}{+}( 1152\,i{t}^{2}{\tau}^{5}{+}2592\,it{\tau}^{3}{+}9504\,{\tau}^{3}t{-}3888\,i\tau{-}972\,\tau) x{-}256\,{\tau}^{6}{t}^{3}\\
&&\hspace{0.5cm}{+}6048\,i{t}^{2}{\tau}^{4}{-}864\,{\tau}^{4}{t}^{2}{+}5832
\,i{\tau}^{2}t{-}11664\,i\tau\,m_{{1}}{+}11664\,i\tau\,n_{{1}}{+}11664\,{\tau}^{2}t{+}11664\,\tau\,m_{{1}}{+}11664\,\tau\,n_{{1}}{-}6561{-}2187\,i,\\
&&\cdots.
\end{eqnarray*}

Through the formulae (\ref{xt-6-11})-(\ref{xt-6-12}), the general expressions of the first-order interactions of localized waves of the three-component coupled DNLS equations (\ref{xt-6-2}) can be derived as
\begin{eqnarray}
&&q_1[1]=d_1e^{-\tfrac{2}{3}\theta}+2i\left(\frac{6d_1\tau^{\tfrac{3}{2}}F_1e^{-\tfrac{2}{3}\theta}+54(1+i)(\alpha d_2+\beta d_3)\tau^{\tfrac{3}{2}} F_2e^{\eta_1}}{-(1+i)\sqrt{2}G_1+324\tau^2(1-i)G_2e^{\eta_2}}\right)_x,\label{xt-6-15}\\
&&q_2[1]=d_2e^{\tfrac{2}{3}\theta}+2i\left(\frac{6d_2\tau^{\tfrac{3}{2}}F_1e^{-\tfrac{2}{3}\theta}-54(1+i)d_1\alpha \tau^{\tfrac{3}{2}} F_2e^{\eta_1}}{-(1+i)\sqrt{2}G_1+324\tau^2(1-i)G_2e^{\eta_2}}\right)_x,\\
&&q_3[1]=d_3e^{\tfrac{2}{3}\theta}+2i\left(\frac{6d_3\tau^{\tfrac{3}{2}}F_1e^{-\tfrac{2}{3}\theta}-54(1+i)d_1\beta \tau^{\tfrac{3}{2}} F_2e^{\eta_1}}{-(1+i)\sqrt{2}G_1+324\tau^2(1-i)G_2e^{\eta_2}}\right)_x,\label{xt-6-16}
\end{eqnarray}
where
\begin{eqnarray*}
&&\eta_1=\frac{\tau}{9}[(-3+3i)x+2i\tau],\quad \eta_2=-\frac{2}{3}\tau x, \\
&&F_1={\frac { \left( 12\,ix\tau+8\,{\tau}^{2}t+9-9\,i \right)  \left( 12\,i
x\tau-8\,{\tau}^{2}t+9+9\,i \right) }{\sqrt { \left( -2-2\,i \right)
\tau}\tau\,\sqrt {-2+2\,i}}},\quad F_2={\frac {12\,ix\tau-8\,{\tau}^{2}t+9+9\,i}{\sqrt { \left( -2+2\,i\right) \tau}}},\\
&& G_1=(72\,i{\tau}^{2}{d_{{1}}}^{2}{+}72\,i{\tau}^{2}{d_{{2}}}^{2}{+}72\,
i{\tau}^{2}{d_{{3}}}^{2}{-}72\,{\tau}^{3}) {x}^{2}{-}108\,\tau\,( i{d_{{1}}}^{2}{+}i{d_{{2}}}^{2}{+}i{d_{{3}}}^{2}{+}\tau) x{+}32\,{t}^{2}{\tau}^{4}( i{d_{{1}}}^{2}{+}i{d_{{2}}}^{2}{+}i{d_{{3}}}^{2}{-}\tau)\\
&&{+}72\,t{\tau}^{2}( i{d_{{1}}}^{2}{+}i{d_{{2}}}^{2}{+}i{d_{{3}}}^{2}{+}\tau) {+}81(i{d_{{1}}}^{2}{+}i{d_{{2}}}^{2}{+}i{d_{{3}}}^{2}{-}\tau),\quad G_2={\alpha}^{2}{d_{{1}}}^{2}+{\alpha}^{2}{d_{{2}}}^{2}+2\,\alpha\,\beta\,
d_{{2}}d_{{3}}+{\beta}^{2}{d_{{1}}}^{2}+{\beta}^{2}{d_{{3}}}^{2}.
\end{eqnarray*}

From the first-order general solutions (\ref{xt-6-15})-(\ref{xt-6-16}), we can find that they are multi-parametric and semi-rational solutions of the three-component coupled DNLS system (\ref{xt-6-1}). Besides, there are five free parameters in the first-order solutions, such as $d_j~(j=1,2,3)$, $\alpha$ and $\beta$. Here, the parameters $\alpha$ and $\beta$ are called  disturbing coefficients, which determine these three disturbing terms $-(\alpha d_2+\beta d_3)e^{ix\lambda^2}$, $\alpha d_1e^{ix\lambda^2}$ and $\beta d_1e^{ix\lambda^2}$ are added or not added in the original RWs solutions. Meanwhile, these three parameters $d_1$, $d_2$ and $d_3$  denote for the backgrounds on which different localized waves emerge, for example, $d_j=0~(j=1,2,3)$ denote for the vanished backgrounds  whereas $d_j\neq0$ denote for the non-vanished backgrounds.  Considering both disturbing coefficients and different backgrounds, these first-order general localized wave solutions are discussed in the following six different mixed cases. Besides, the disturbing coefficients $\alpha$ and $\beta$ play an important role in controlling the dynamics of interactions of different nonlinear waves.

\begin{figure}[H]
\renewcommand{\figurename}{{Fig.}}
\subfigure[]{\includegraphics[height=0.25\textwidth]{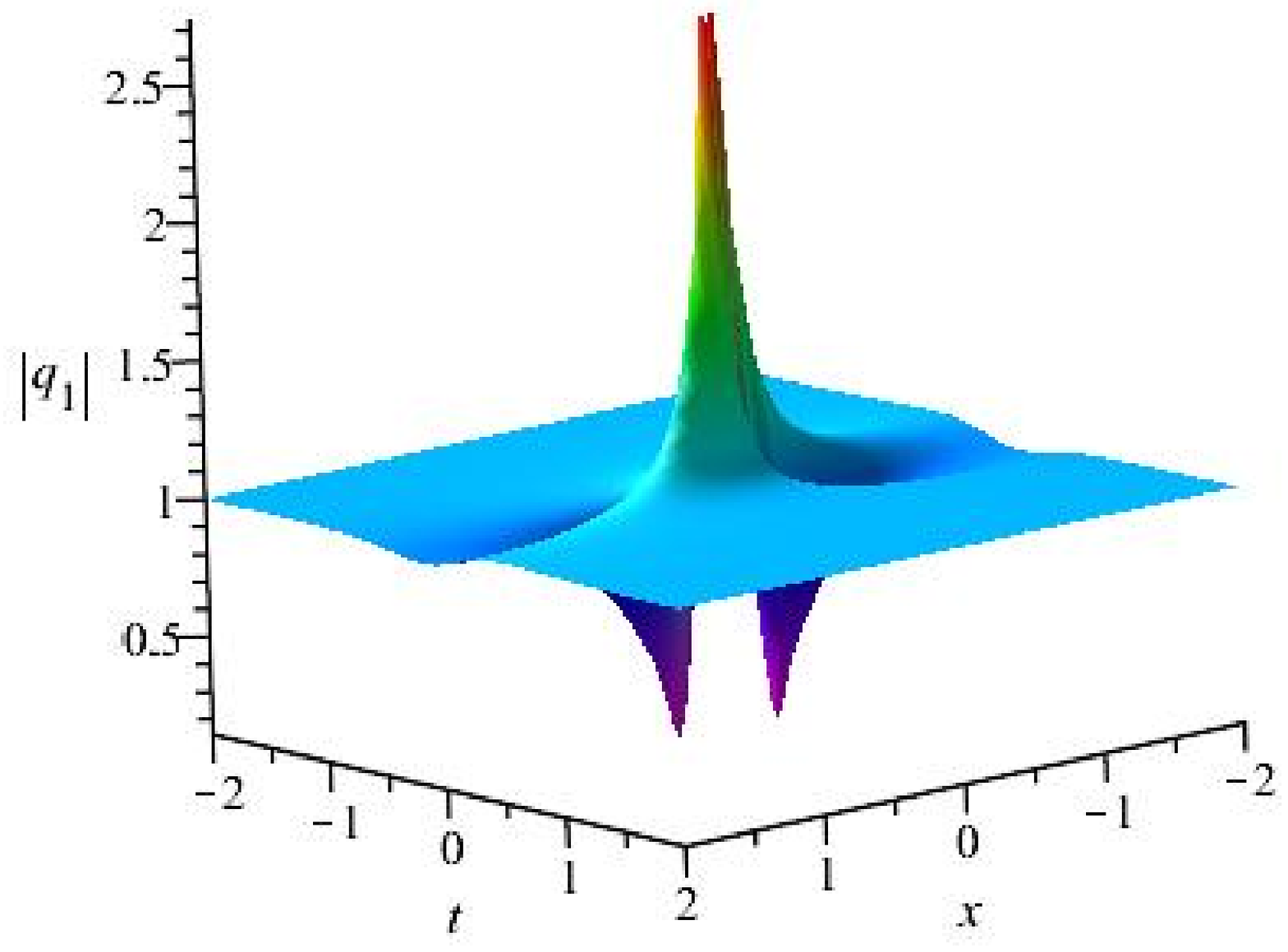}}
\centering
\subfigure[]{\includegraphics[height=0.25\textwidth]{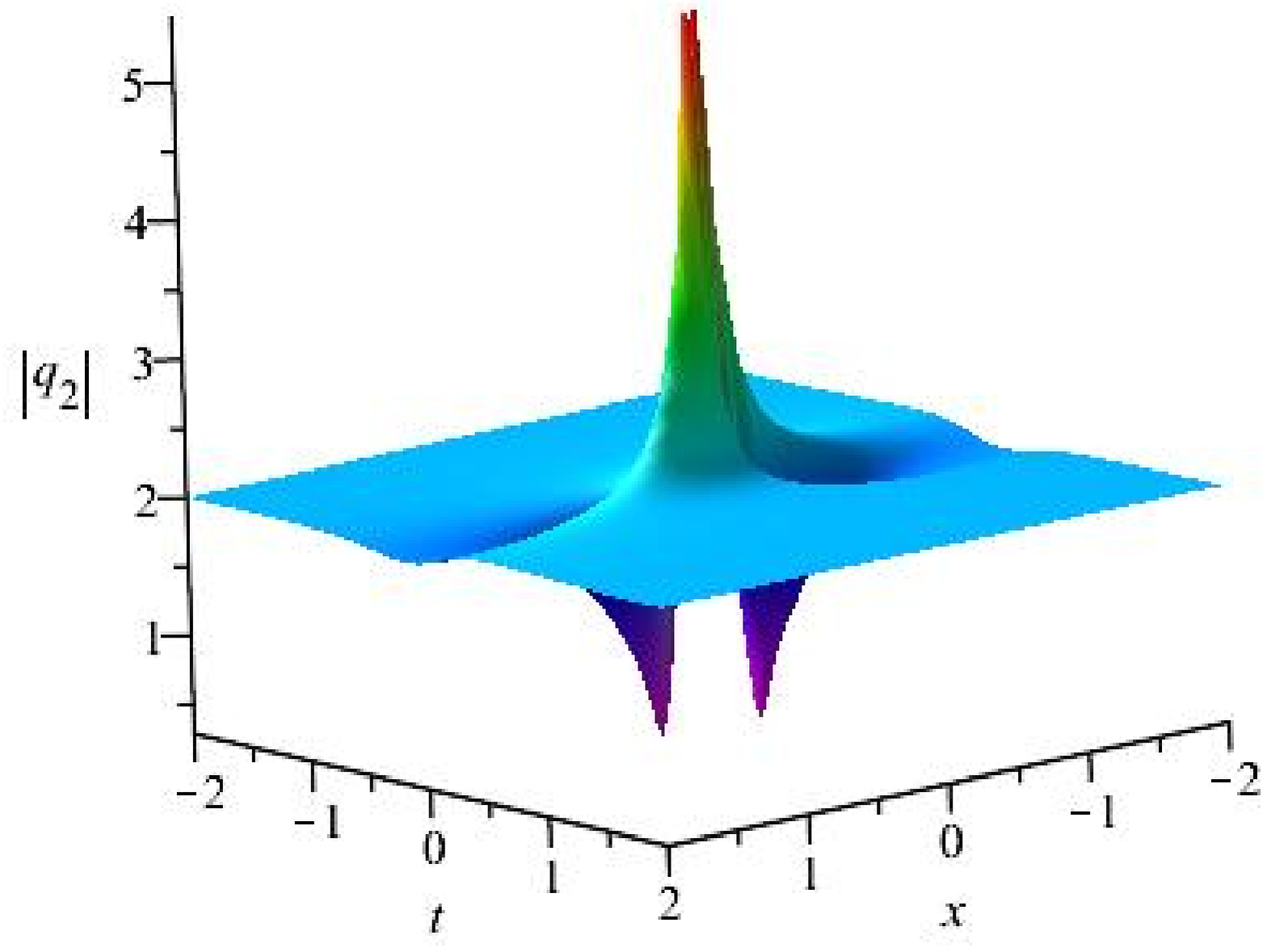}}
\centering
\subfigure[]{\includegraphics[height=0.25\textwidth]{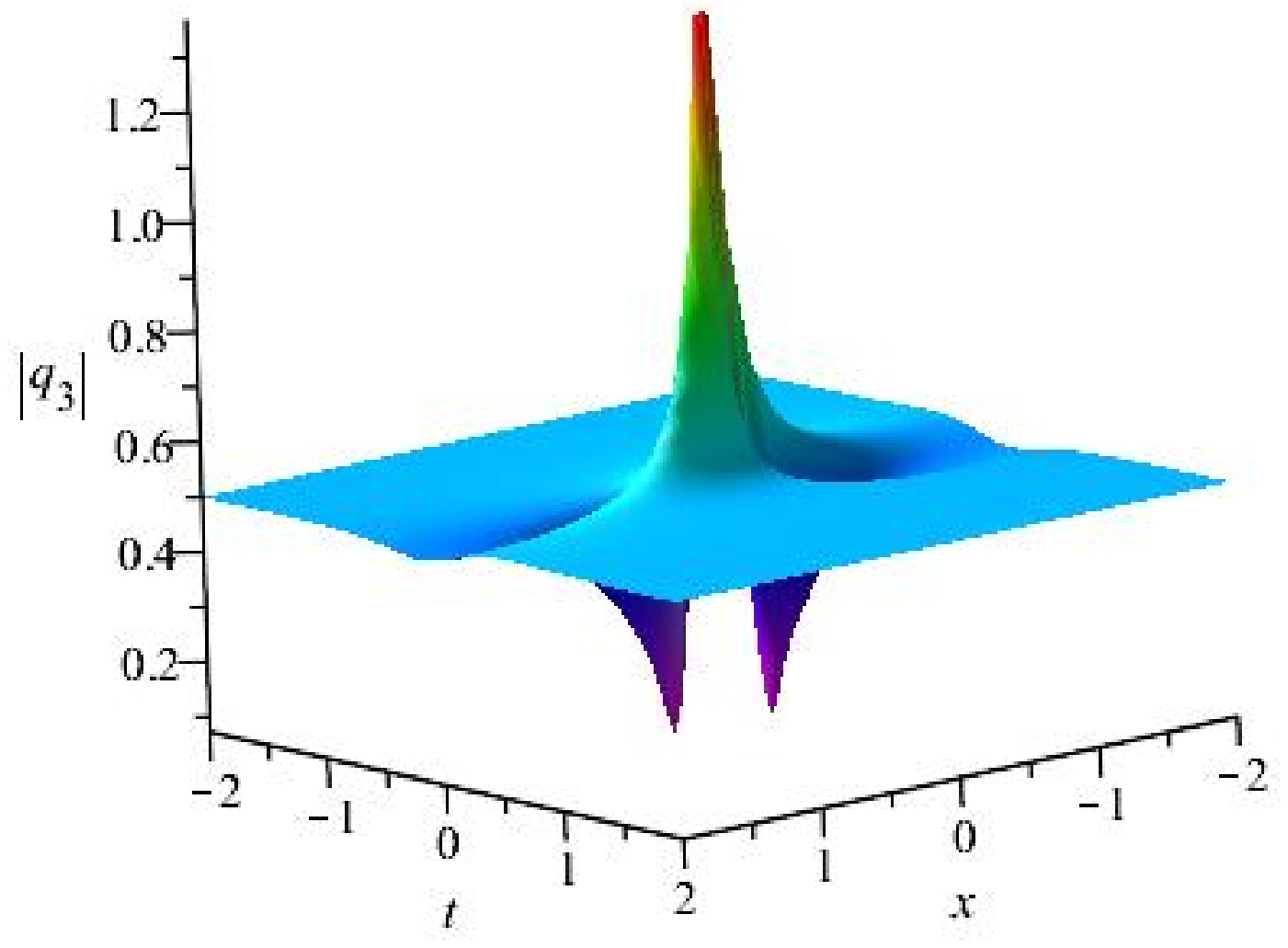}}
\centering
\caption{\small(Color online) Evolution plot of the first-order RW in the three-component coupled DNLS equations with the parameters chosen by $d_1=1, d_2=-2, d_3=\tfrac{1}{2},\alpha=\beta=0$: (a) $q_1$, (b) $q_2$, (c) $q_3$.\label{xt-6f-1}}
\end{figure}

(\textrm{i}) Getting ride of all  the disturbing terms in  (\ref{xt-6-15})-(\ref{xt-6-16}), namely $\alpha=\beta=0$, the first-order interactional solutions degenerate to rational solutions. Choosing $d_j\neq0~(j=1,2,3)$ , these three components are proportional to each other and they are all the first-order RWs, see Fig. \ref{xt-6f-1}.

\begin{figure}[H]
\renewcommand{\figurename}{{Fig.}}
\subfigure[]{\includegraphics[height=0.3\textwidth]{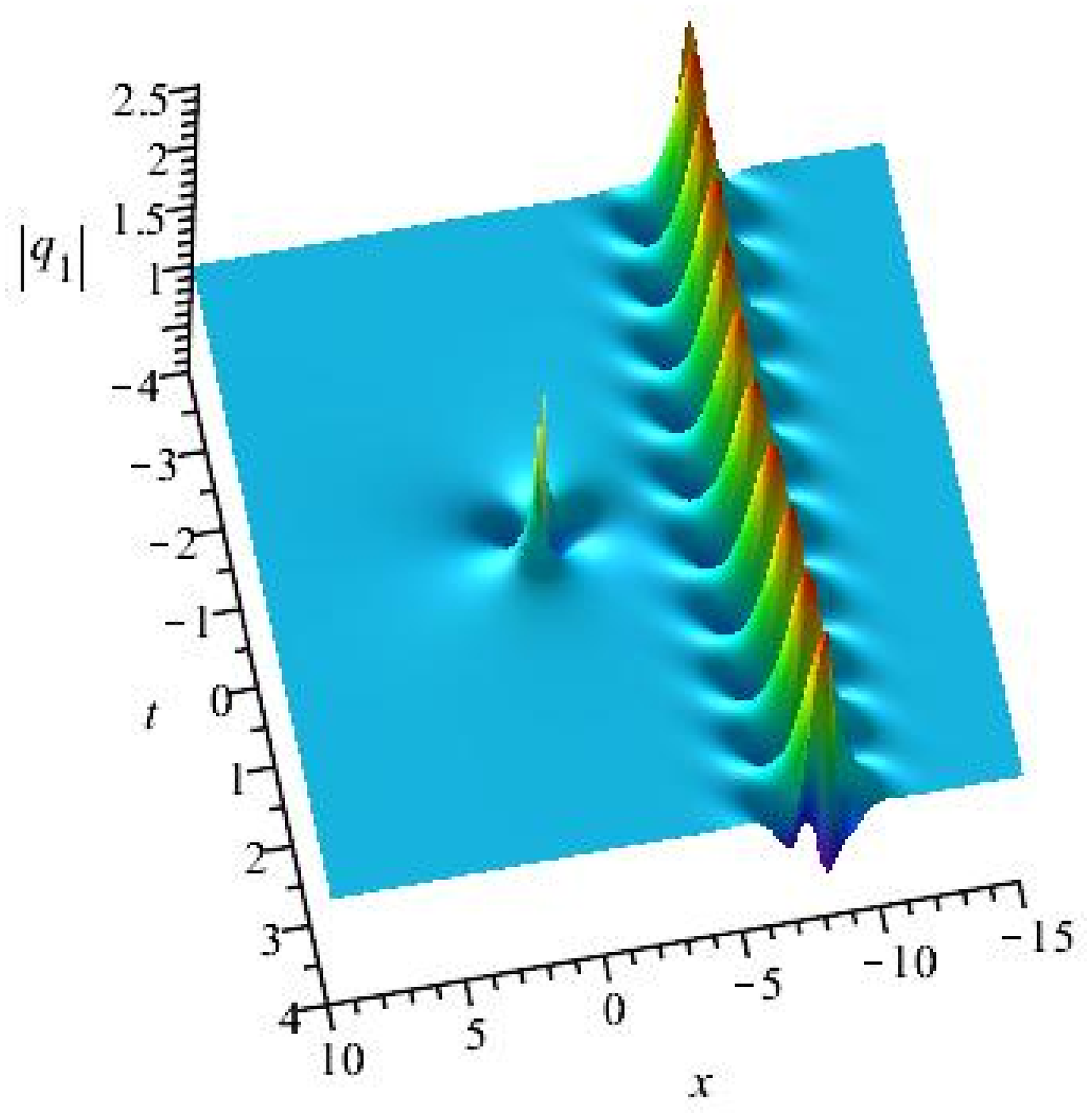}}
\centering
\subfigure[]{\includegraphics[height=0.3\textwidth]{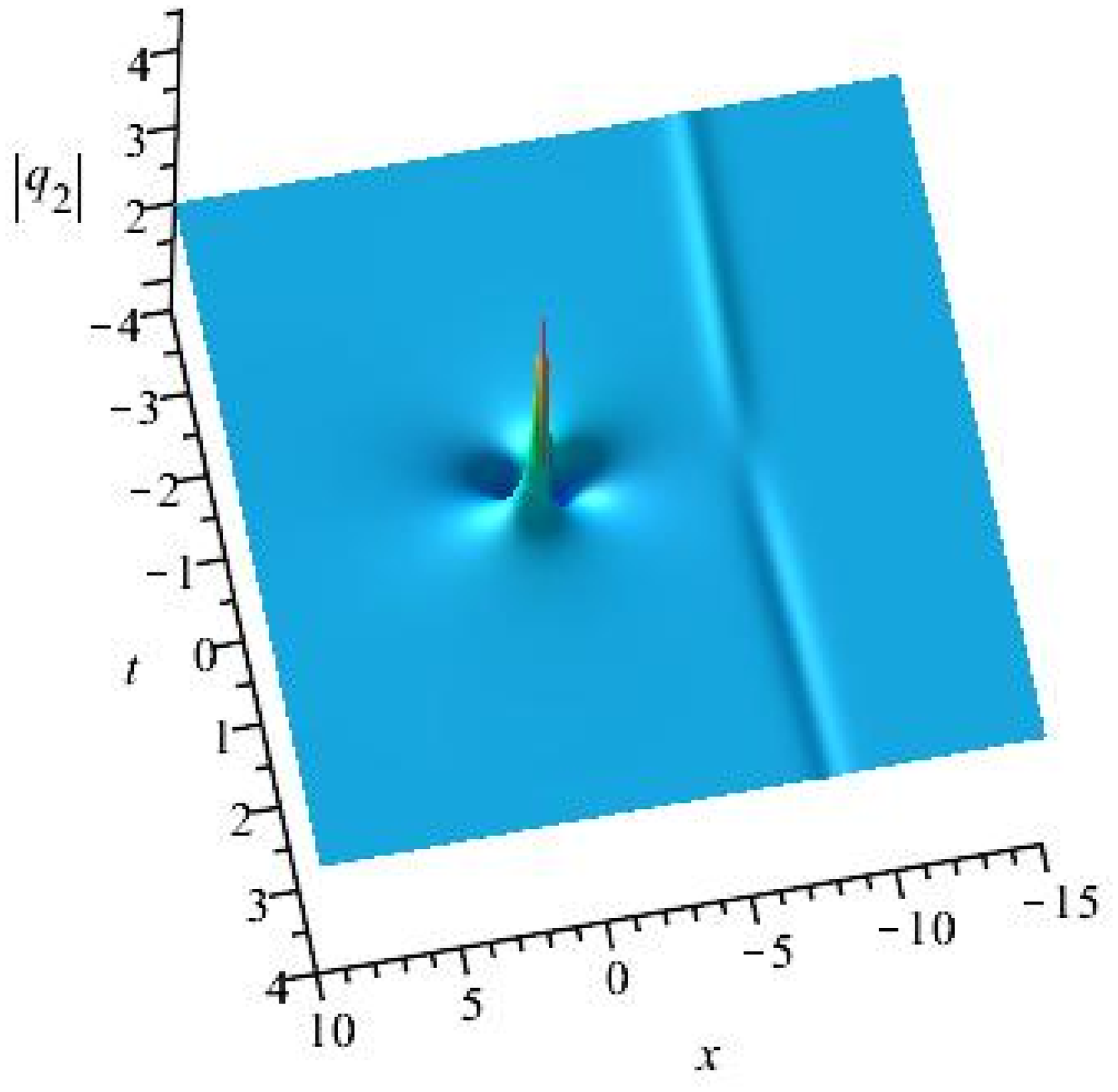}}
\centering
\subfigure[]{\includegraphics[height=0.3\textwidth]{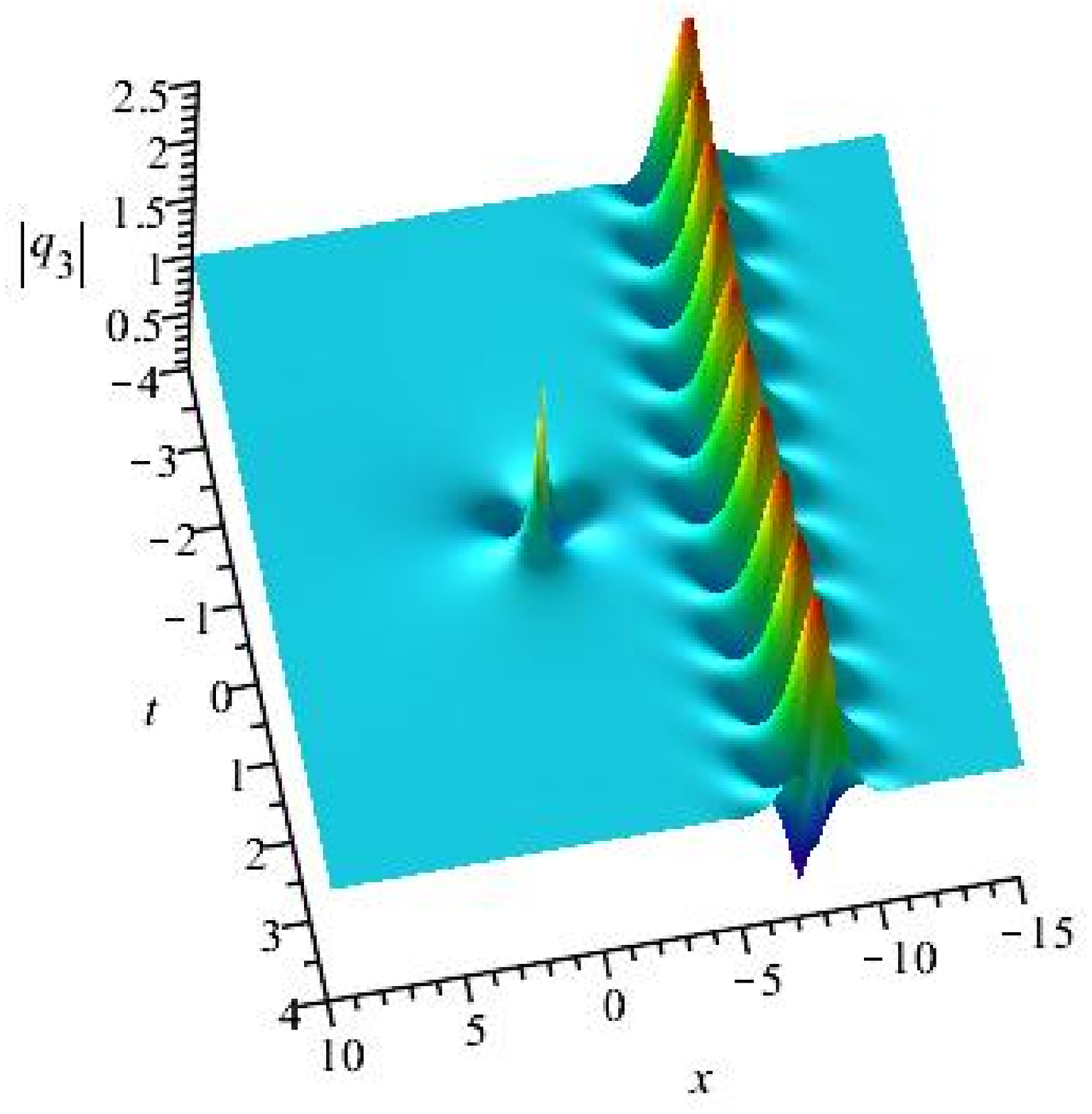}}
\centering
\caption{\small(Color online) Evolution plot of the interactional solution between the first-order RW and one-breather or one-amplitude-varying soliton in the three-component coupled DNLS equations with the parameters chosen by $d_1=1, d_2=-2, d_3=-1,\alpha=0,\beta=\tfrac{1}{200000}$: (a) a first-order RW and a breather separate in $q_1$ componet;  (b) a first-order RW and a amplitude-varying soliton separate in $q_2$ componet; (c) a first-order RW and a breather separate in $q_2$ componet.\label{xt-6f-2}}
\end{figure}

\begin{figure}[H]
\renewcommand{\figurename}{{Fig.}}
\subfigure[]{\includegraphics[height=0.24\textwidth]{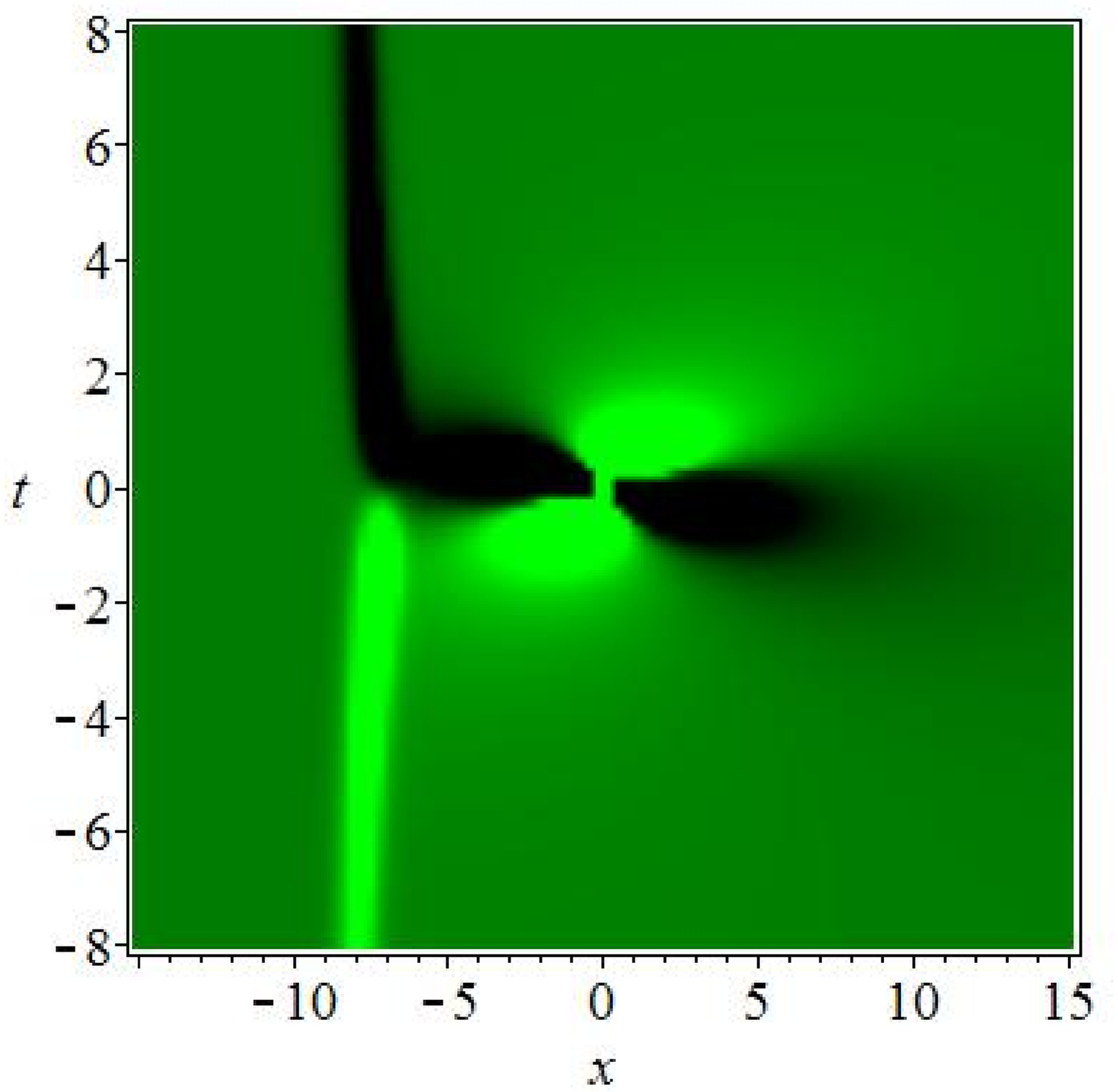}}
\centering
\subfigure[]{\includegraphics[height=0.24\textwidth]{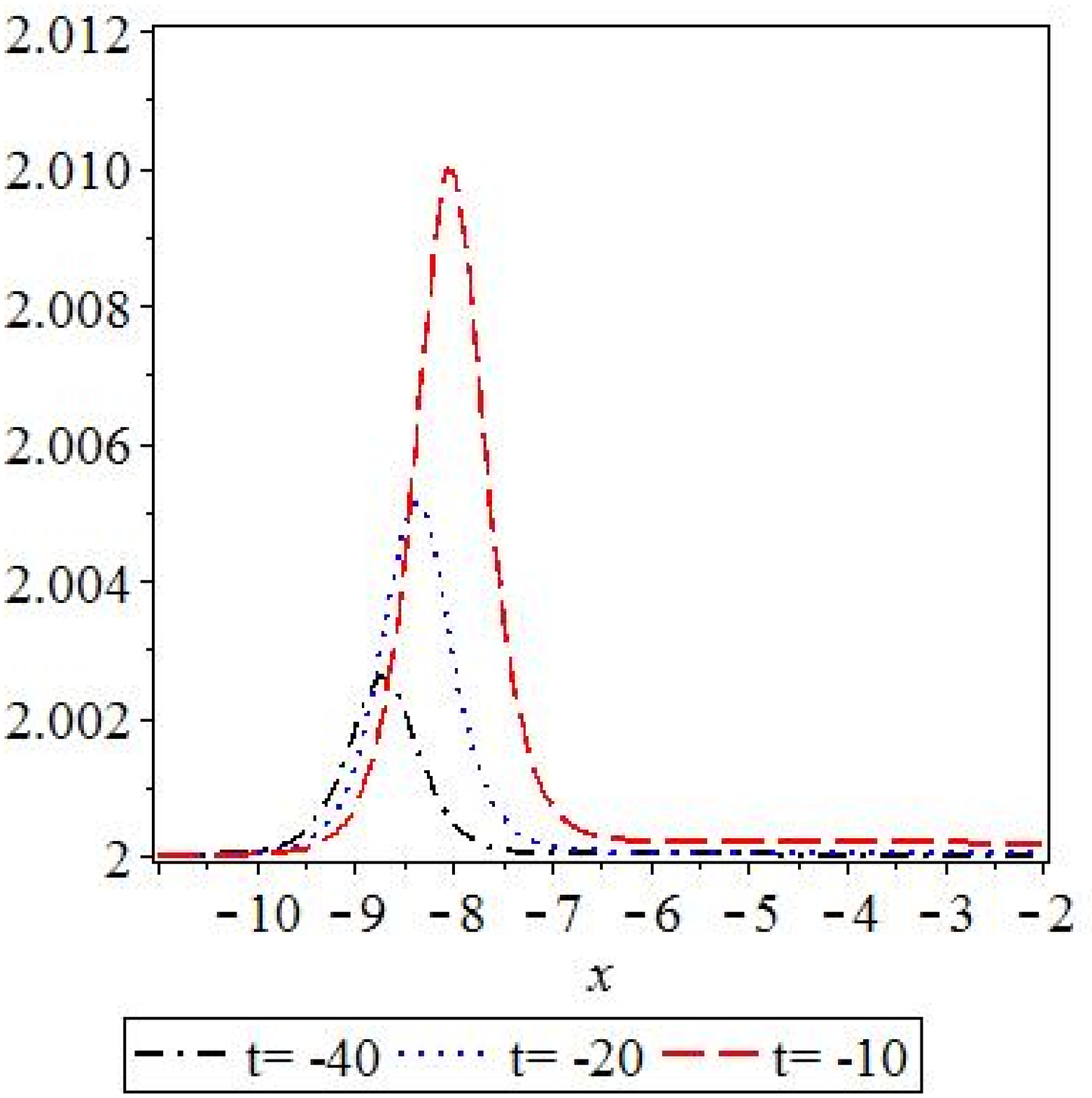}}
\centering
\subfigure[]{\includegraphics[height=0.24\textwidth]{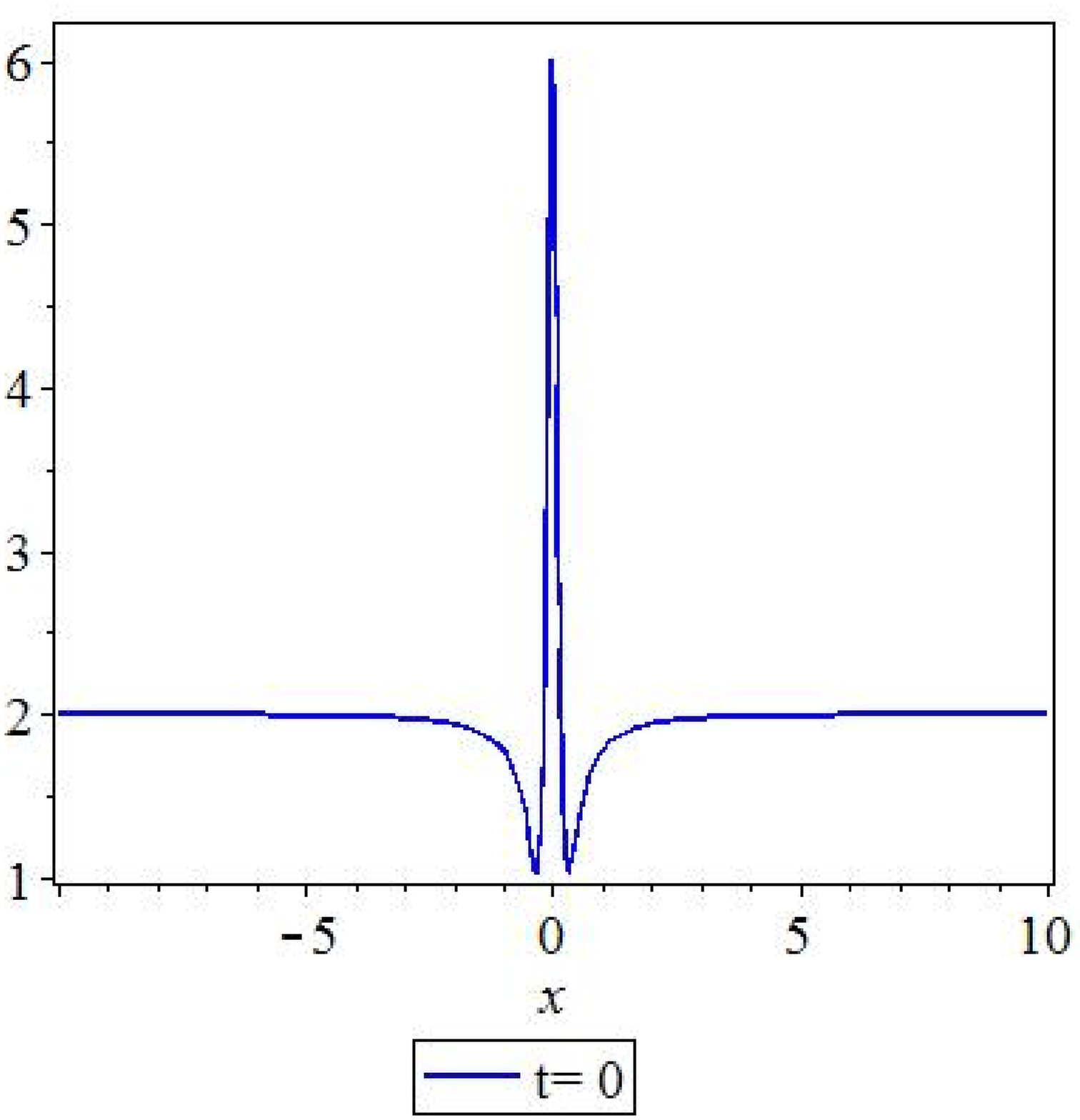}}
\centering
\subfigure[]{\includegraphics[height=0.24\textwidth]{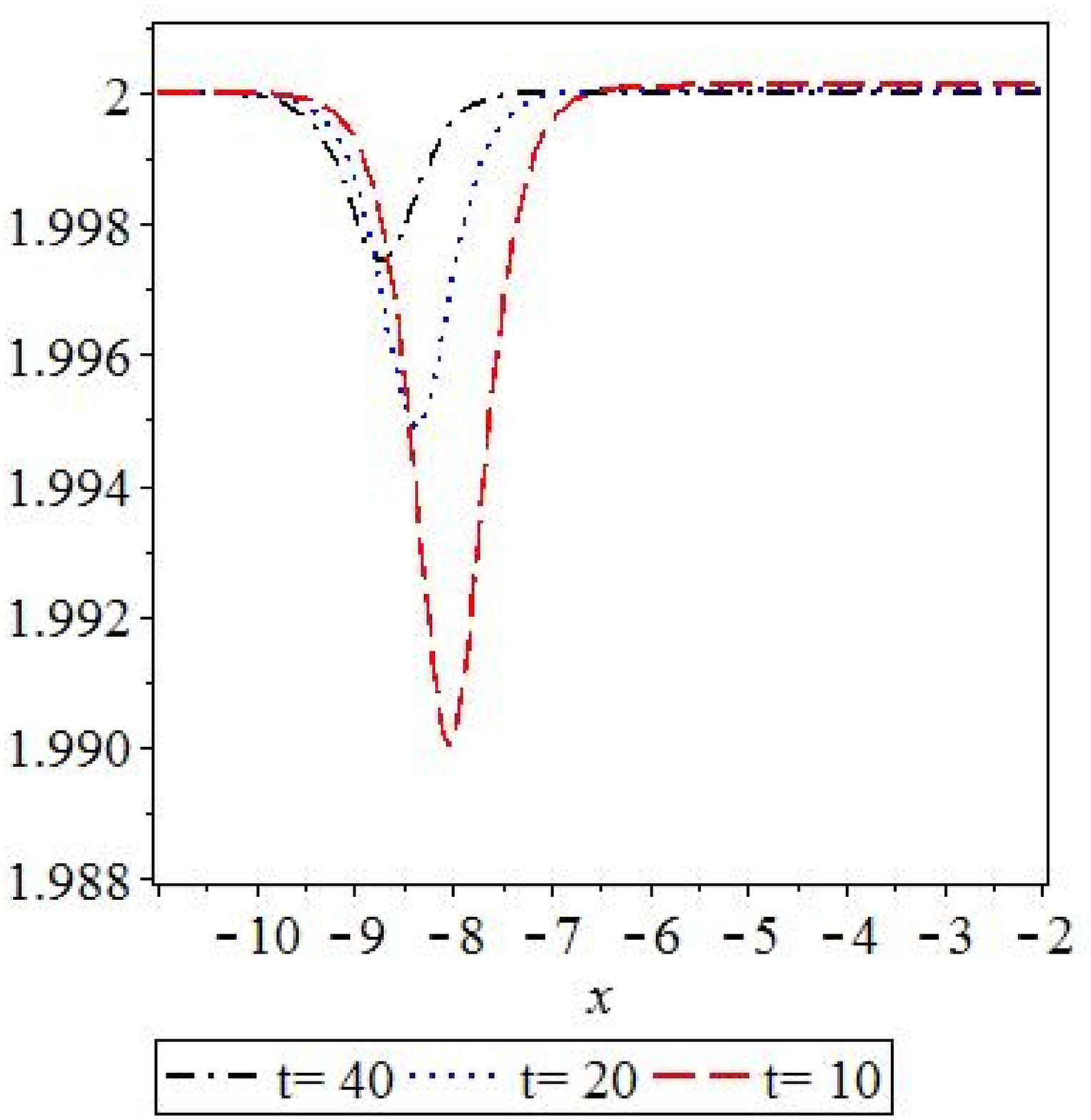}}
\centering
\caption{\small (Color online) (a) Evolution density plot  of $q_2$ component in Fig. \ref{xt-6f-2} (b). Plane evolution plot of the interactional process between the first-order RW and the amplitude-varying soliton of  $q_2$ component in Fig. \ref{xt-6f-2} (b) in  different moments: (b) $t<0$; (c) $t=0$; (d) $t>0$. \label{xt-6f-3}}
\end{figure}

\begin{figure}[H]
\renewcommand{\figurename}{{Fig.}}
\subfigure[]{\includegraphics[height=0.3\textwidth]{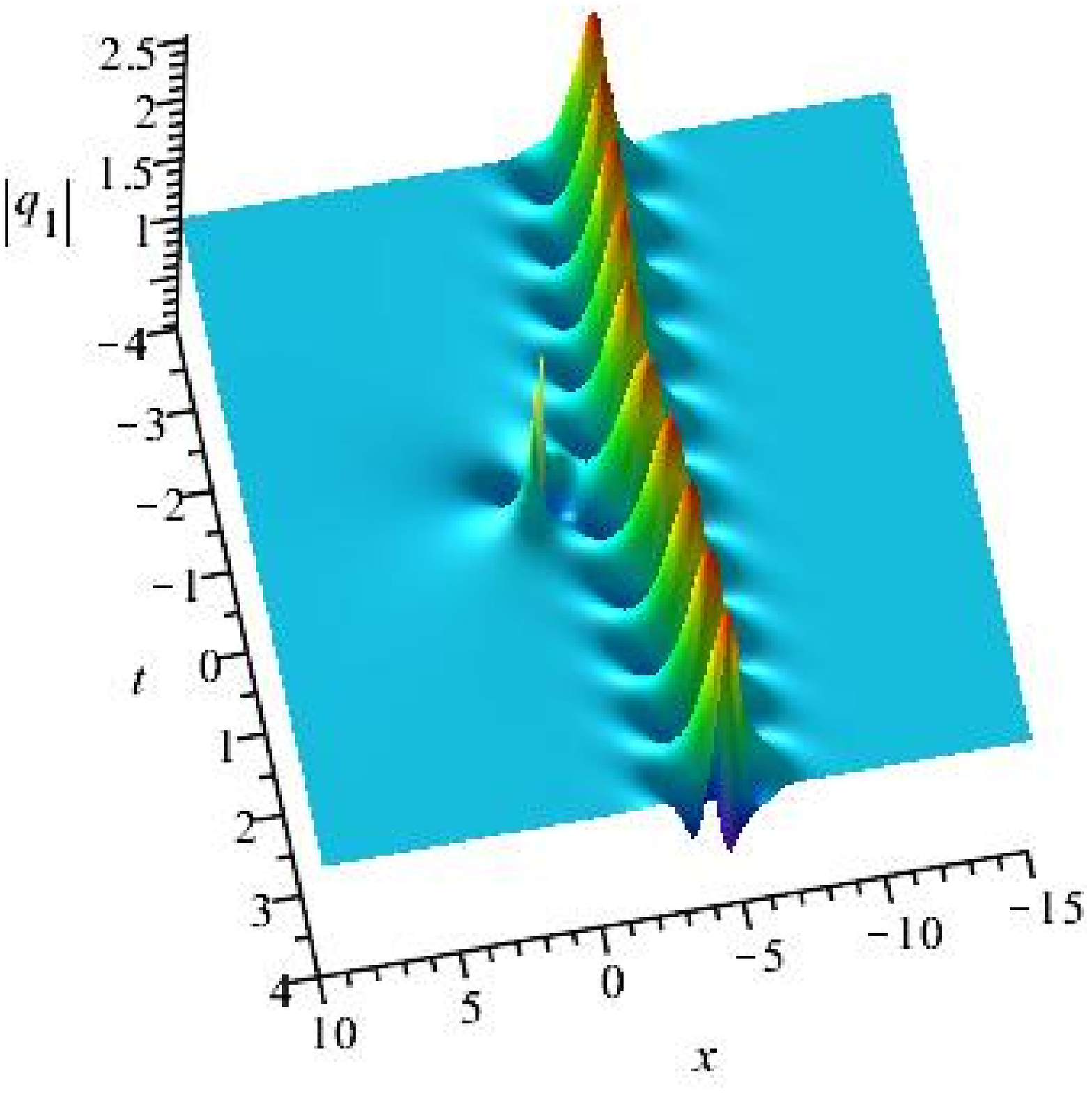}}
\centering
\subfigure[]{\includegraphics[height=0.3\textwidth]{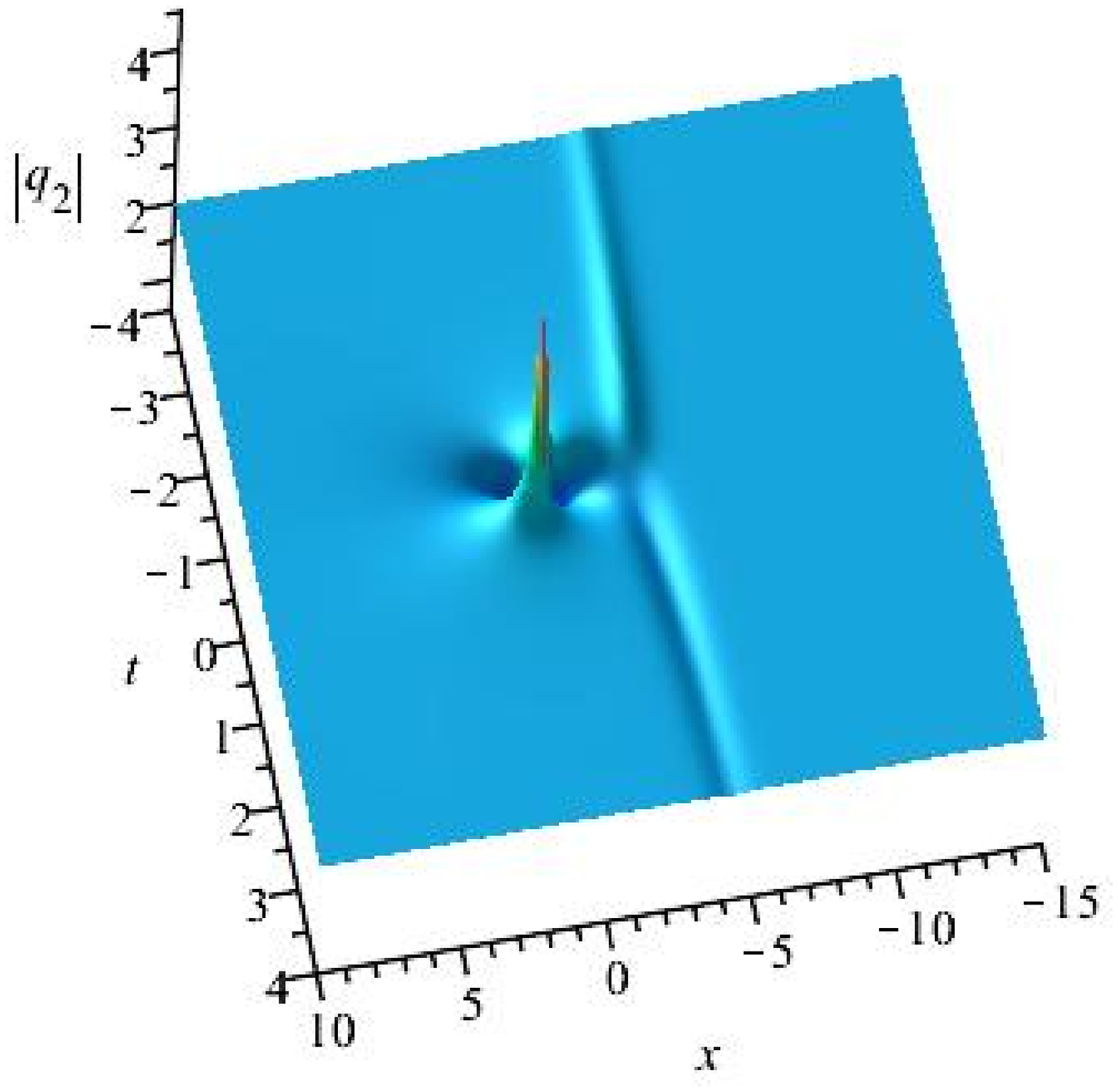}}
\centering
\subfigure[]{\includegraphics[height=0.3\textwidth]{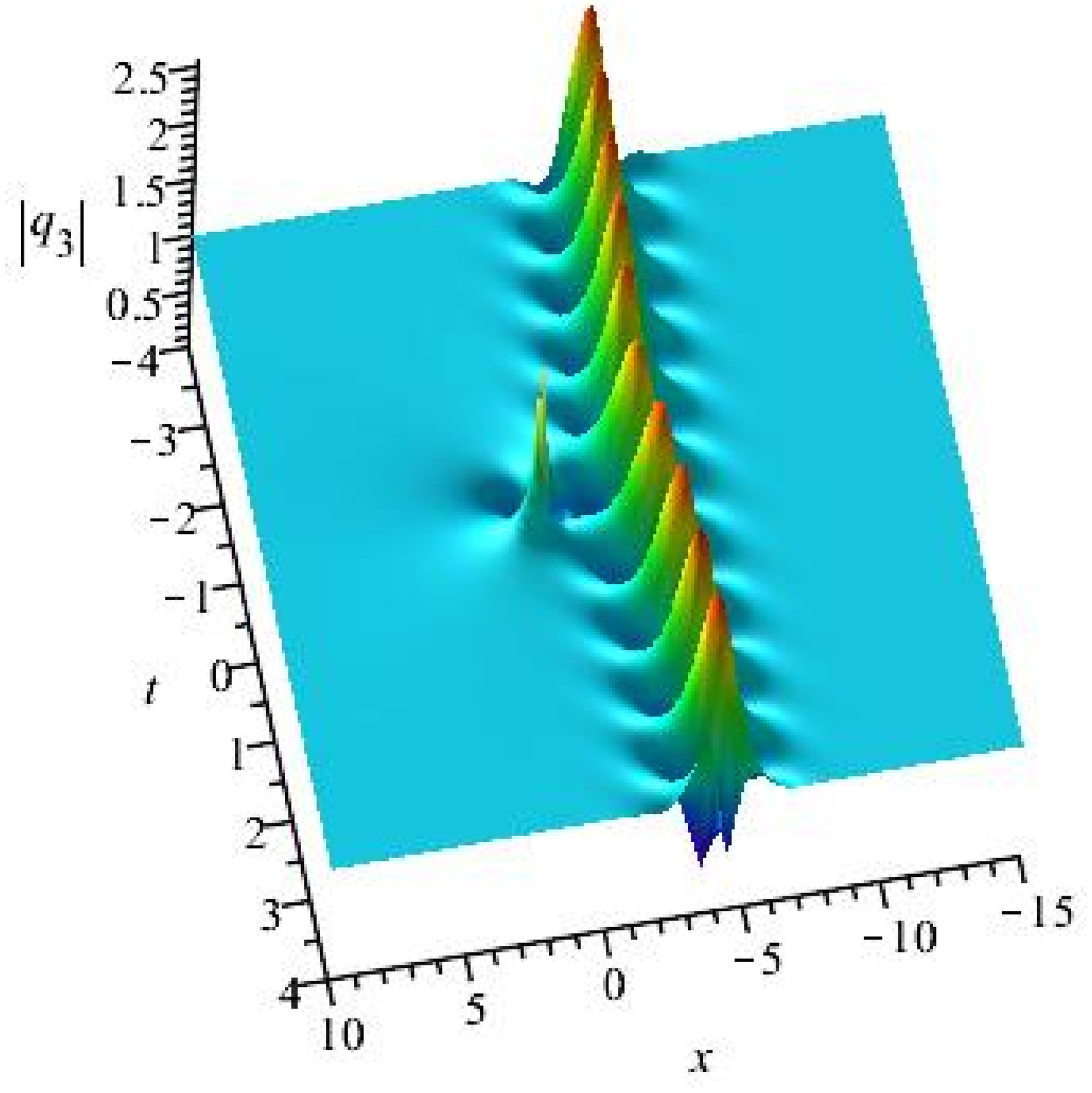}}
\centering
\caption{\small(Color online) Evolution plot of interactional solution between the first-order RW and one-breather or one-amplitude-varying soliton in the three-component coupled DNLS equations with the parameters chosen by $d_1=1, d_2=-2, d_3=-1,\alpha=0,\beta=\tfrac{1}{200}$: (a) a first-order RW merges with a breather  in $q_1$ component;  (b)  a first-order RW merges with a amplitude-varying soliton in $q_2$ component; (c) a first-order RW merges with a breather  in $q_3$ component.\label{xt-6f-4}}
\end{figure}

(\textrm{ii}) If one of the disturbing terms is zero and the backgrounds of  three components are all non-vanished, we can get  the interactional solutions  that two components are interactional solutions between a first-order RW and a breather, and one component is interactional solution between a first-order RW and an amplitude-varying soliton, see Fig. (\ref{xt-6f-2})-(\ref{xt-6f-4}). Without loss of of generality, choosing $\alpha=0,\beta\neq0,$ and $d_j\neq0~(j=1,2,3)$, it can be shown that a first-order RW and a breather separate in $q_1$ and $q_3$ components, and a first-order RW and an amplitude-varying soliton separate in $q_2$ component  from Fig. \ref{xt-6f-2}. Choosing different cases among $\alpha,\beta,d_1,d_2$ and $d_3$, we can get different various arrangements of three components $q_1$, $q_2$ and $q_3$. However, the combination is same, namely  two components are RW and breather, one component is RW and  amplitude-varying soliton. Here, we define the same combination as the same type solution.

The density plot of  $q_2$ in Fig. \ref{xt-6f-3}(a) shows that  the soliton in $q_2$ component is anti-dark soliton if $t<0$ and becomes dark soliton if $t>0$. Besides, this kind of amplitude-varying soliton annihilates if $t=0$. When $t<0$, the amplitude of anti-dark soliton in $q_2$ component becomes big with $t$ increasing, see Fig. \ref{xt-6f-3}(b); otherwise, if $t>0$,  the amplitude of dark soliton in $q_2$ component becomes small with $t$ increasing, see Fig. \ref{xt-6f-3}(d). Besides, it demonstrates that the soliton annihilates and only a first-order RW exist at $t=0$ in $q_2$ component from Fig. \ref{xt-6f-3}(c). It is shown that this kind of amplitude-varying soliton appears on the plane background  at $t={-}\infty$ and annihilates at $t=0$, then disappears on the plane background at $t=\infty$. By increasing the absolute values of $\beta$, we can find that the first-order RW emerges with one-breather or one-amplitude-varying soliton in Fig. \ref{xt-6f-4}.

\begin{figure}[H]
\renewcommand{\figurename}{{Fig.}}
\subfigure[]{\includegraphics[height=0.3\textwidth]{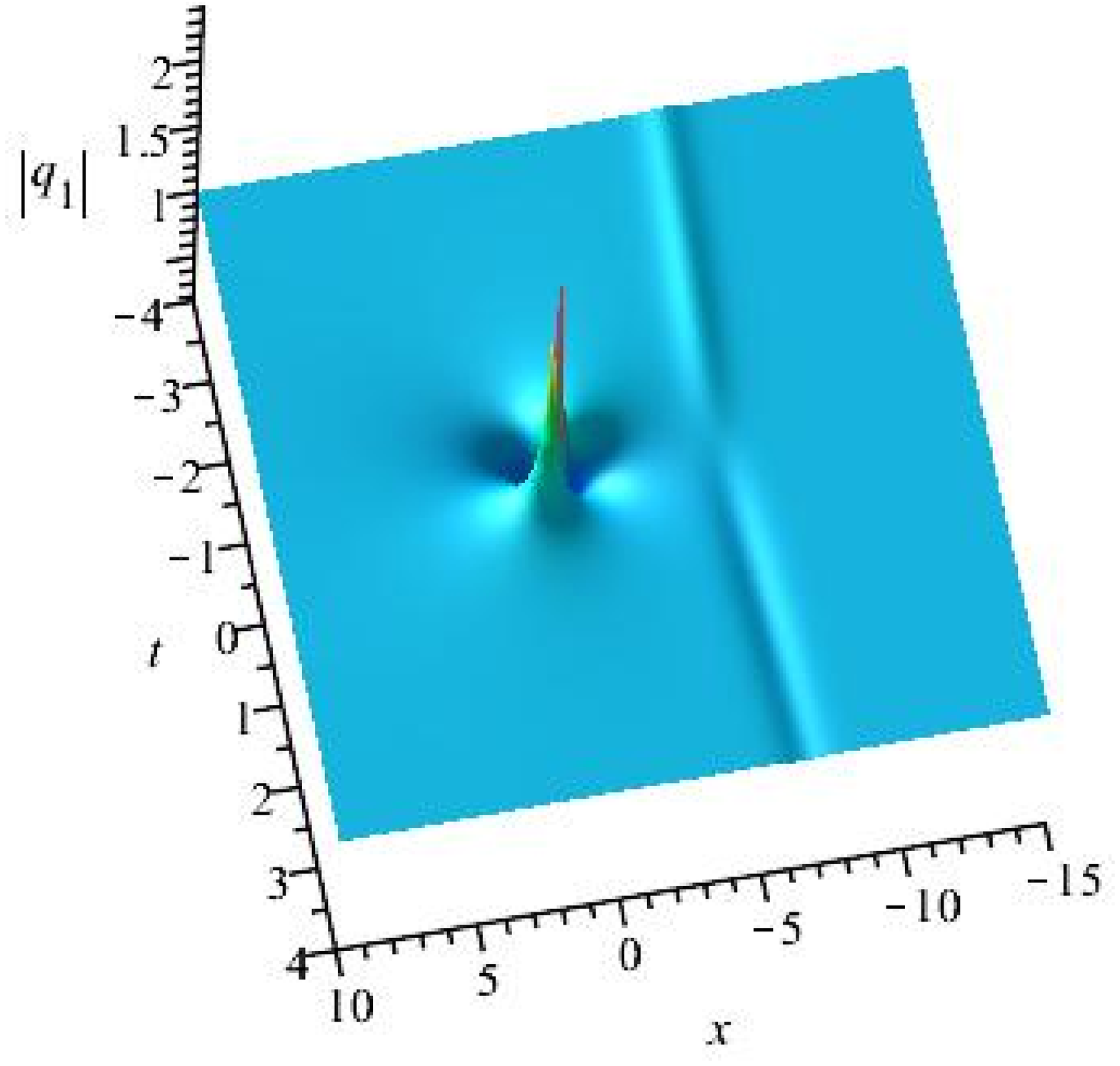}}
\centering
\subfigure[]{\includegraphics[height=0.3\textwidth]{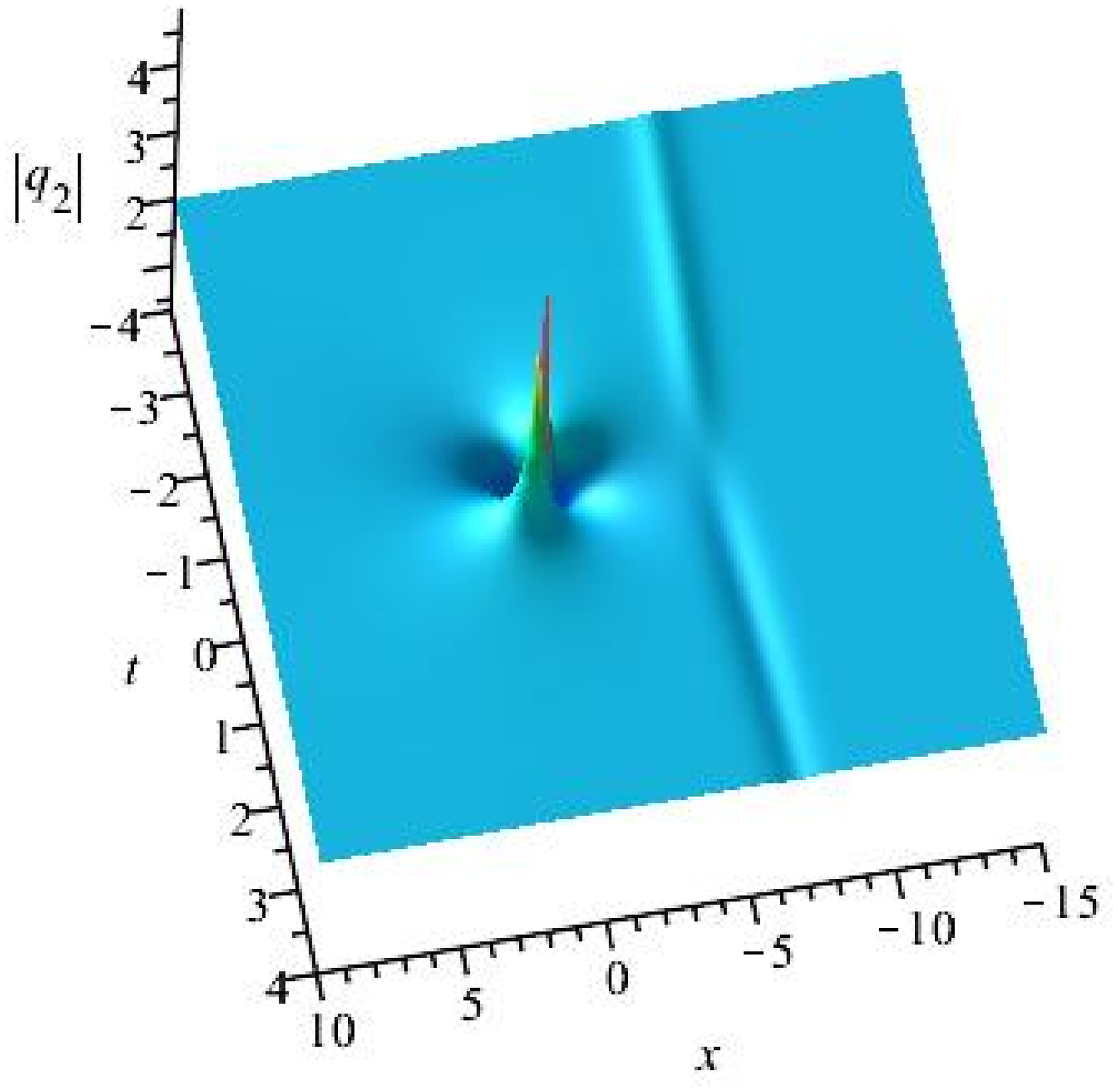}}
\centering
\subfigure[]{\includegraphics[height=0.3\textwidth]{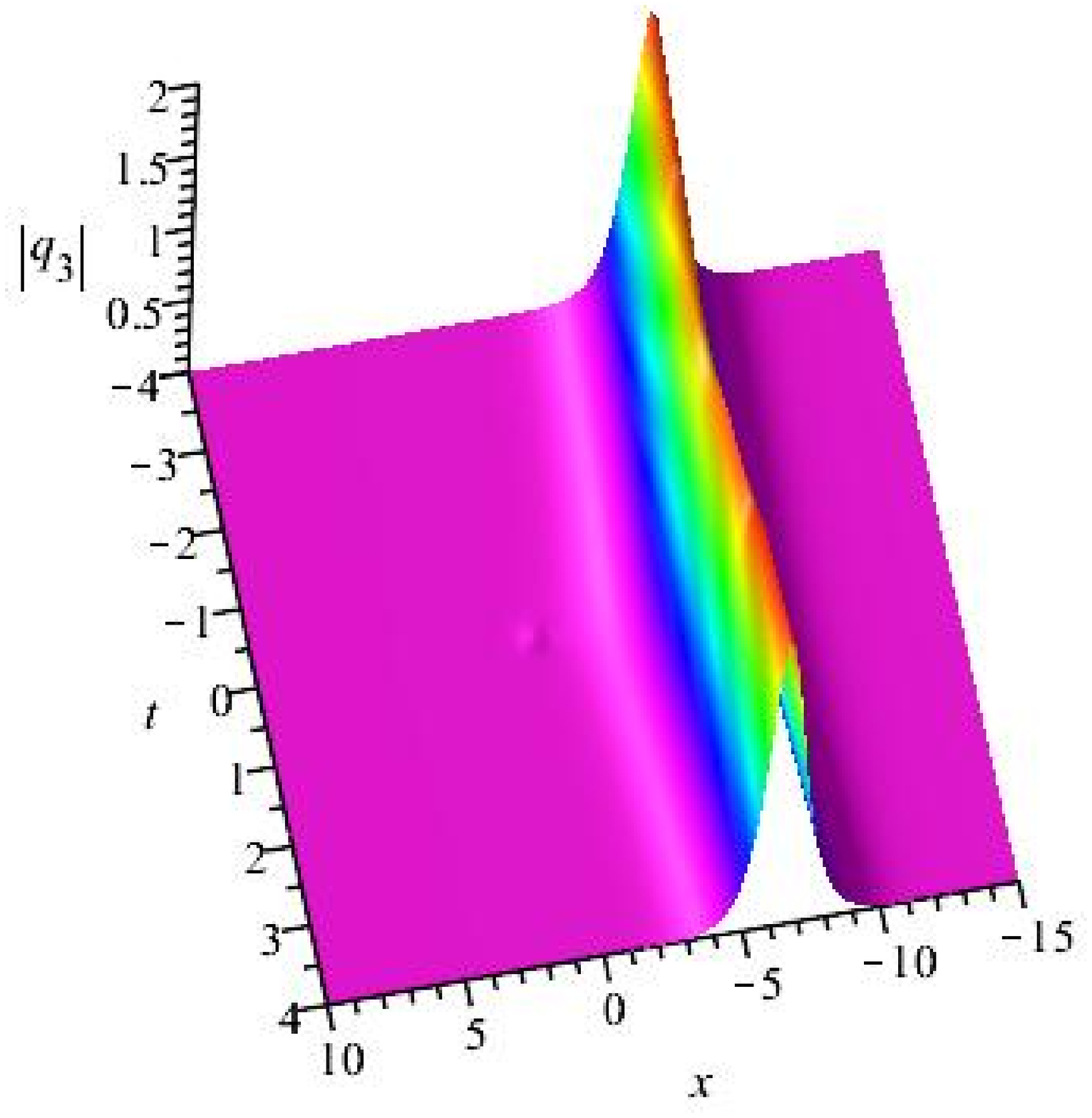}}
\centering
\caption{\small(Color online) Evolution plot of interactional solution between the first-order RW and one-breather or one-amplitude-varying soliton in the three-component coupled DNLS equations with the parameters chosen by $d_1=1, d_2=-2, d_3=0,\alpha=0,\beta=\tfrac{1}{2000}$: (a) a first-order RW and an amplitude-varying soliton separate in $q_1$ component; (b) a first-order RW and an amplitude-varying soliton separate in $q_2$ component; (c) a first-order RW and a  bright soliton separate in $q_3$ component. \label{xt-6f-5}}
\end{figure}

\begin{figure}[H]
\renewcommand{\figurename}{{Fig.}}
\subfigure[]{\includegraphics[height=0.3\textwidth]{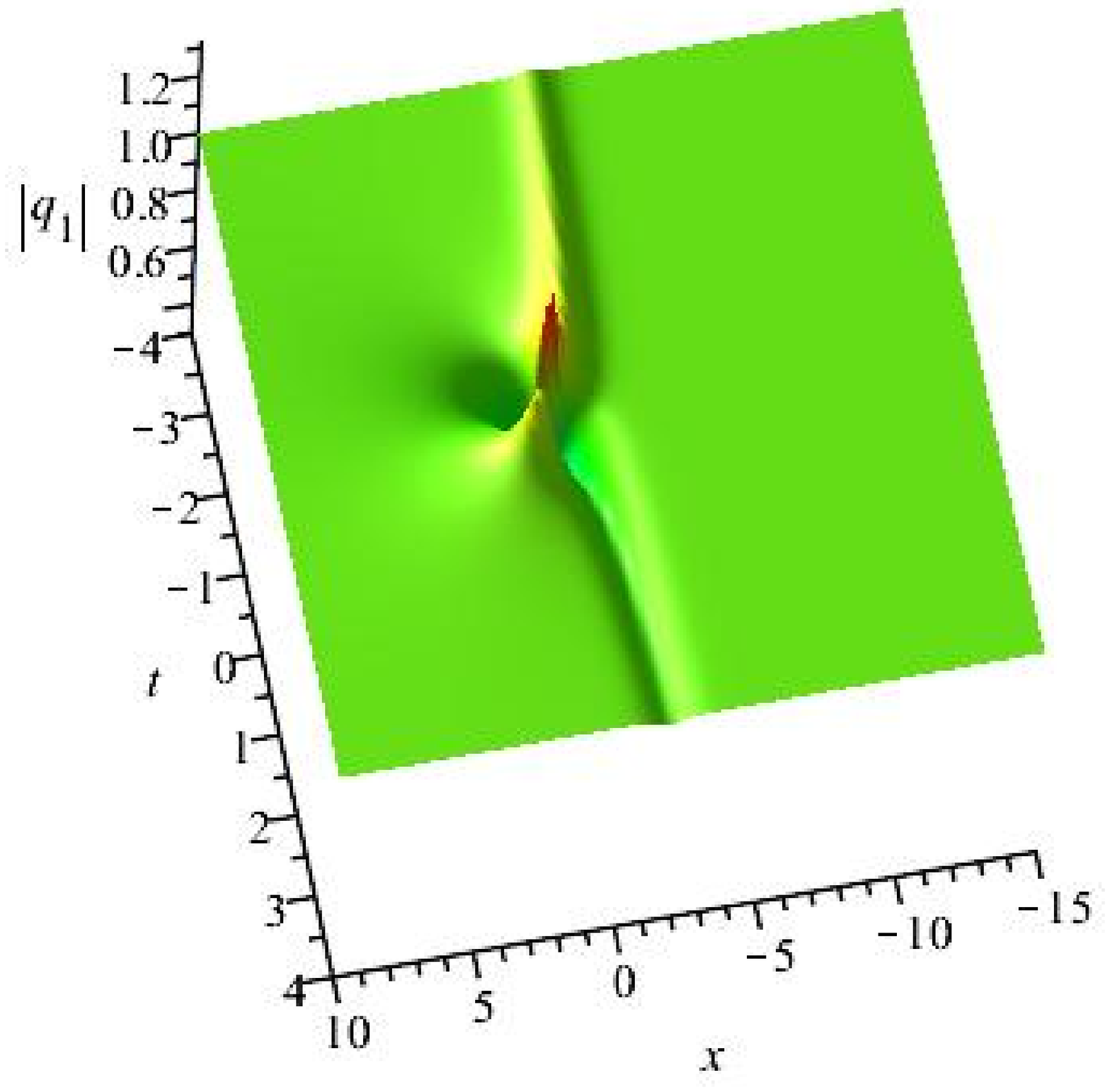}}
\centering
\subfigure[]{\includegraphics[height=0.3\textwidth]{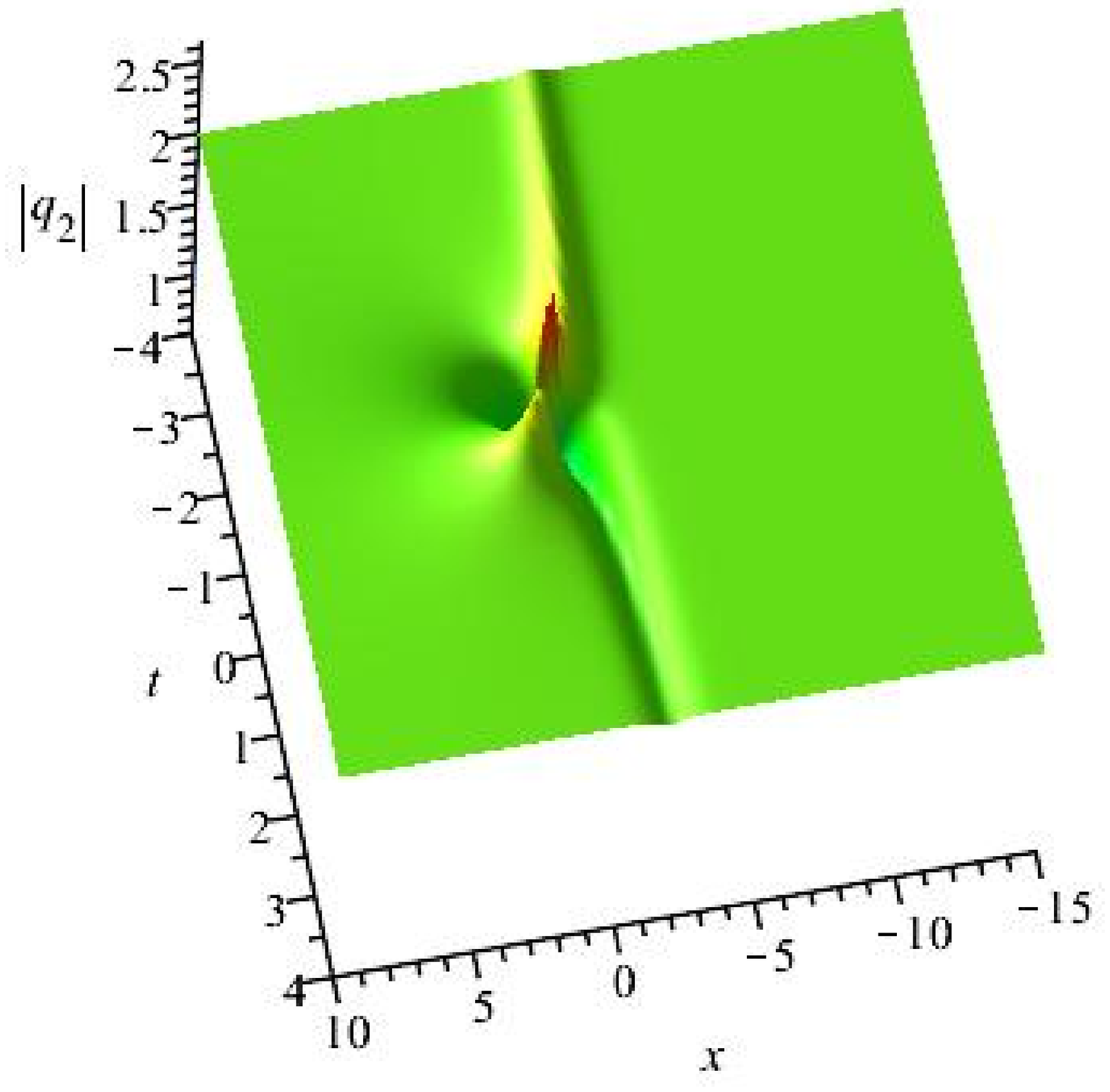}}
\centering
\subfigure[]{\includegraphics[height=0.3\textwidth]{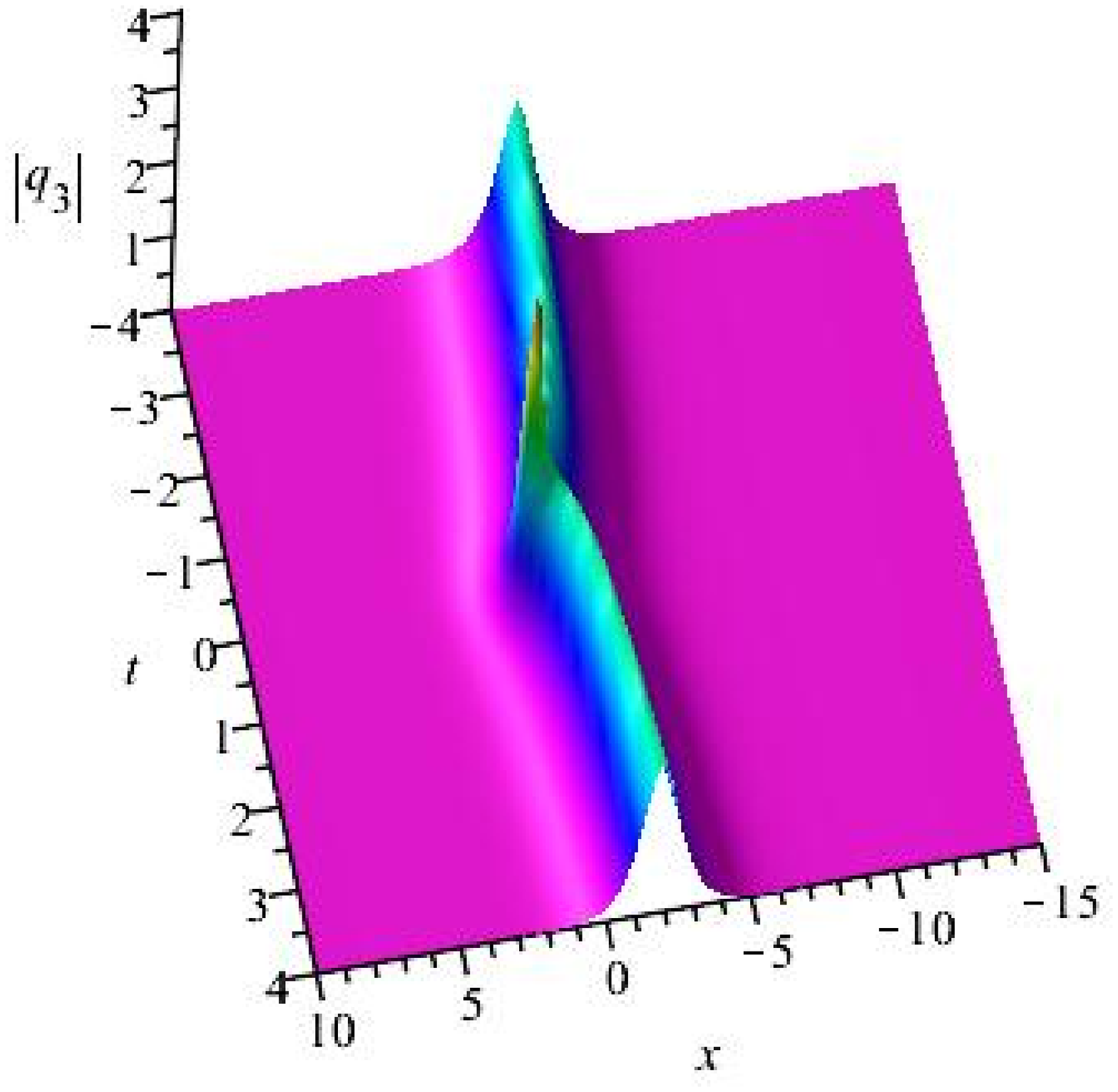}}
\centering
\caption{\small(Color online) Evolution plot of interactional solution between the first-order RW and one-amplitude-varying soliton  or one-bright soliton in the three-component coupled DNLS equations with the parameters chosen by  $d_1=1, d_2=-2, d_3=0,\alpha=0,\beta=1$: (a) a first-order RW merges with an amplitude-varying soliton  in $q_1$ component; (b)  a first-order RW merges with an amplitude-varying soliton  in $q_2$ component; (c) a first-order RW merges with a bright soliton  in $q_3$ component. \label{xt-6f-6} }
\end{figure}

(\textrm{iii}) Setting one of the disturbing terms be zero and one of the backgrounds in the three components be vanished, we can construct the following interactional solutions that two components are  a first-order RW and an amplitude-varying soliton, and one component is  a first-order RW and a breather, see Fig. \ref{xt-6f-5}-\ref{xt-6f-6}. Choosing $\alpha=0$, $\beta\neq0$, $d_1\neq0$,$d_2\neq0$, and $d_3=0$, it demonstrates that a first-order RW and  an amplitude-varying soliton separate in $q_1$ and $q_2$ components, and a first-order RW and a bright soliton separate in $q_3$ component from Fig. \ref{xt-6f-5}. It can be found that the first-order RW is not easily observable in Fig. \ref{xt-6f-5}(c). Here, the amplitude of the background of $q_3$ component is zero, so it is not easily to observe the RW emerging on zero background. In Fig. \ref{xt-6f-6}, the first-order RW emerges with one-amplitude-varying soliton  or one-bright soliton by increasing the absolute values of two disturbing coefficients $\alpha$ and $\beta$. In Fig.  \ref{xt-6f-6}(c), the RW can be easily found because the amplitude of the part of the plane wave background where the first-order RW emerges is not zero. Under the condition $\alpha=0,\beta\neq0$, the three backgrounds $d_1$, $d_2$ and $d_3$ can be chosen as two or three being zero, but there exist some zero solutions in three components $q_1[1]$, $q_2[1]$ and $q_3[1]$.  Therefore, we leave these two cases out.

\begin{figure}[H]
\renewcommand{\figurename}{{Fig.}}
\subfigure[]{\includegraphics[height=0.3\textwidth]{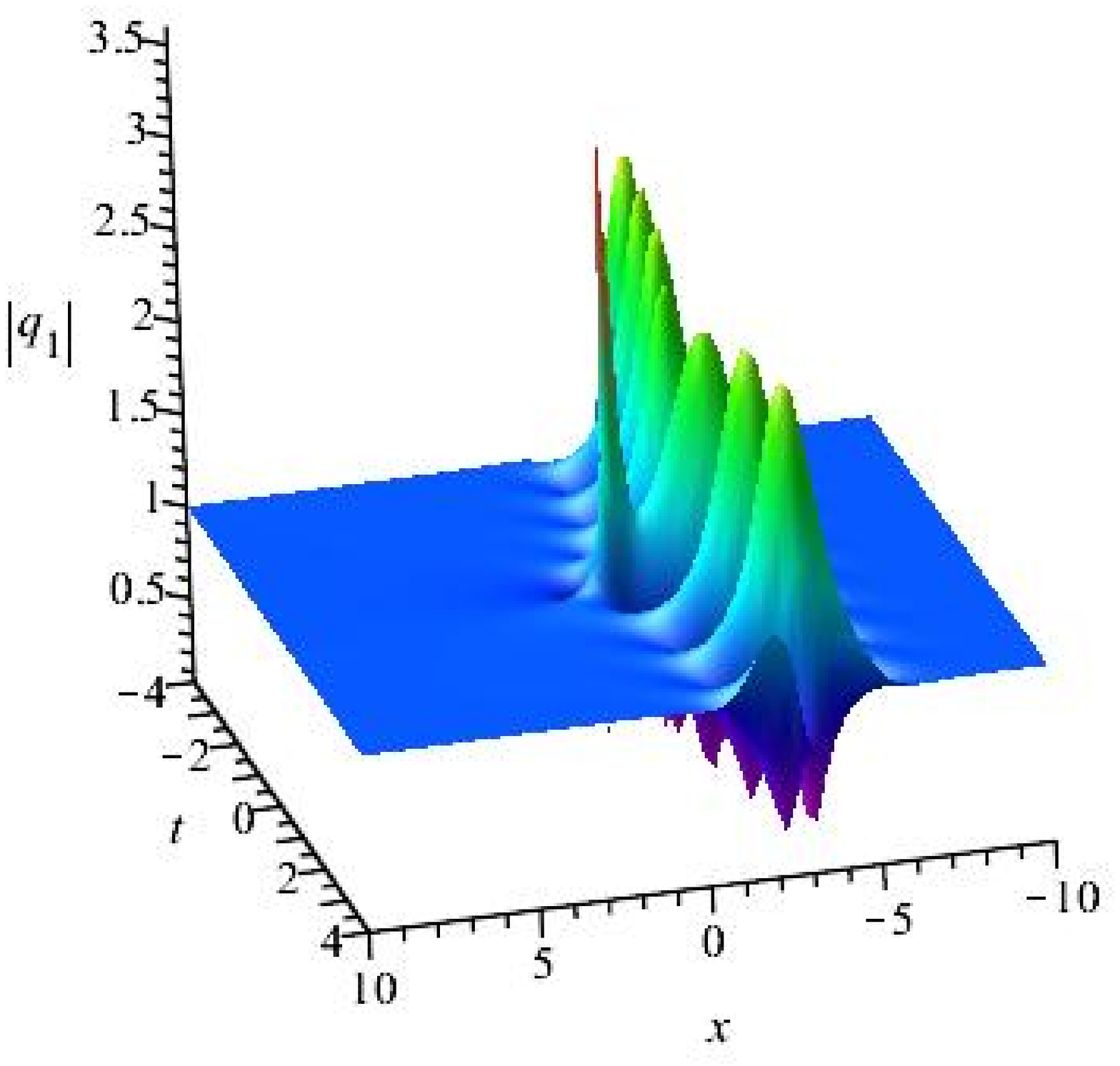}}
\centering
\subfigure[]{\includegraphics[height=0.3\textwidth]{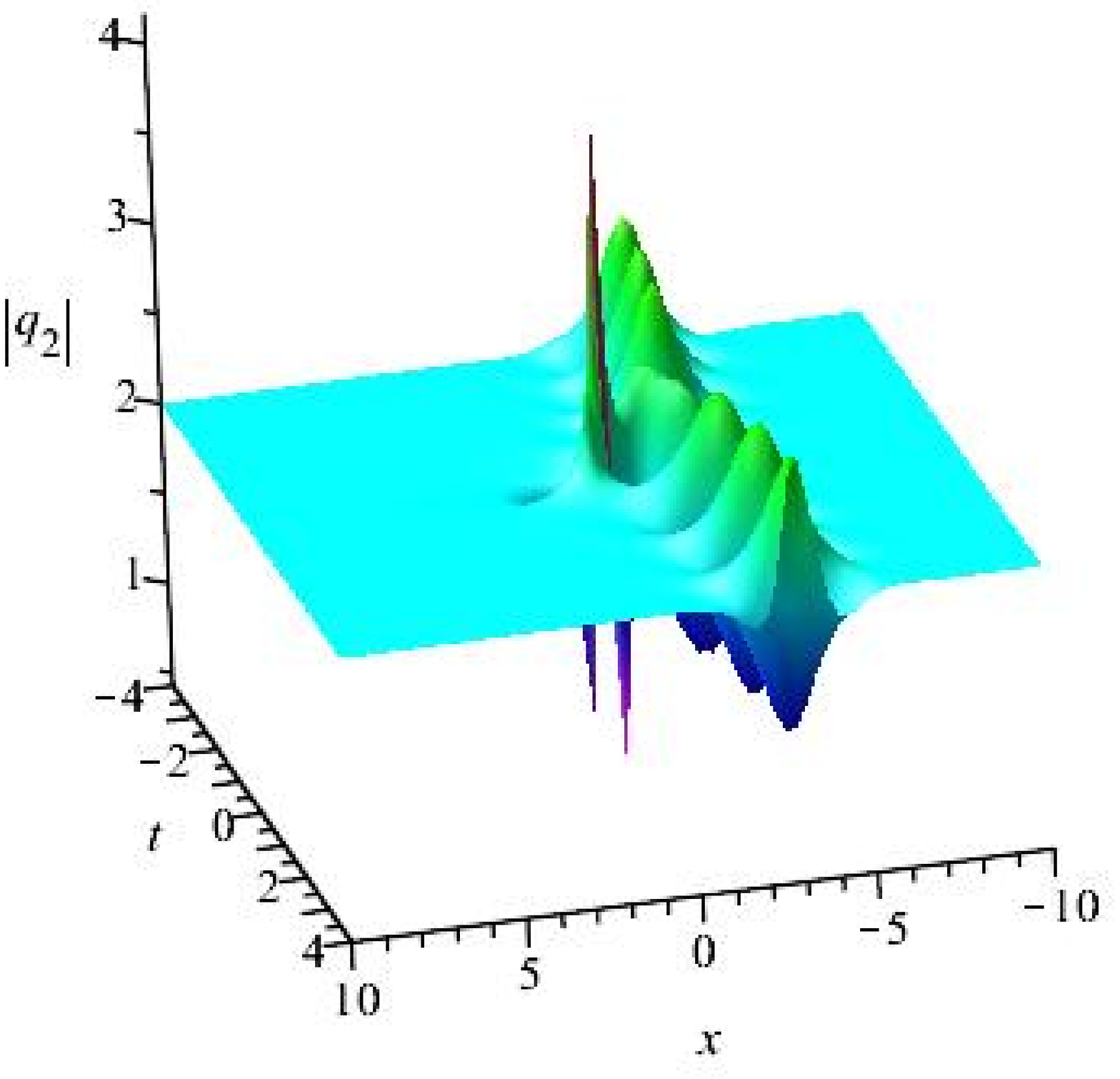}}
\centering
\subfigure[]{\includegraphics[height=0.3\textwidth]{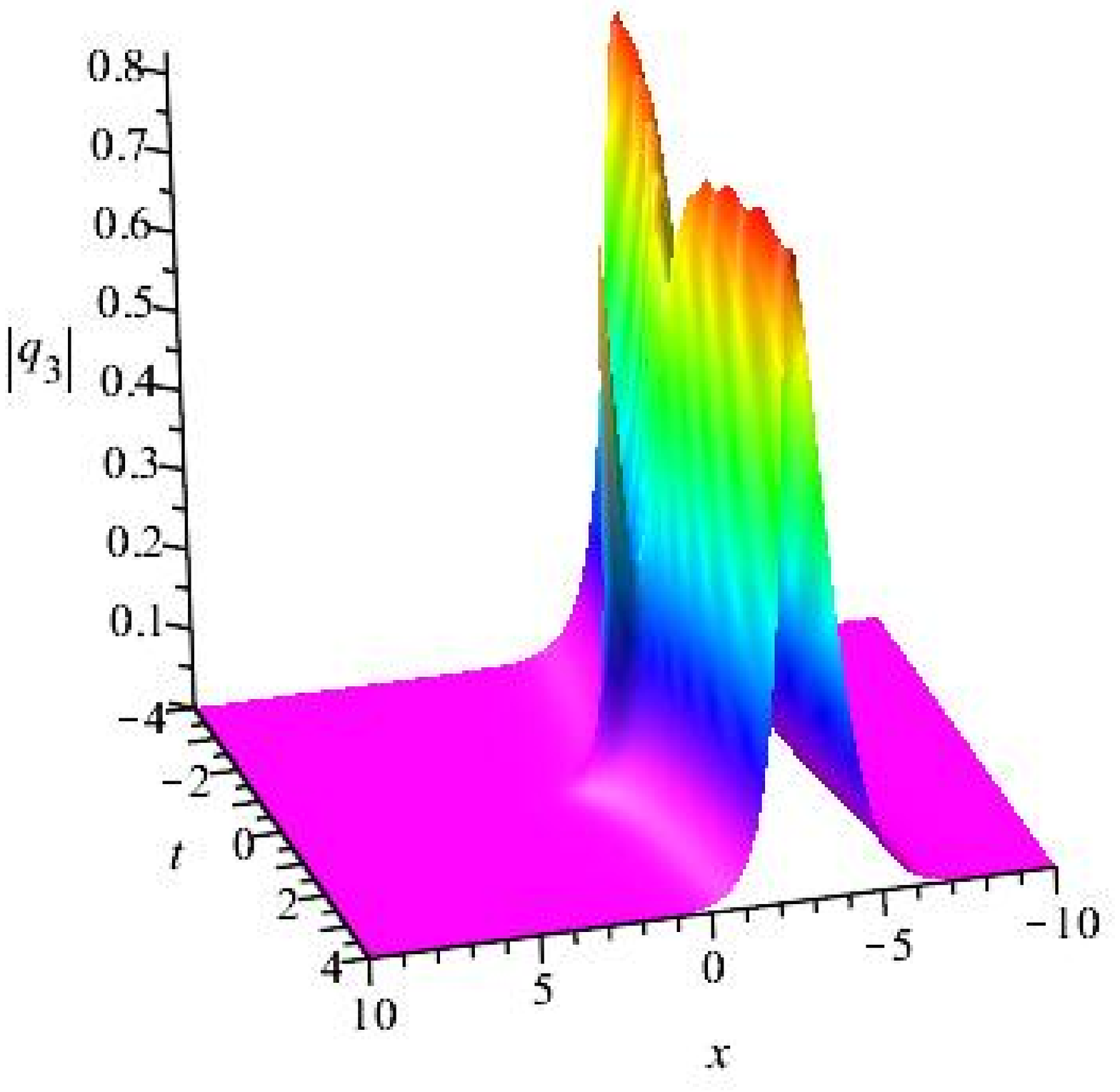}}
\centering
\caption{\small(Color online)  Evolution plot of interactional solution between a first-order RW and one-breather or one-bright soliton in the three-component coupled DNLS equations with the parameters chosen by $d_1=1, d_2=-2, d_3=0,\alpha=\tfrac{1}{20},\beta=-\tfrac{1}{20}$: (a) a first-order RW merges with a breather in $q_1$ component; (b) a first-order RW merges with a breather in $q_2$ component;(c) a first-order RW merges with a bright soliton in $q_3$ component. \label{xt-6f-7}}
\end{figure}

(\textrm{iv}) Choosing all the disturbing terms are not zero, namely $\alpha\neq0,\beta\neq0$, and one of the three backgrounds is vanished, we can generate the following interactional solutions that two components are a first-order RW and a breather, one component is a first-order RW and a bright soliton. Fixing $d_1\neq0$, $d_2\neq0$ and $d_3=0$, we can see that a first-order RW interacts with a breathe in $q_1$ and $q_2$ component, and a first-order RW interacts with a bright soliton in $q_3$ component form Fig. \ref{xt-6f-7}. Similar with case (\textrm{iii}), the nonlinear localized waves can separate in the three components $q_1$, $q_2$ and $q_3$ by decreasing the absolute values of disturbing coefficients $\alpha$ and $\beta$ and we omit these figures.

\begin{figure}[H]
\renewcommand{\figurename}{{Fig.}}
\subfigure[]{\includegraphics[height=0.3\textwidth]{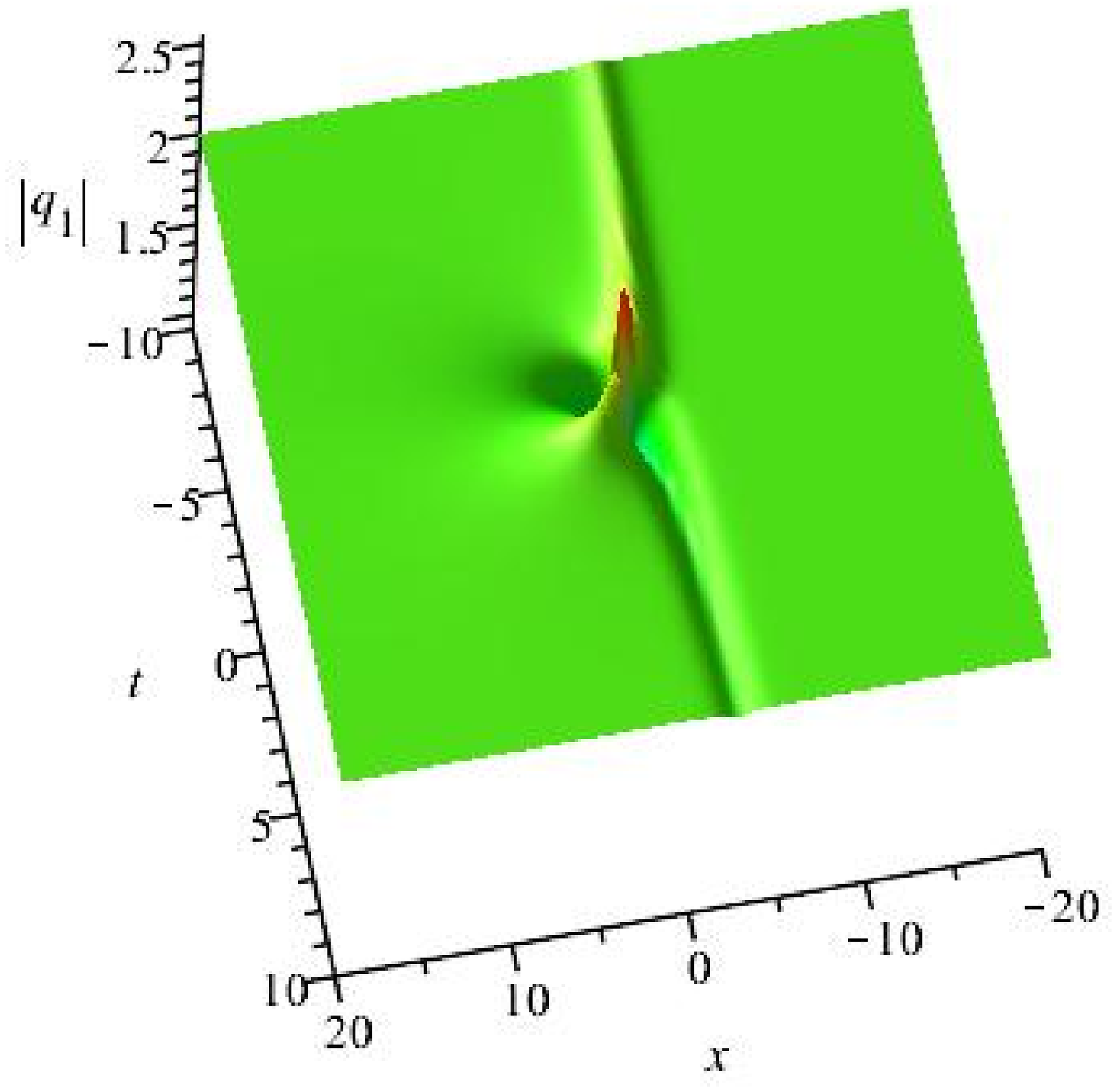}}
\centering
\subfigure[]{\includegraphics[height=0.3\textwidth]{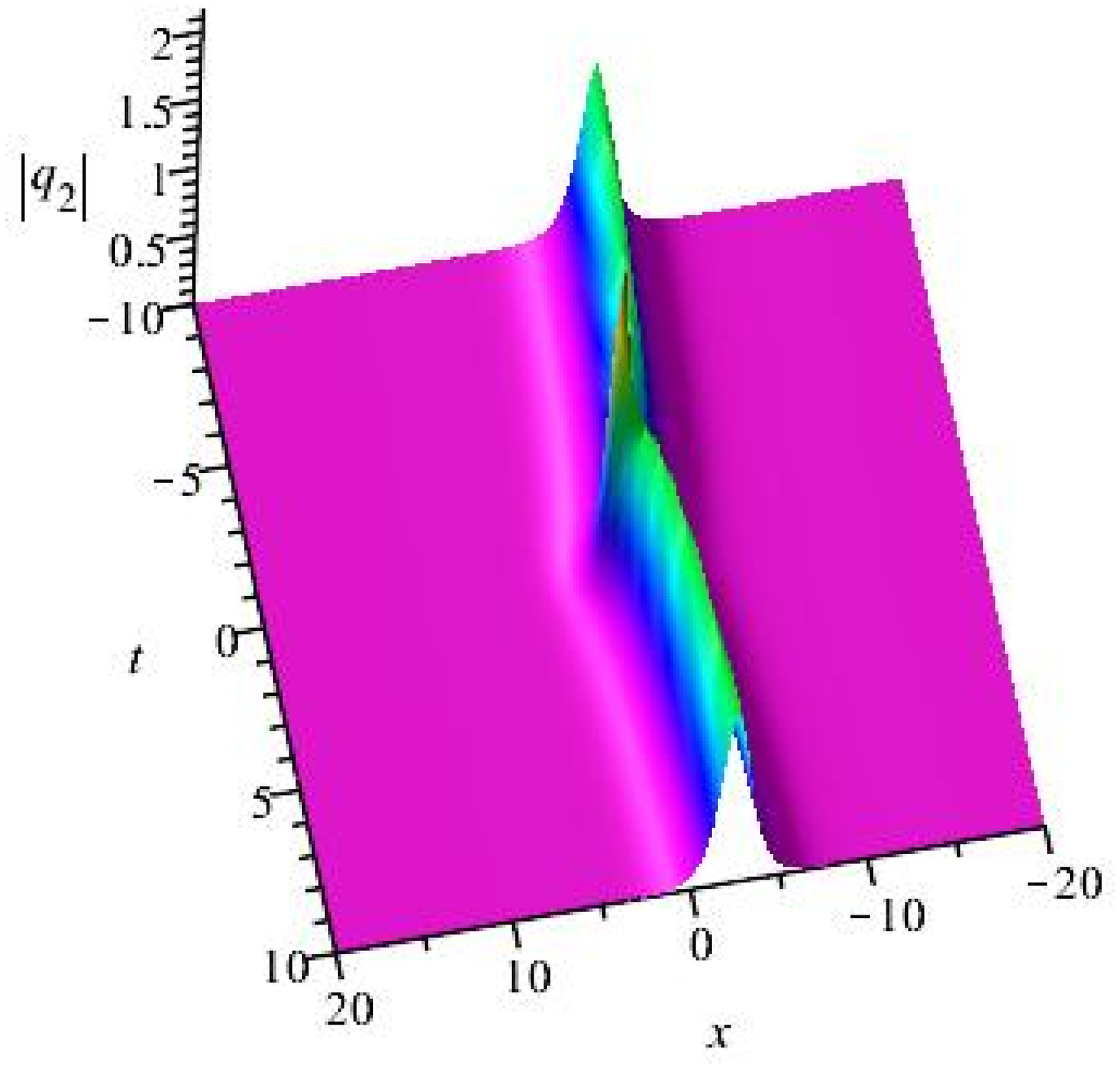}}
\centering
\subfigure[]{\includegraphics[height=0.3\textwidth]{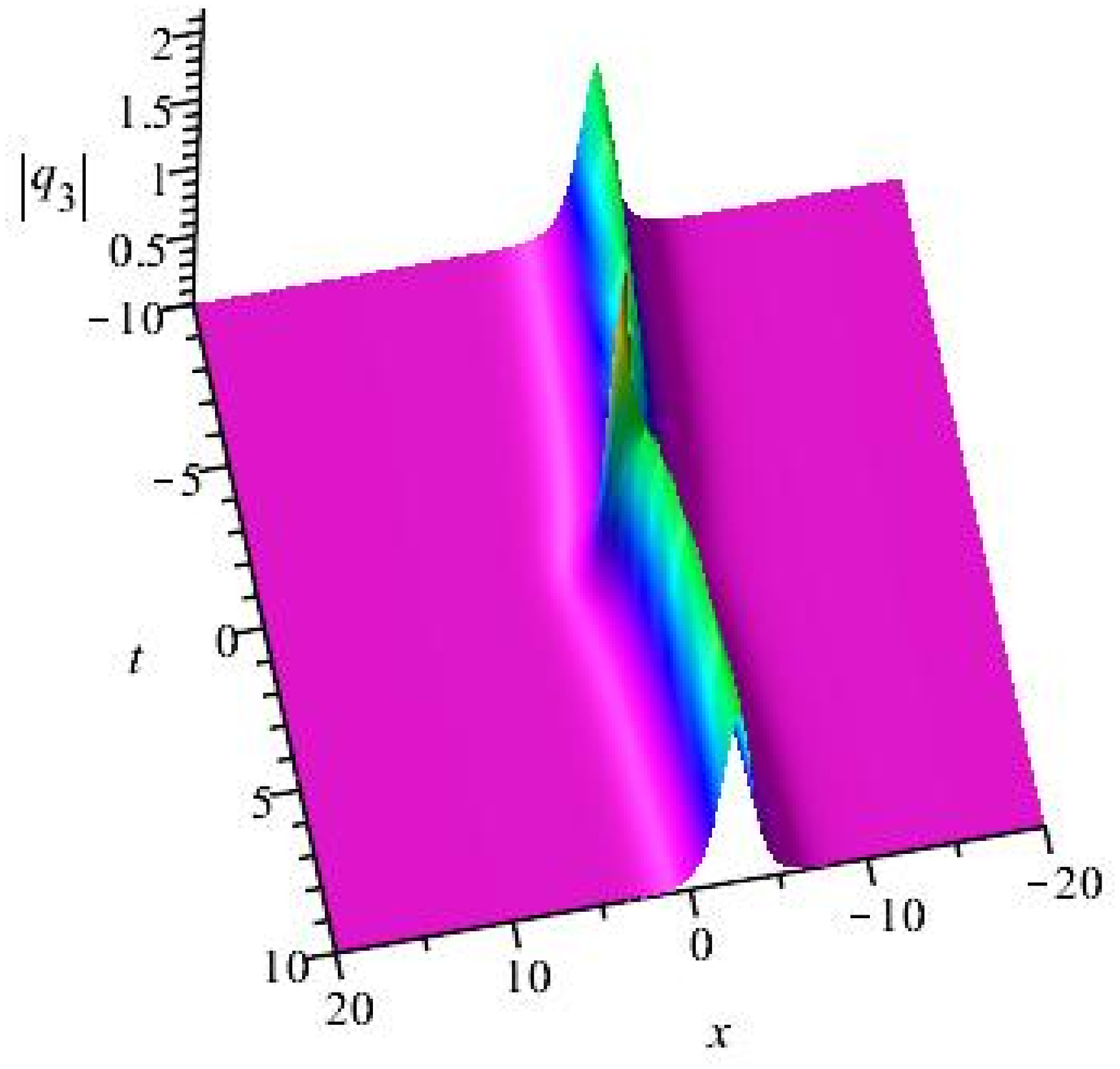}}
\centering
\caption{\small (Color online) Evolution plot of interactional solution between the first-order RW  and one-amplitude-varying soliton or one-bright soliton in the three-component coupled DNLS equations with the parameters chosen by $d_1=2, d_2=0, d_3=0,\alpha=\tfrac{1}{2},\beta=-\tfrac{1}{2}$: (a) a first-order RW merges with an amplitude-varying soliton in $q_1$ component; (b) a first-order RW merges with a bright soliton in $q_2$ component; (c) a first-order RW merges with a bright soliton in $q_3$ component. \label{xt-6f-8}}
\end{figure}

(\textrm{v}) Similar with case (\textrm{iv}), choosing all the disturbing terms are not zero ($\alpha\neq0$, $\beta\neq0$), and two of the three backgrounds are zero,  we can get the following interactional solutions that two components are a first-order RW and a bright soliton, one component is a first-order RW and an amplitude-varying soliton. Setting $d_1\neq0$, $d_2=d_3=0$, a first-order RW merges with a bright soliton in $q_2$ and $q_3$ components, and a first-order RW merges with an amplitude-varying soliton in $q_1$ component, see Fig. \ref{xt-6f-8}.  Analogously, a first-order RW can separate with a bright soliton or an amplitude-varying soliton by decreasing the absolute values of $\alpha$ and $\beta$.

\begin{figure}[H]
\renewcommand{\figurename}{{Fig.}}
\subfigure[]{\includegraphics[height=0.3\textwidth]{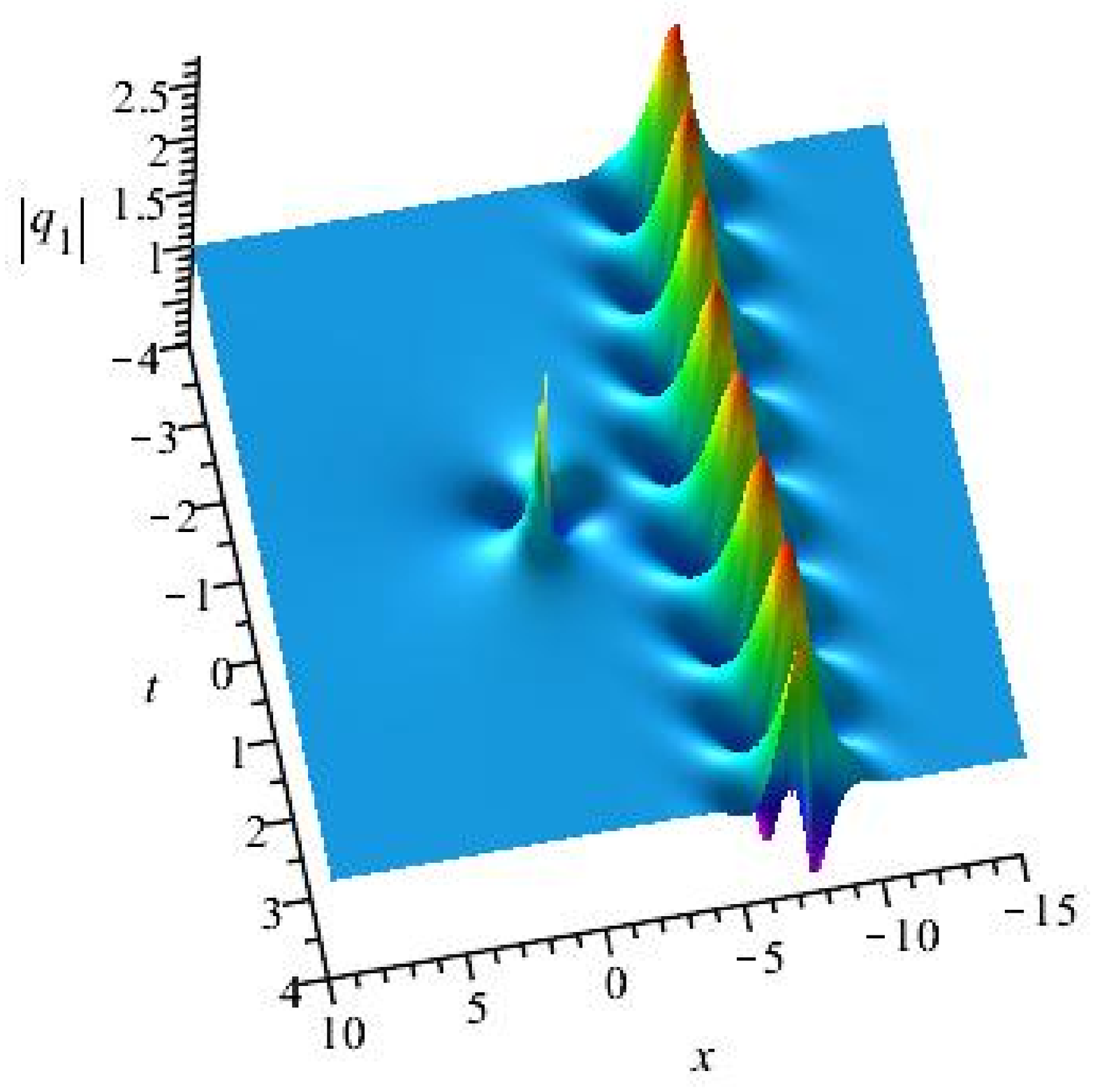}}
\centering
\subfigure[]{\includegraphics[height=0.3\textwidth]{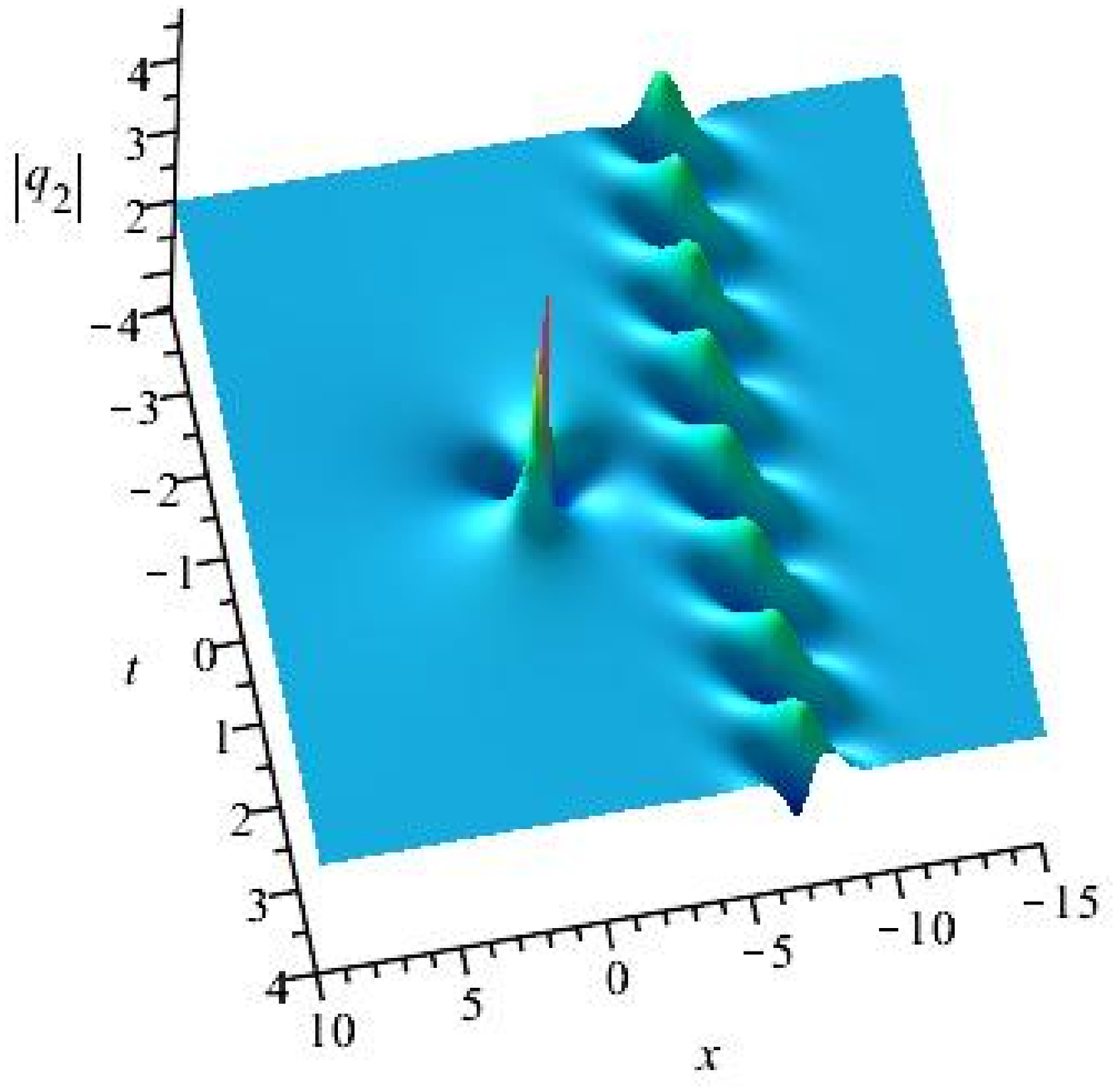}}
\centering
\subfigure[]{\includegraphics[height=0.3\textwidth]{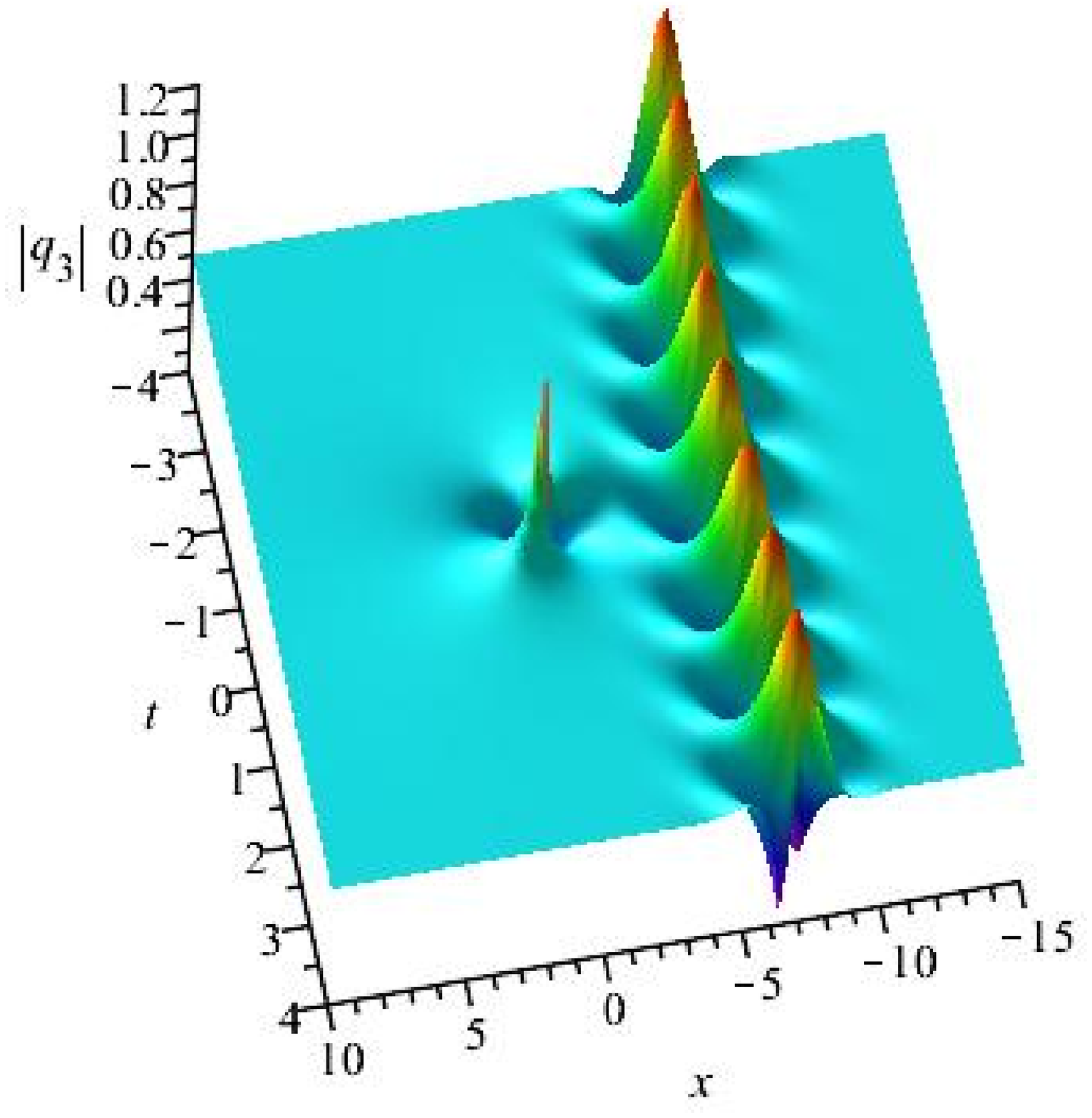}}
\centering
\caption{\small(Color online) Evolution plot of interactional solution between the first-order RW and one-breather in the three-component coupled DNLS equations with the parameters chosen by $d_1=1, d_2=-2, d_3=\tfrac{1}{2},\alpha=\tfrac{1}{20000},\beta=-\tfrac{1}{20000}$, three components are all a first-order separate with a breather: (a) $q_1$, (b) $q_2$, (c) $q_3$. \label{xt-6f-9}}
\end{figure}

(\textrm{vi}) Considering the case that all the disturbing terms are not zero ($\alpha\neq0$, $\beta\neq0$) and all the backgrounds are non-vanished ($d_1\neq0,d_2\neq0,d_3\neq0$), we can obtain the interactional solutions including a first-order and a breather in three components $q_1$, $q_2$ and $q_3$. In Fig. \ref{xt-6f-9}, it demonstrates that a first-order separate with a breather in the three components. In the same way, a breather merges with a breather by increasing the absolute values of $\alpha$ and $\beta$. Besides, it is interesting that the breather in $q_2$ component is different from ones in other two components.

Instead of considering different  arrangements among  the three components $q_1$, $q_2$ and $q_3$, we define the same combination as the same type interactional solution. Considering  the  disturbing coefficients $\alpha$ and $\beta$  as well as three backgrounds ($d_1$, $d_2$, $d_3$)  are zero or not, the first-order interactions of localized waves of the three-component coupled system (\ref{xt-6-1}) can be classified into six types using our method and definition, see Table 1. It is shown that Type 2-5 are four mixed interactions of localized waves among three components $q_1$, $q_2$ and $q_3$. These four mixed interactions of localized waves were also constructed in three-component Hirota equations by us in \cite{6-41} and they cannot be constructed in two-component system \cite{6-38,6-39,6-42} using the same method. Additionally, we can draw a conclusion that these kinds of mixed interactions of localized waves may only be obtained by DT in the nonlinear systems,  whose components are more than 3 with the corresponding Lax pair including the matrices larger than $3\times3$.

\begin{center}
{\footnotesize{Table 1.} Six types of the first-order interactions of localized waves\\
\vspace{2mm}
\begin{tabular}{cccc}
  \hline
  Types & $q_j~(j=1,2,3)$\\
  \hline
  Type 1& three components are all first-order RW \\
  Type 2& the two components are  RW and one-breather, and another one is  RW and one-amplitude-varying soliton \\
  Type 3& the two components are RW and one-amplitude-varying soliton, and another one is  RW and one-bright soliton\\
  Type 4& the two components are RW and one-breather, and another one is  RW and one-bright soliton\\
  Type 5& the two components are  RW and one-bright soliton, and another one is RW and one-amplitude-varying soliton\\
  Type 6& three components are all the first-order  RW and one-breather\\
  \hline
\end{tabular}}
\end{center}

Iterating one time of DT for the three-component coupled DNLS equations (\ref{xt-6-1}), the second-order interactions of localized waves can be generated in the following content through the formulae (\ref{xt-6-17})-(\ref{xt-6-18}). However, the concrete general expressions of $q_1[2]$, $q_2[2]$ and $q_3[2]$ are very tedious and complicated, we omit these equations and only give the corresponding figures. Similar to the first-order case, the second-order interactional  solutions are also classified into six types. In the same way, considering the disturbing coefficients $\alpha$ and $\beta$, and the different plane backgrounds (vanished or non-vanished), the second-order interactional solutions of the three-component coupled DNLS equations (\ref{xt-6-1}) are discussed in detail.

\begin{figure}[H]
\renewcommand{\figurename}{{Fig.}}
\subfigure[]{\includegraphics[height=0.3\textwidth]{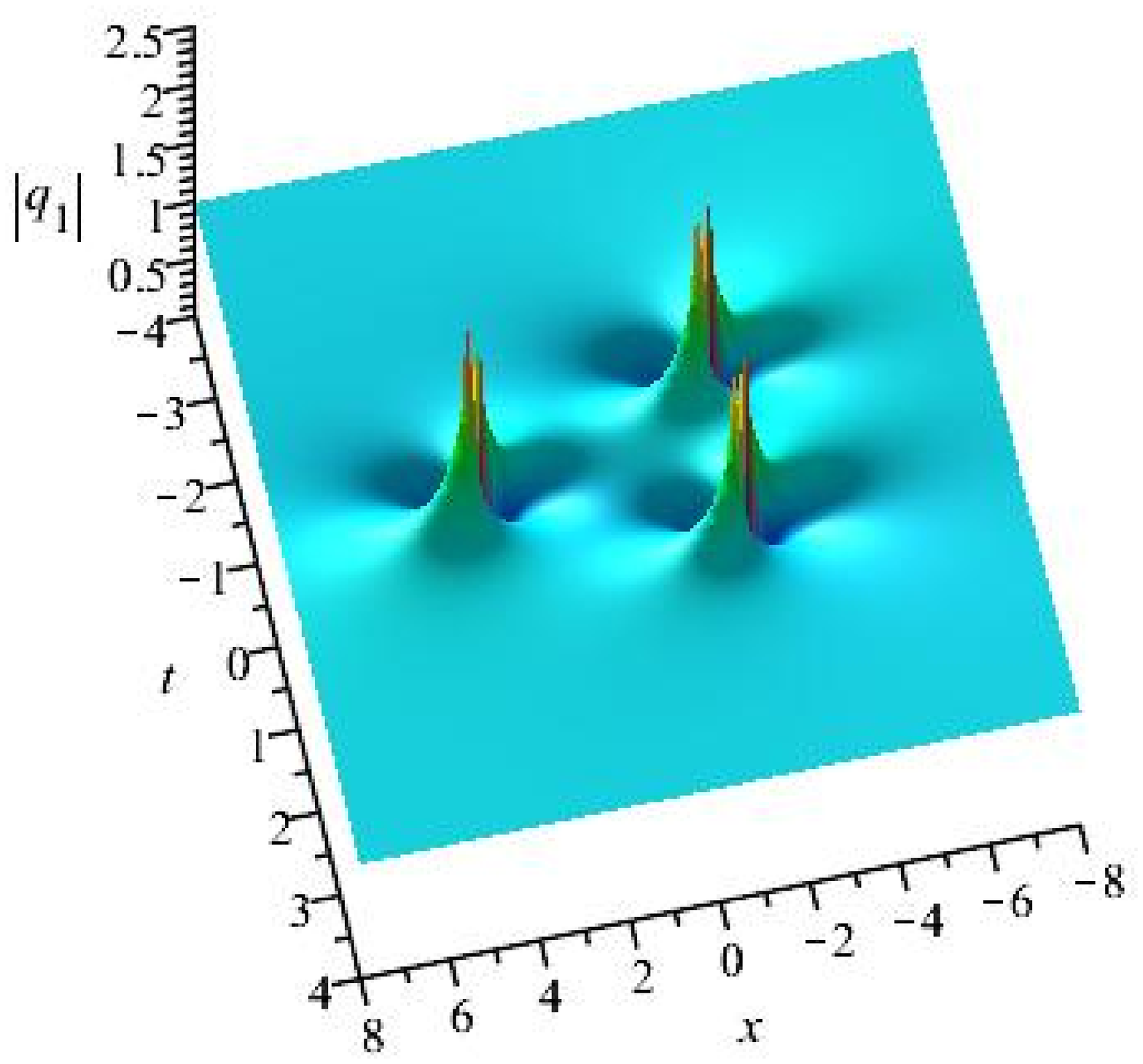}}
\centering
\subfigure[]{\includegraphics[height=0.3\textwidth]{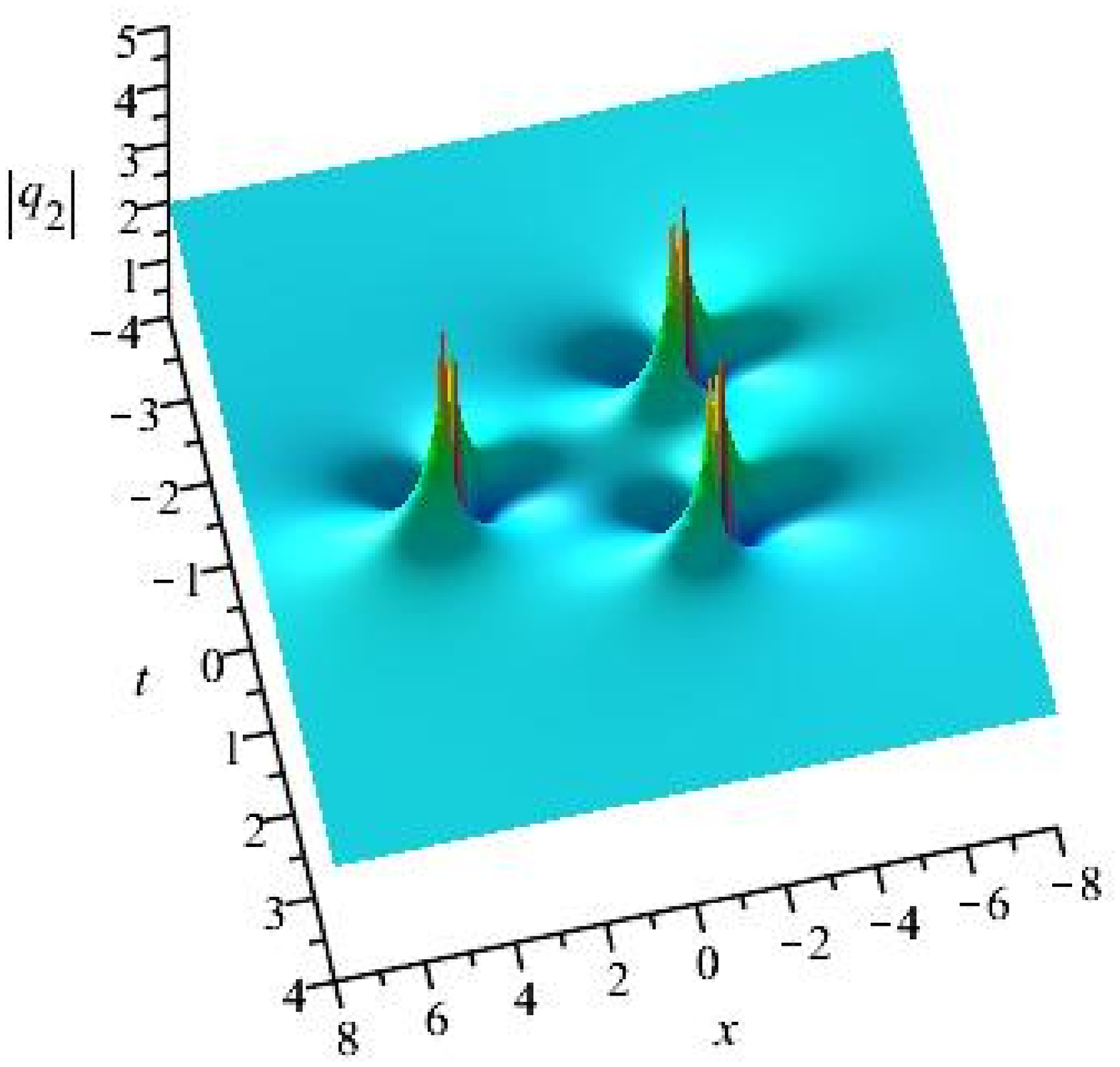}}
\centering
\subfigure[]{\includegraphics[height=0.3\textwidth]{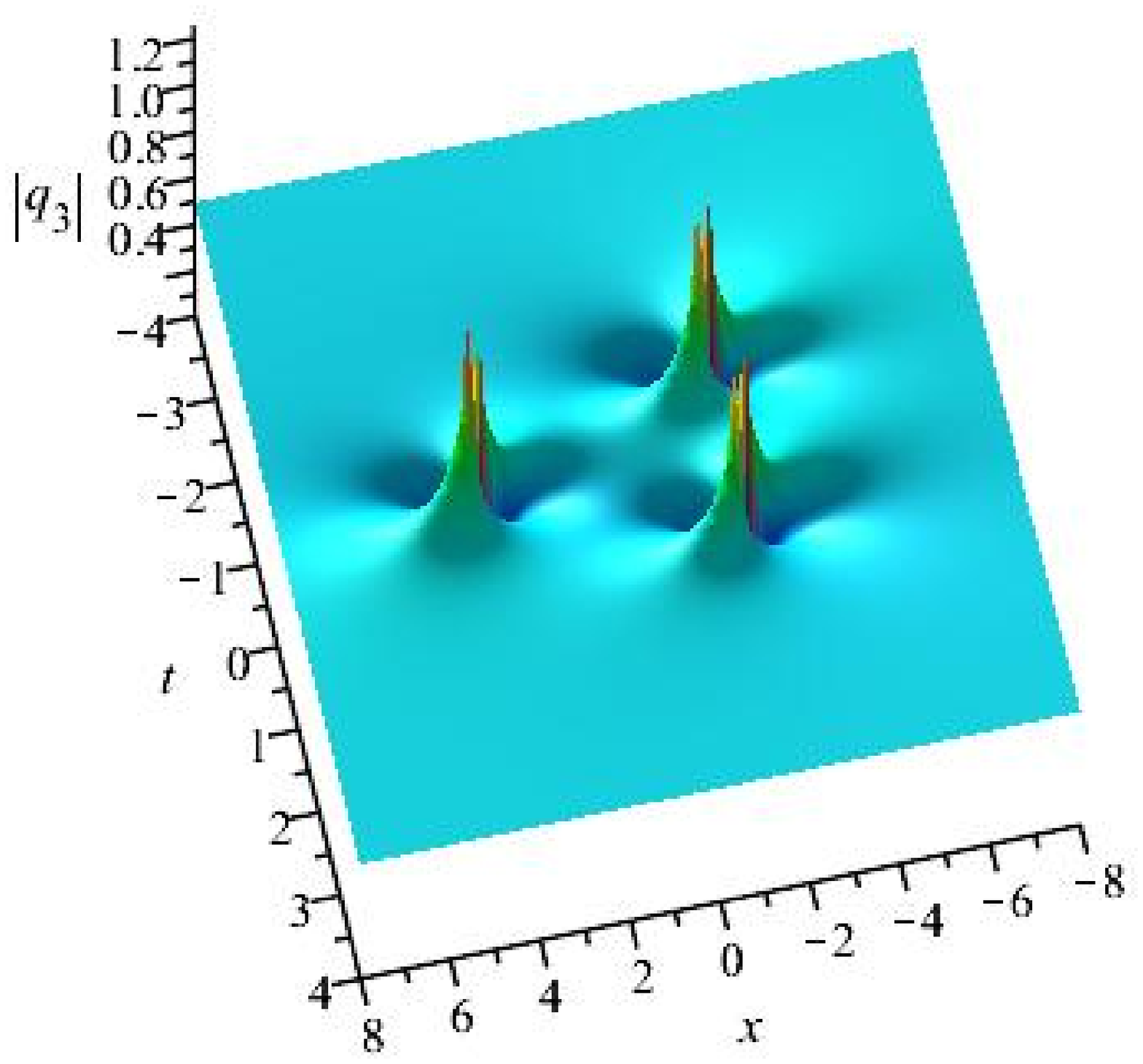}}
\centering
\caption{\small(Color online) Evolution plot of the second-order RW of triangular pattern  with the parameters chosen by $d_1=1, d_2=-2, d_3=\tfrac{1}{2},\alpha=\beta=0,m_1=100,n_1=-100$: (a) $q_1$, (b) $q_2$, (c) $q_3$.\label{xt-6f-10}}
\end{figure}

(\textrm{i}) When $\alpha=\beta=0$ and $d_j\neq0~(j=1,2,3)$, all the disturbing terms are removed and three plane backgrounds are all non-vanished, the interactional solutions degenerate to the second-order rational ones. In this case, the three components $q_1$, $q_2$ and $q_3$ are all proportionable to each other, and they are the second-order RW. If $m_1=n_1=0$, they are the fundamental second-order RW; whereas $m_1\neq0,n_1\neq0$, it is shown that the fundamental second-order RW splits into three first-order ones and they form a triangle pattern in Fig. \ref{xt-6f-10}.

\begin{figure}[H]
\renewcommand{\figurename}{{Fig.}}
\subfigure[]{\includegraphics[height=0.3\textwidth]{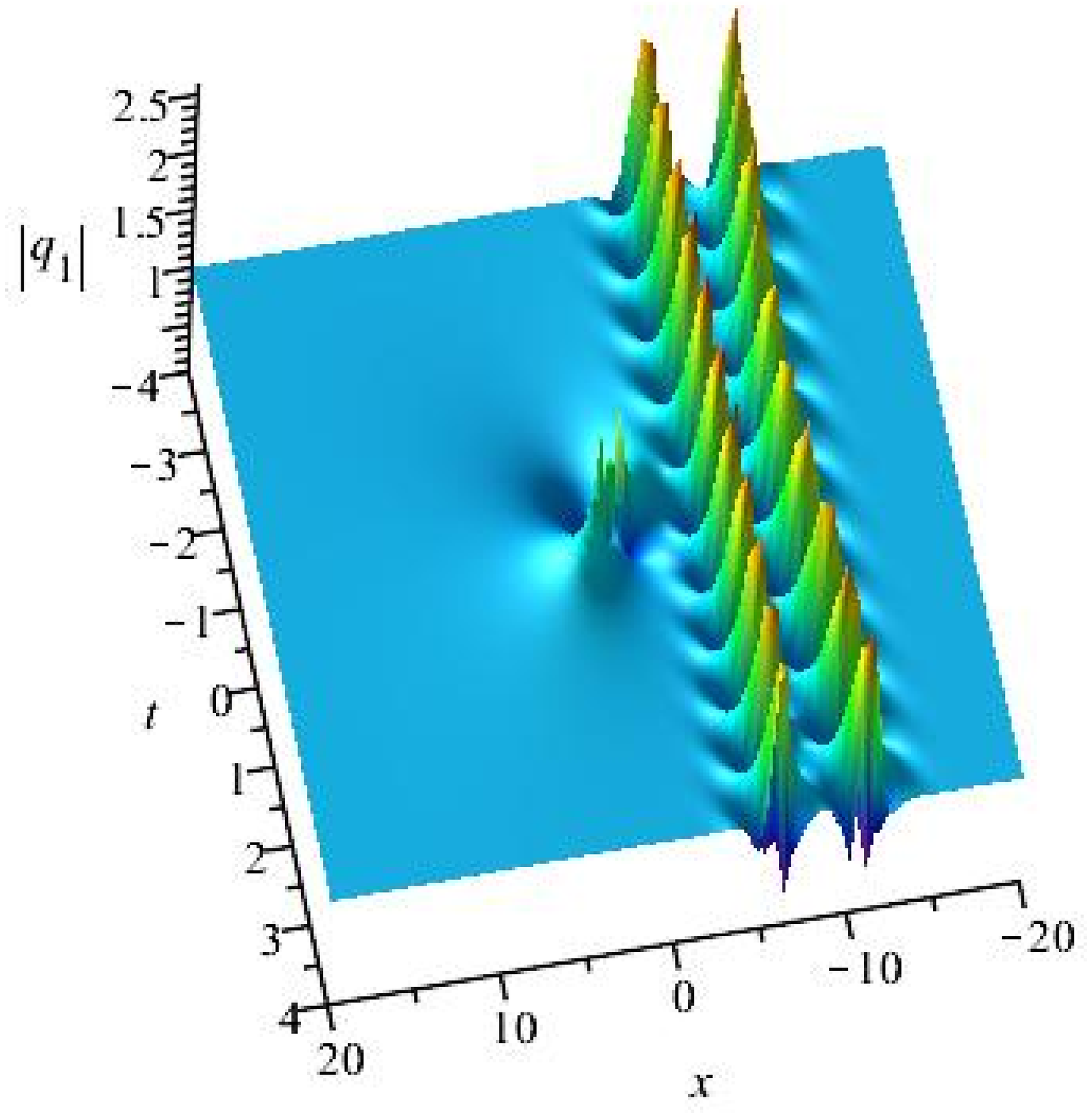}}
\centering
\subfigure[]{\includegraphics[height=0.3\textwidth]{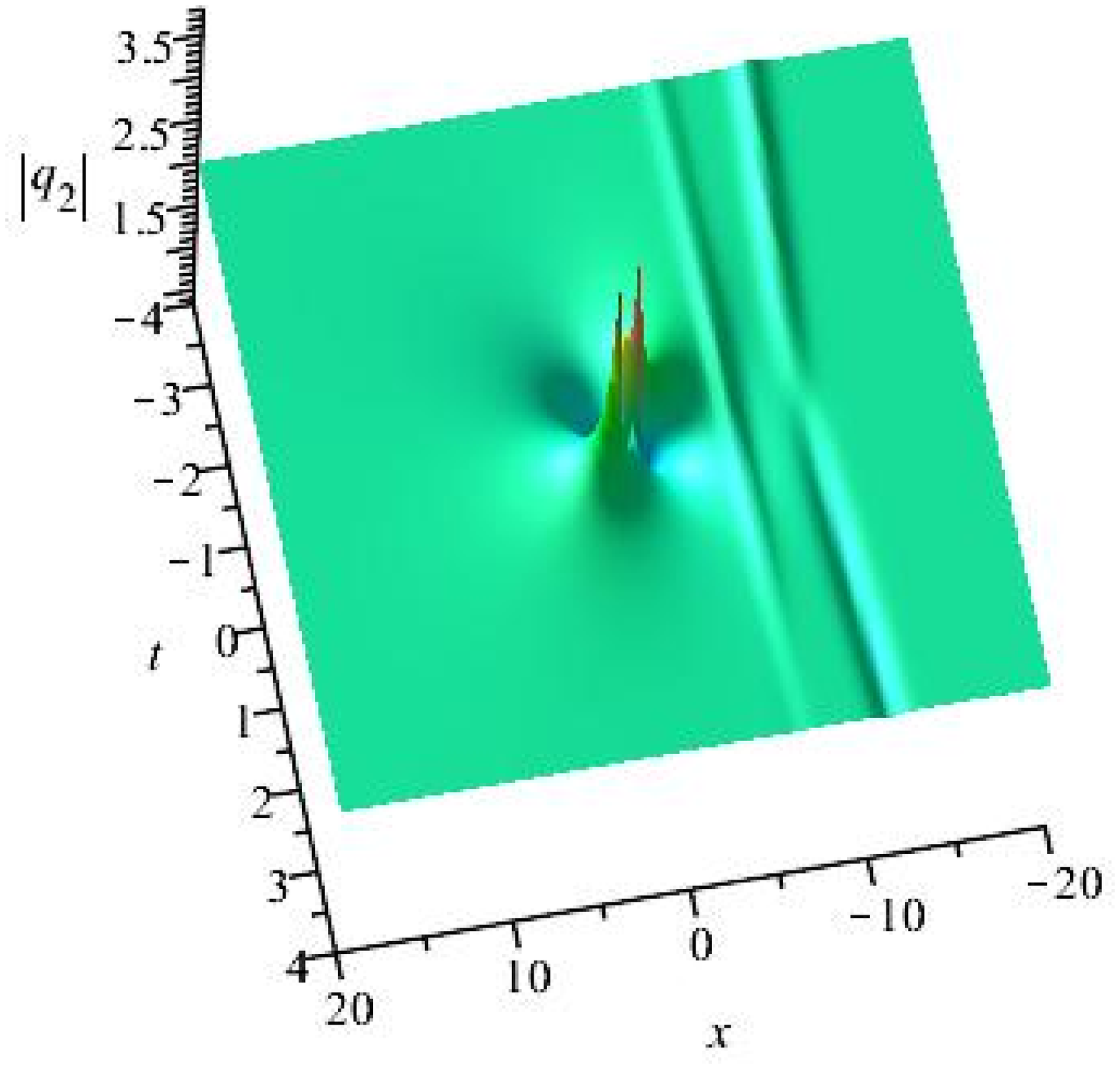}}
\centering
\subfigure[]{\includegraphics[height=0.3\textwidth]{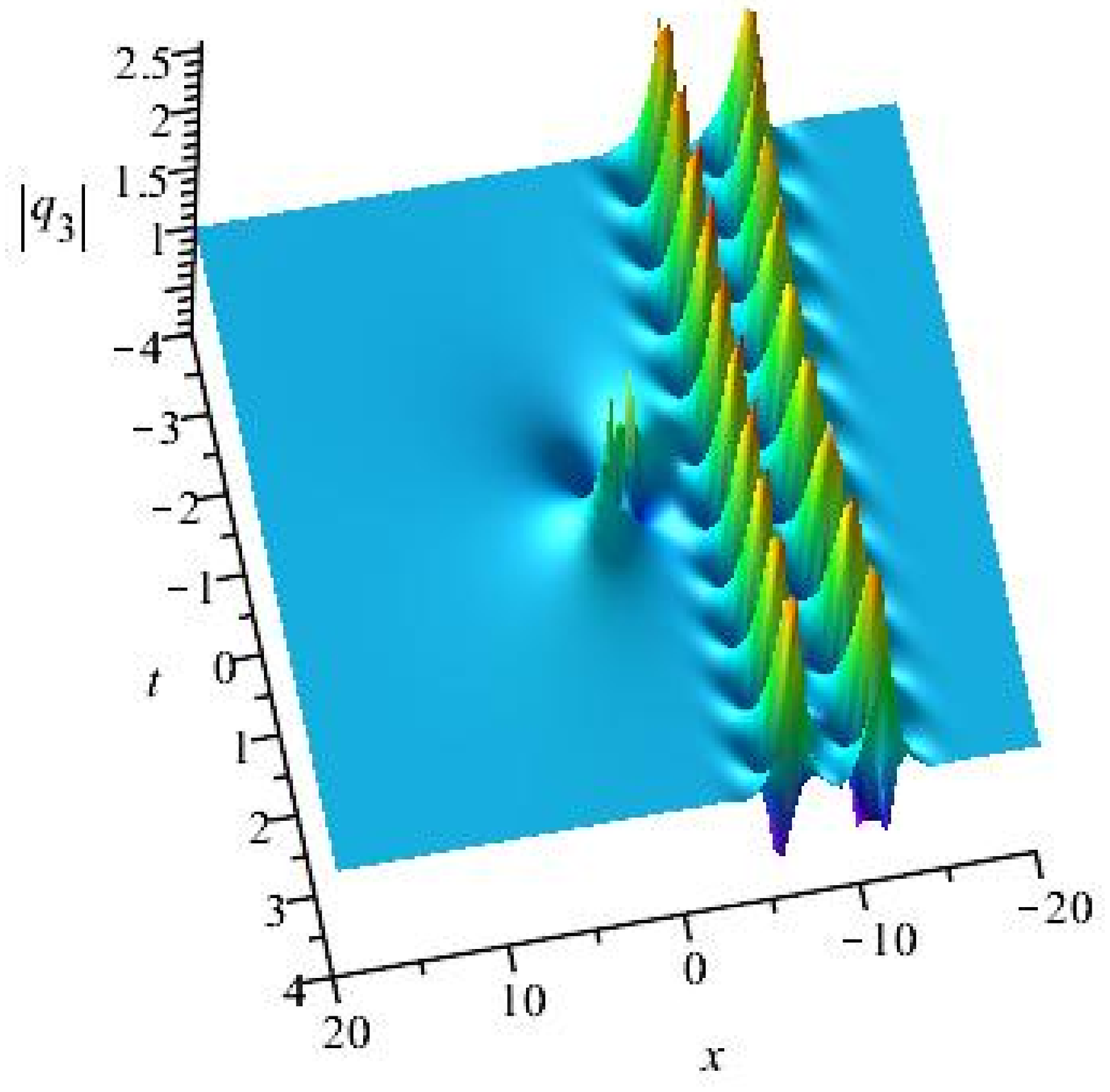}}
\centering
\caption{\small(Color online) Evolution plot of the interactional solution between the fundamental second-order RW and two-breather or two-amplitude-varying soliton in the three-component coupled DNLS equations with the parameters chosen by $d_1=1, d_2=-2, d_3=-1,\alpha=0,\beta=\tfrac{1}{200000},m_1=0,n_1=0$: (a) a fundamental second-order RW and two breathers split in $q_1$ component; (b) a fundamental second-order  RW and two amplitude-varying solitons split in $q_2$ component; (c) a fundamental second-order RW and two breathers split in $q_3$ component.\label{xt-6f-11}}
\end{figure}

\begin{figure}[H]
\renewcommand{\figurename}{{Fig.}}
\subfigure[]{\includegraphics[height=0.24\textwidth]{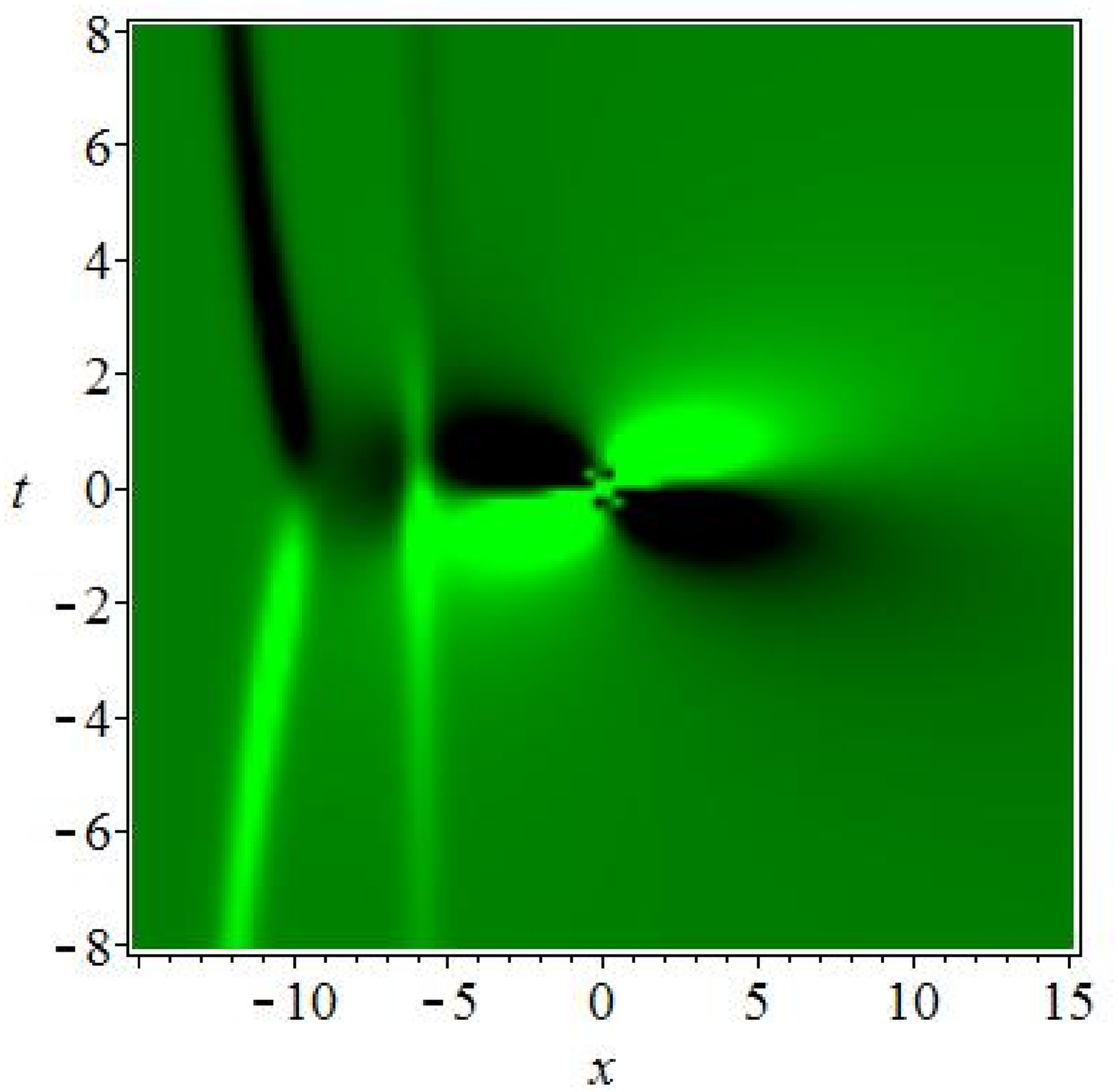}}
\centering
\subfigure[]{\includegraphics[height=0.24\textwidth]{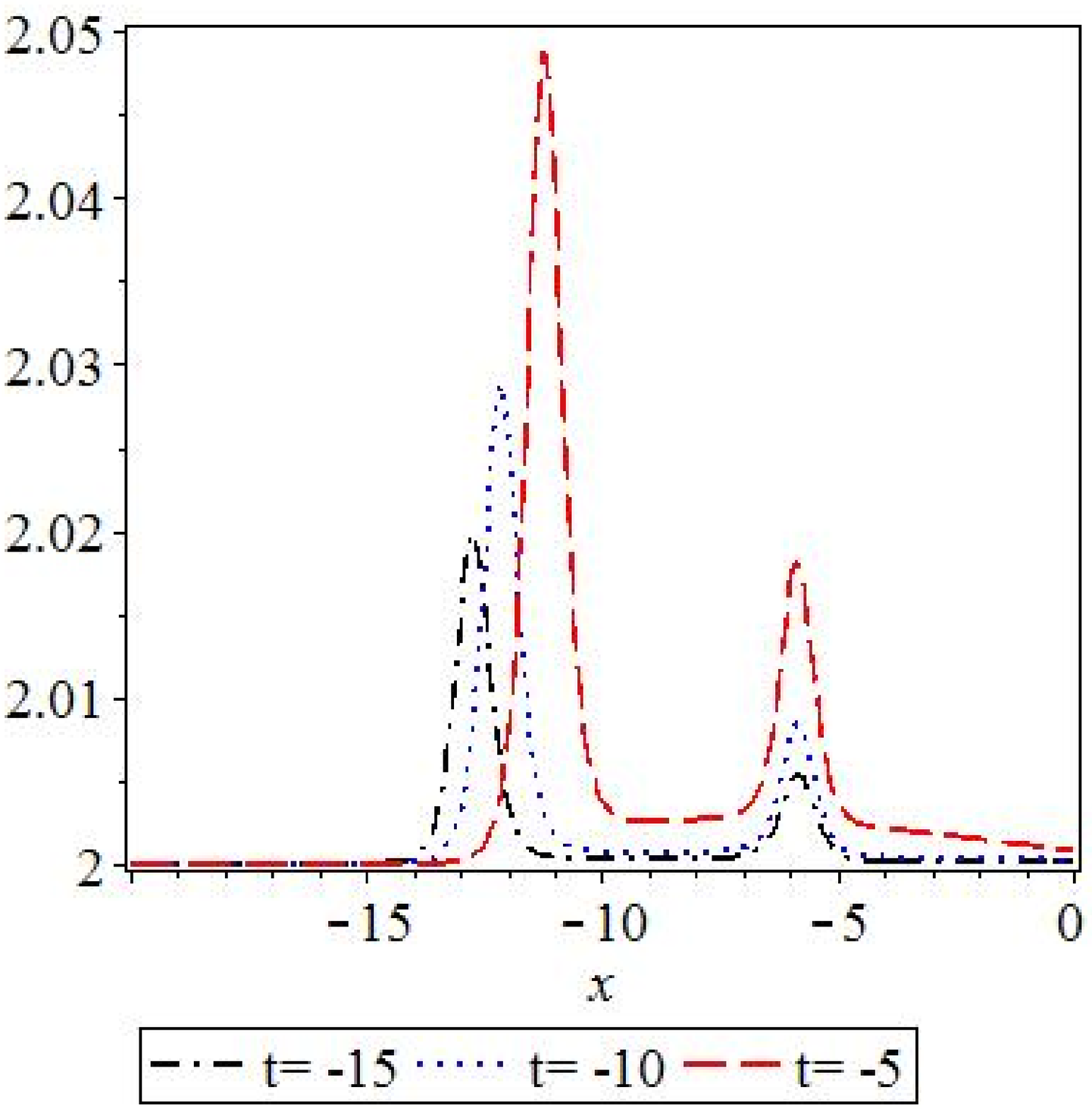}}
\centering
\subfigure[]{\includegraphics[height=0.24\textwidth]{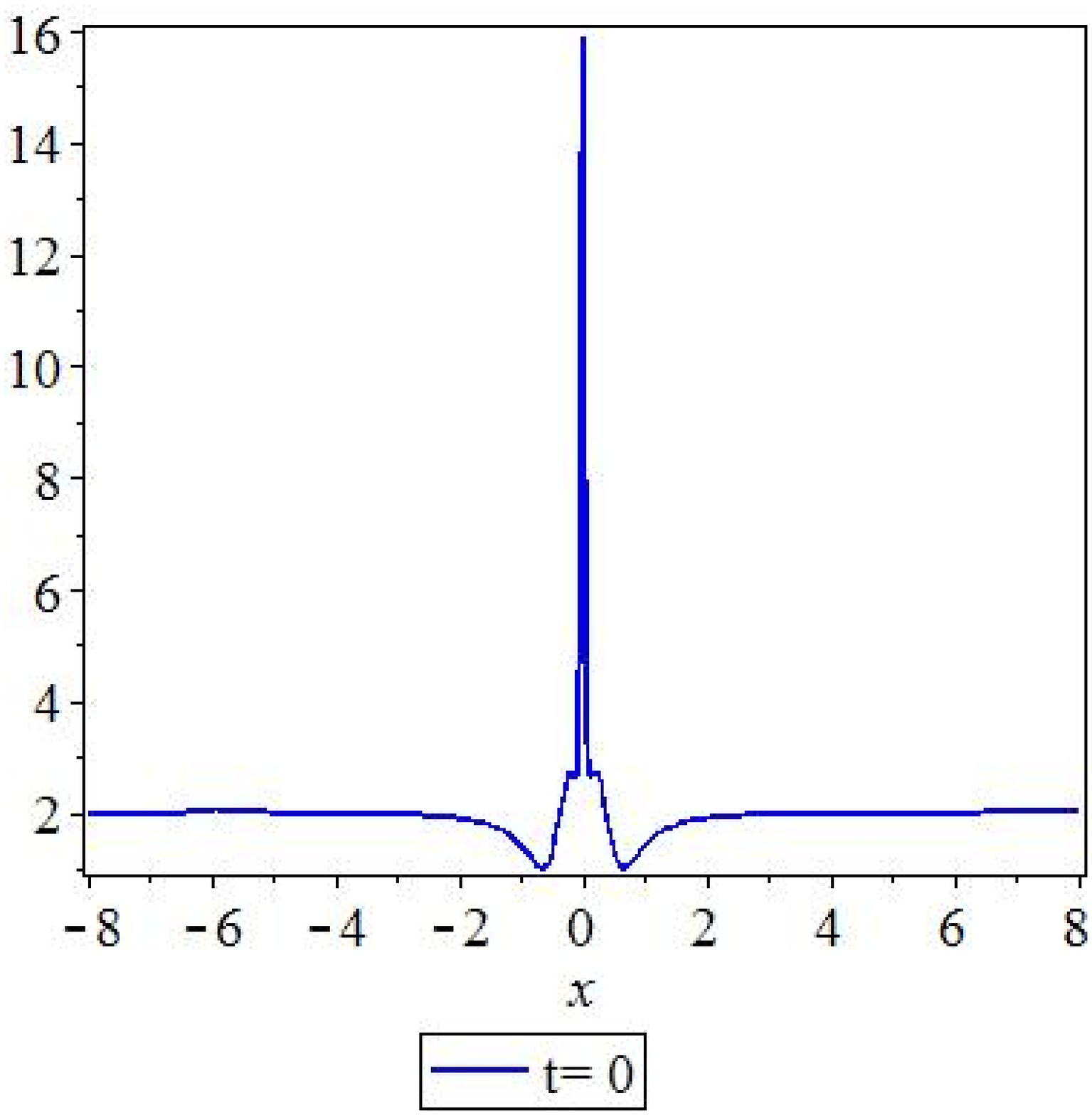}}
\centering
\subfigure[]{\includegraphics[height=0.24\textwidth]{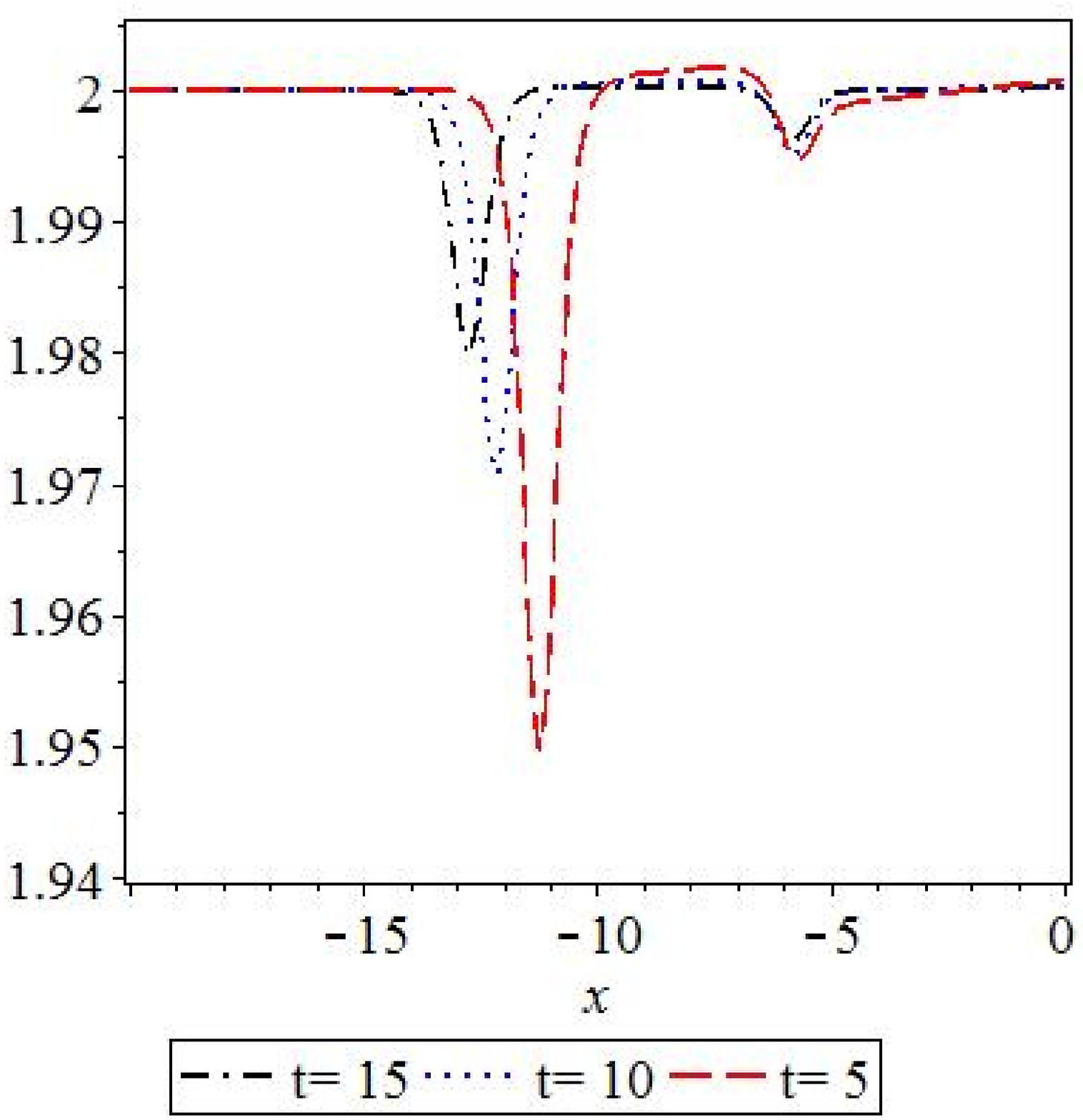}}
\centering
\caption{\small(Color online) (a) Evolution density plot  of $q_2$ component in Fig. \ref{xt-6f-11}(b). Plane evolution plot of the interactional process between the fundamental second-order RW and the two-amplitude-varying soliton of  $q_2$ component in Fig. \ref{xt-6f-2}(b) in different moments: (b) $t<0$; (c) $t=0$; (d) $t>0$.\label{xt-6f-12}}
\end{figure}

(\textrm{ii}) When $\alpha=0,\beta\neq0$ and $d_j\neq0~(j=1,2,3)$, one of the three disturbing terms is zero and three plane backgrounds are all non-vanished. If $m_1=n_1=0$, $q_1$ and $q_3$ components are all the hybrid solution between a fundamental RW and two breathers, and $q_2$ is the hybrid solution between a fundamental RW and two amplitude-varying solitons, see Fig. \ref{xt-6f-11}. In Fig. \ref{xt-6f-12}(a), the density plot of $q_2$ component indicates that the right amplitude-varying soliton in Fig. \ref{xt-6f-11}(b) is the same as the one in Fig. \ref{xt-6f-2}(b) and it annihilates at $t=0$. But, the left amplitude-varying soliton in Fig. \ref{xt-6f-11}(b) is very different from the one in the first-order case. It demonstrates that the left amplitude-varying soliton in Fig. \ref{xt-6f-11}(b) annihilates at about $t=2$ from Fig. \ref{xt-6f-12}(a). Furthermore, it can be seen that the above mentioned left amplitude-varying soliton in Fig. \ref{xt-6f-11}(b) is anti-dark soliton if $t<2$ and becomes dark soliton if $t>2$ from Fig. \ref{xt-6f-12}(a).

\begin{figure}[H]
\renewcommand{\figurename}{{Fig.}}
\subfigure[]{\includegraphics[height=0.3\textwidth]{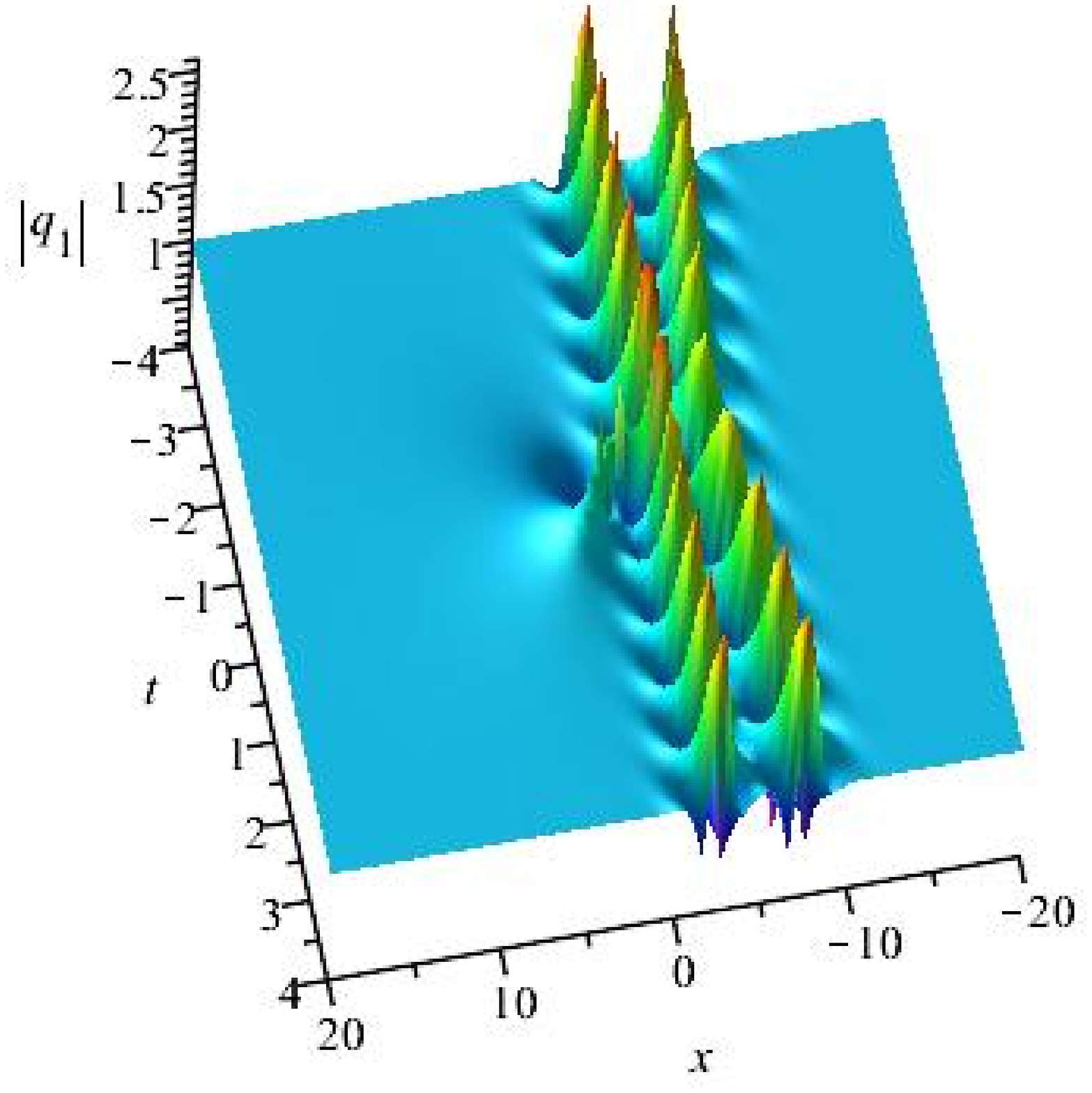}}
\centering
\subfigure[]{\includegraphics[height=0.3\textwidth]{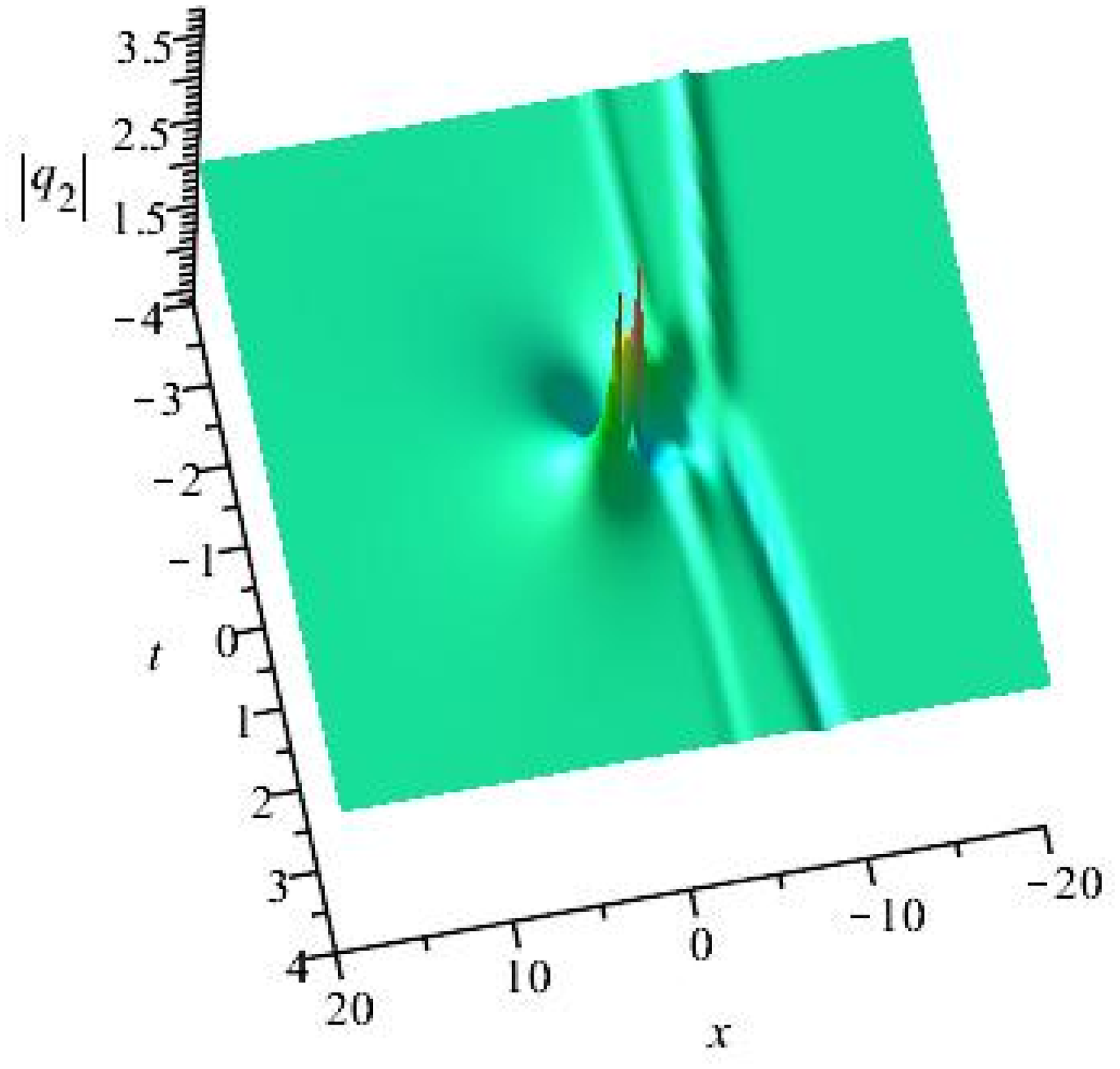}}
\centering
\subfigure[]{\includegraphics[height=0.3\textwidth]{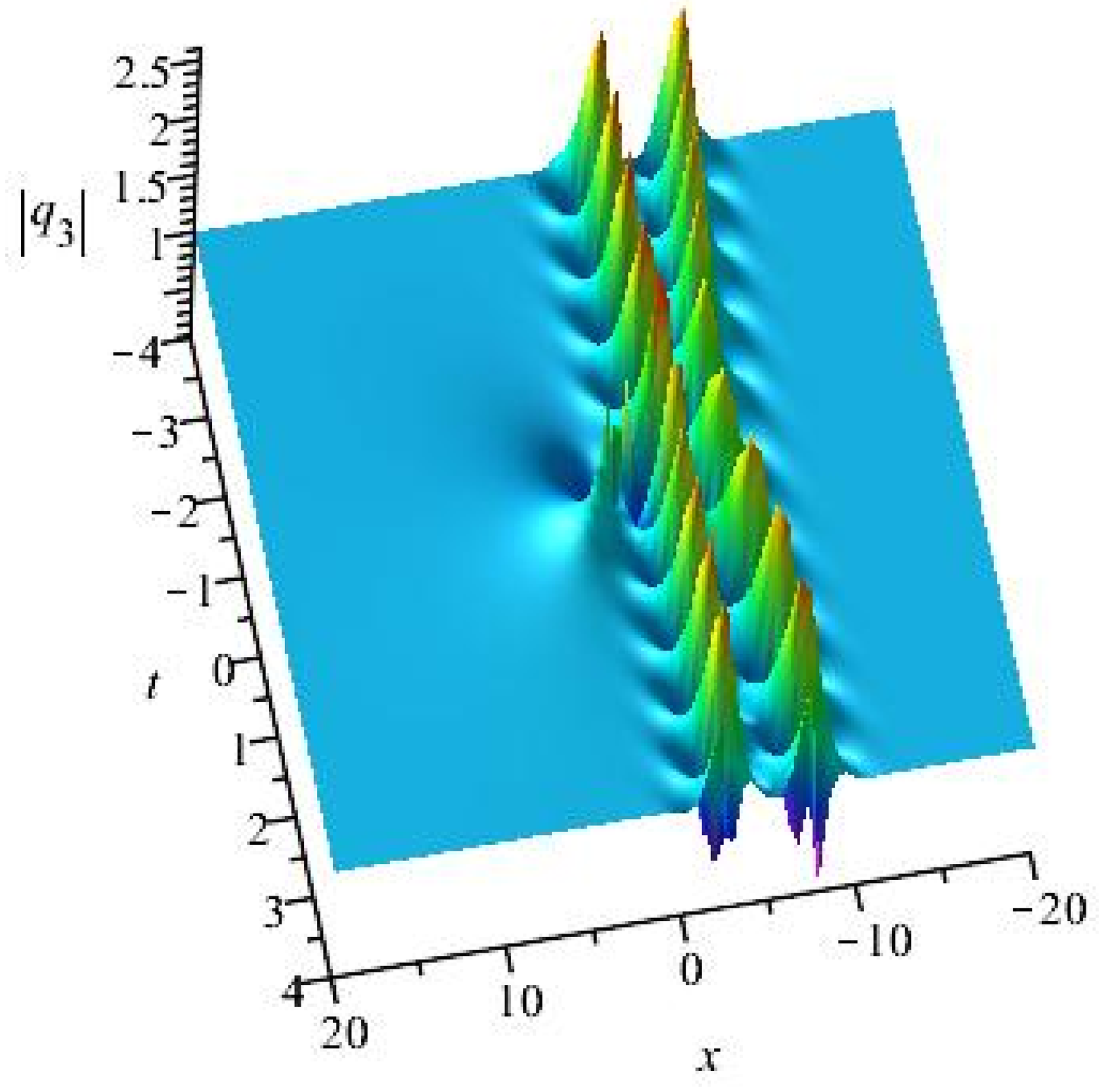}}
\centering
\caption{\small(Color online) Evolution plot of the interactional solution between the fundamental second-order RW and two-breather or two-amplitude-varying soliton in the three-component coupled DNLS equations with the parameters chosen by $d_1=1, d_2=-2, d_3=-1,\alpha=0,\beta=\tfrac{1}{200},m_1=0,n_1=0$: (a) a fundamental second-order  RW merges with two breathers in $q_1$ component; (b) a fundamental second-order RW merges with two amplitude-varying solitons in $q_2$ component; (c) a fundamental second-order RW merges with two breathers in $q_3$ component.\label{xt-6f-13}}
\end{figure}

\begin{figure}[H]
\renewcommand{\figurename}{{Fig.}}
\subfigure[]{\includegraphics[height=0.3\textwidth]{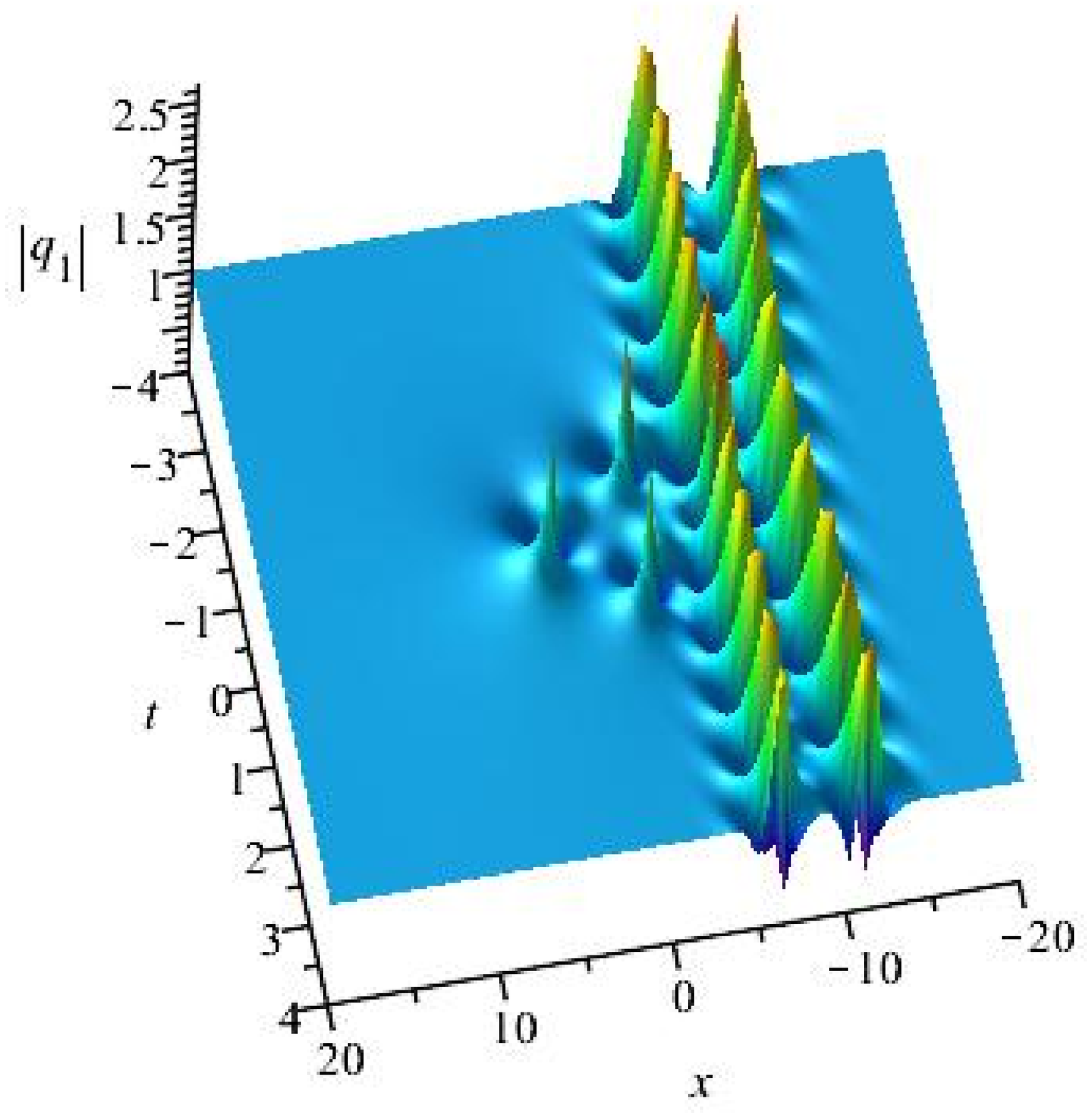}}
\centering
\subfigure[]{\includegraphics[height=0.3\textwidth]{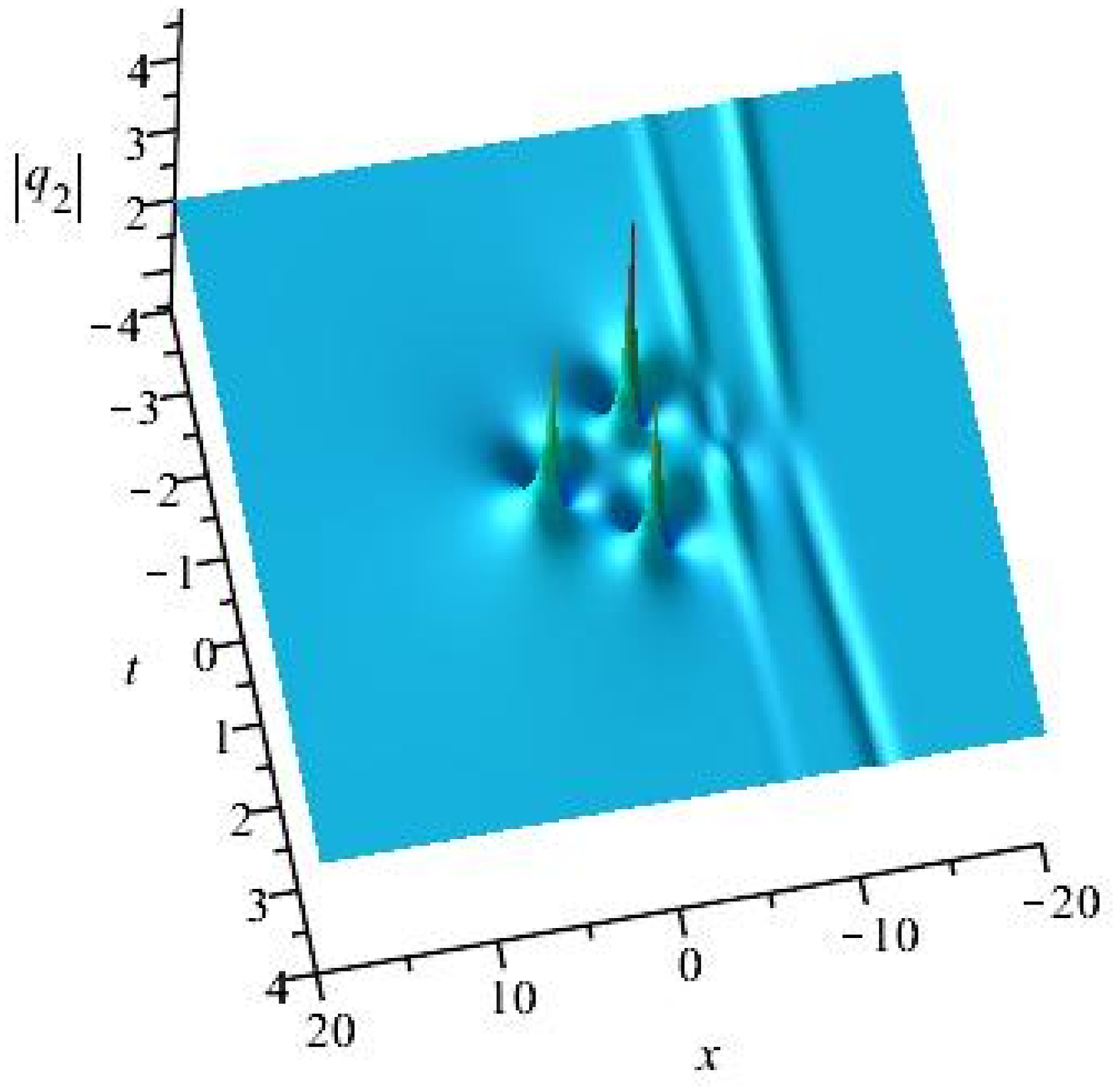}}
\centering
\subfigure[]{\includegraphics[height=0.3\textwidth]{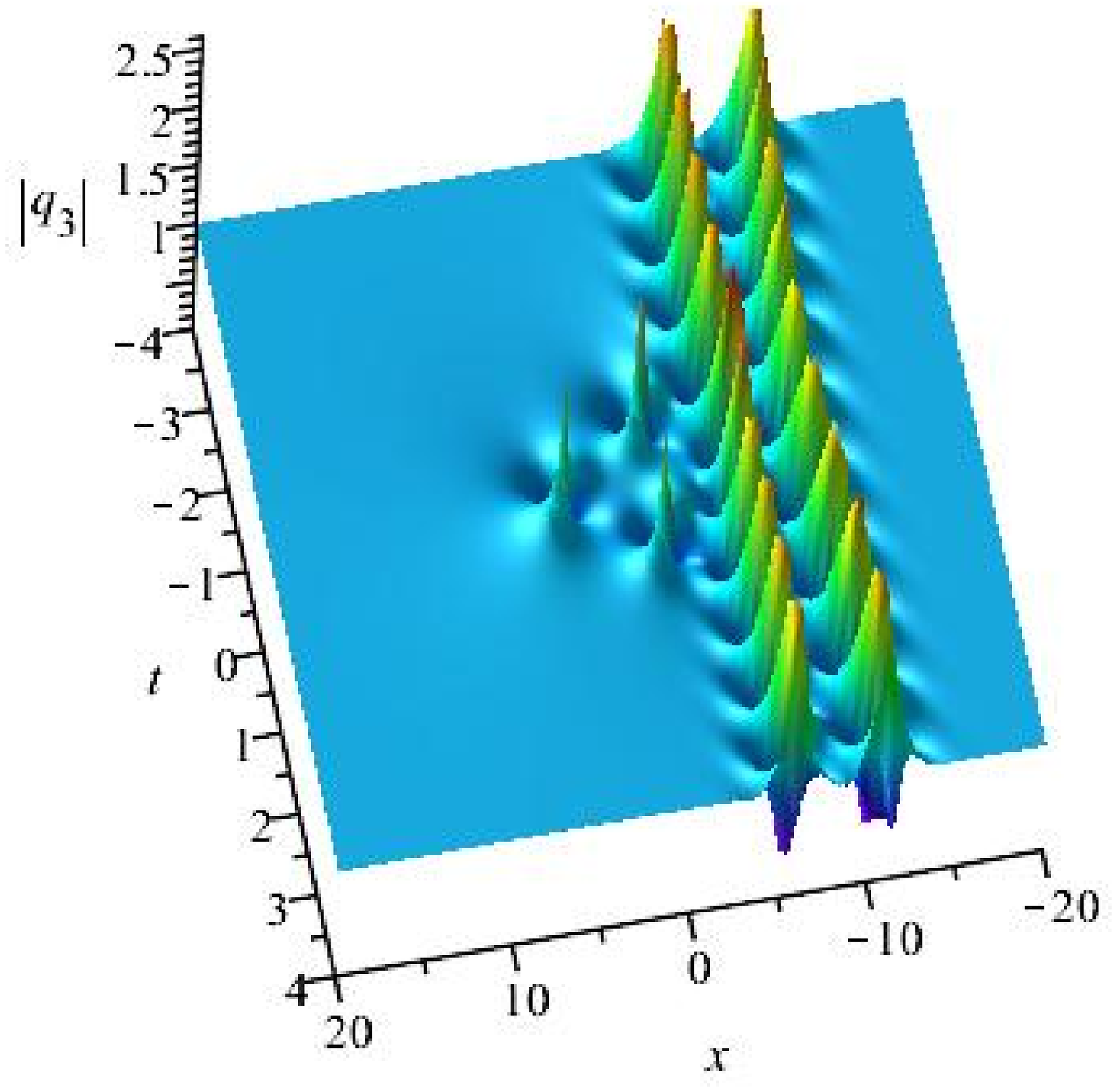}}
\centering
\caption{\small(Color online) Evolution plot of the interactional solution between the second-order RW of triangular pattern and two-breather or two-amplitude-varying soliton in the three-component coupled DNLS equations with the parameters chosen by $d_1=1, d_2=-2, d_3=-1,\alpha=0,\beta=\tfrac{1}{200000},m_1=100,n_1=-100$: (a) a second-order RW of triangular pattern and two breathers split in $q_1$ component; (b) a second-order  RW of triangular pattern and two amplitude-varying solitons split in $q_2$ component; (c) a second-order RW of triangular pattern and two breathers split in $q_3$ component.\label{xt-6f-14}}
\end{figure}

For discussing the two amplitude-varying soliton together, we choose the following time periods $t<0$ and $t>2$. When $t<0$, the amplitudes of two anti-dark solitons in $q_2$ component becomes big with $t$ increasing, see Fig. \ref{xt-6f-12}(b); otherwise, if $t>2$,  the amplitudes of dark soliton in $q_2$ component becomes small with $t$ increasing, see Fig. \ref{xt-6f-13}(d). In Fig. \ref{xt-6f-13}(c), the fundamental second-order RW appear at $t=0$ and its amplitude is about 8 times of the plane background since it is nonlinear superposition of three first-order RWs. When $t=0$,  the left amplitude-varying soliton in Fig. \ref{xt-6f-11}(b) becomes anti-dark soliton and coexists with the fundamental second-order RW in Fig. \ref{xt-6f-12}(b). At this moment, the amplitude of this  anti-dark soliton is small, so we cannot observe it from   Fig. \ref{xt-6f-12}(b). Additionally, the fundamental second-order RW merges with two breathers or two amplitude-varying solitons by increasing the absolute values of  disturbing coefficients $\alpha$ and $\beta$, see Fig. \ref{xt-6f-13}. When $m_1\neq0,n_1\neq0$, the fundamental second-order RW in Fig. \ref{xt-6f-11} can split into  three first-order RWs and form a  triangle pattern in Fig. \ref{xt-6f-14}.

\begin{figure}[H]
\renewcommand{\figurename}{{Fig.}}
\subfigure[]{\includegraphics[height=0.3\textwidth]{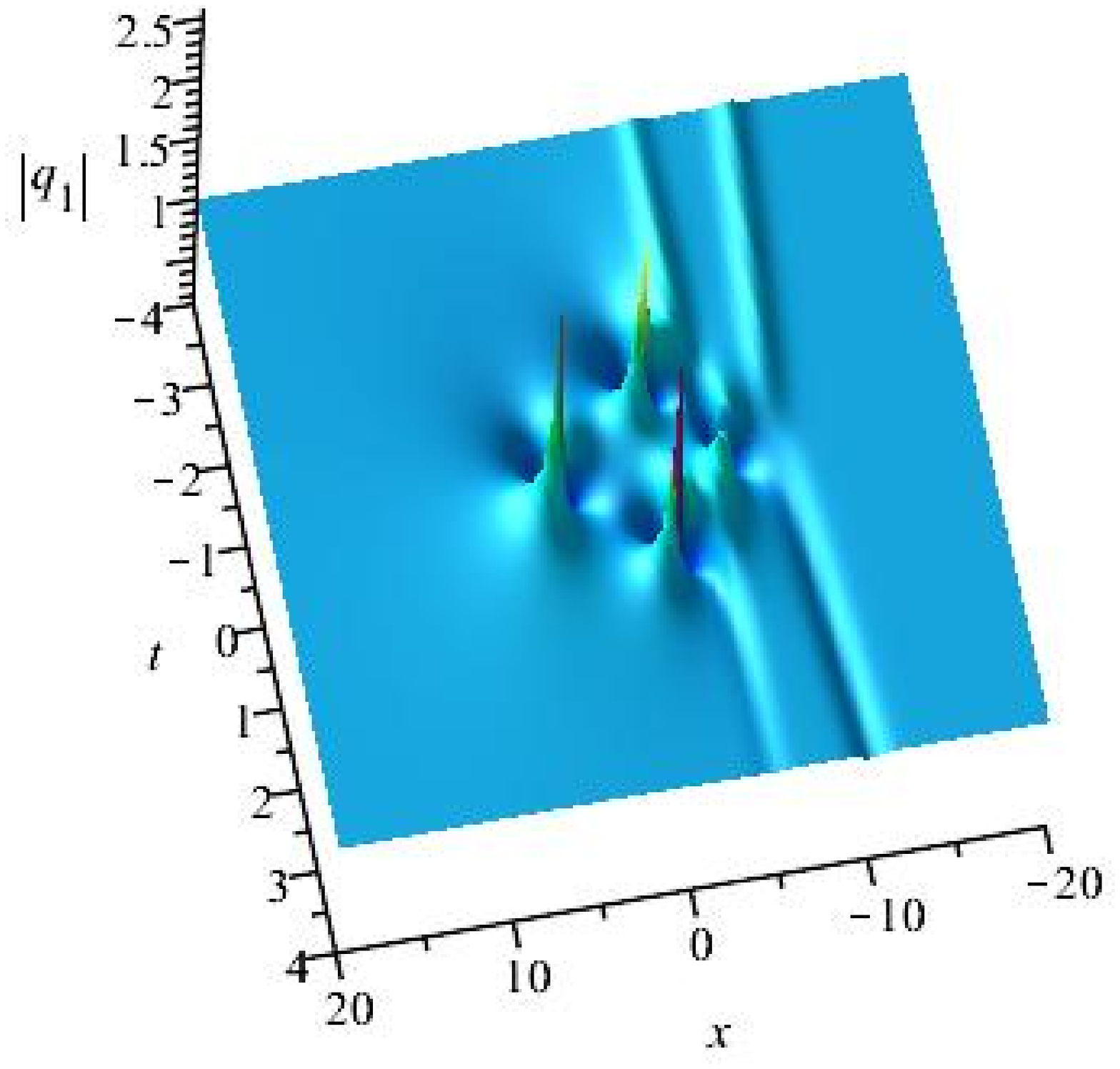}}
\centering
\subfigure[]{\includegraphics[height=0.3\textwidth]{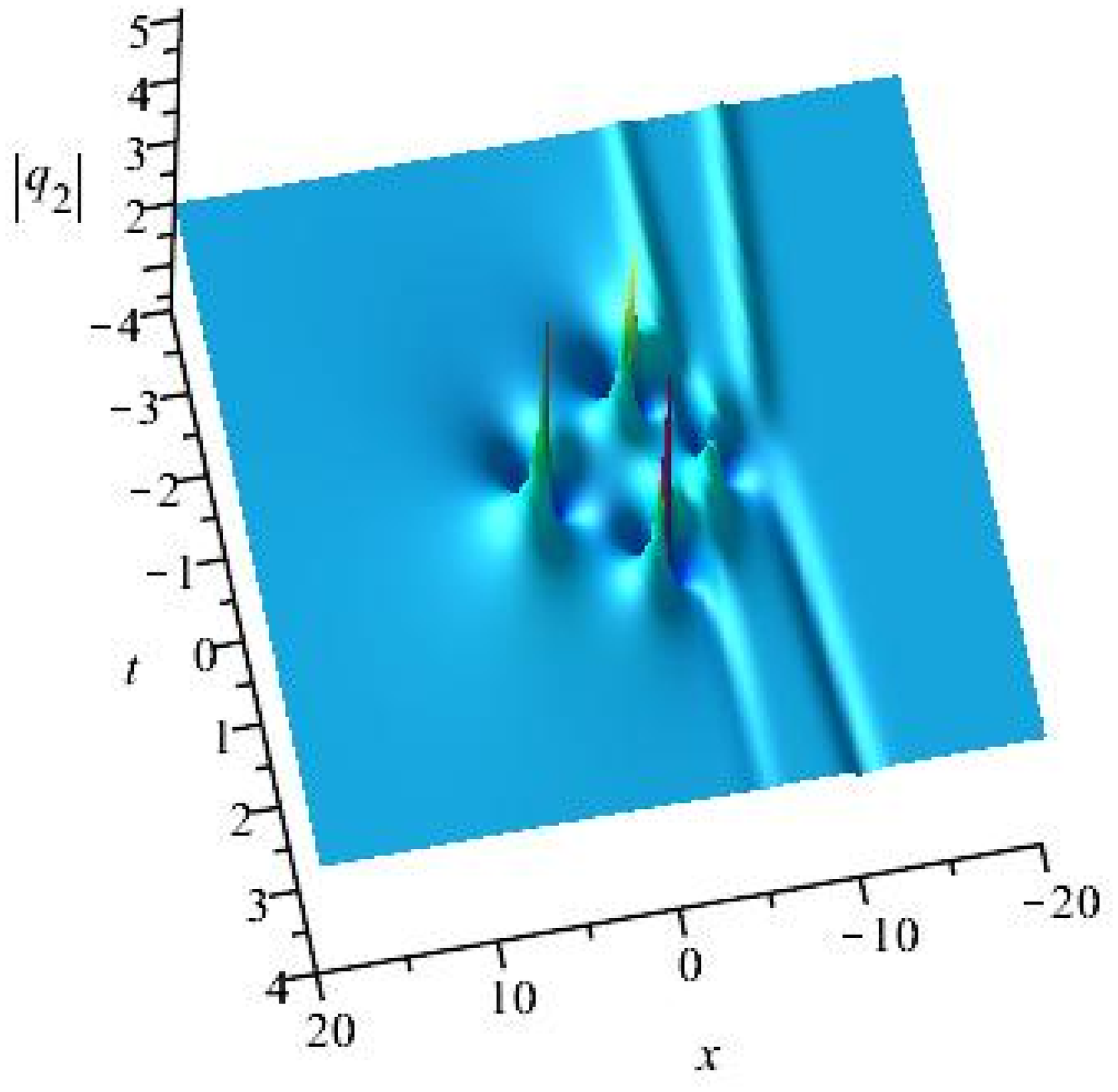}}
\centering
\subfigure[]{\includegraphics[height=0.3\textwidth]{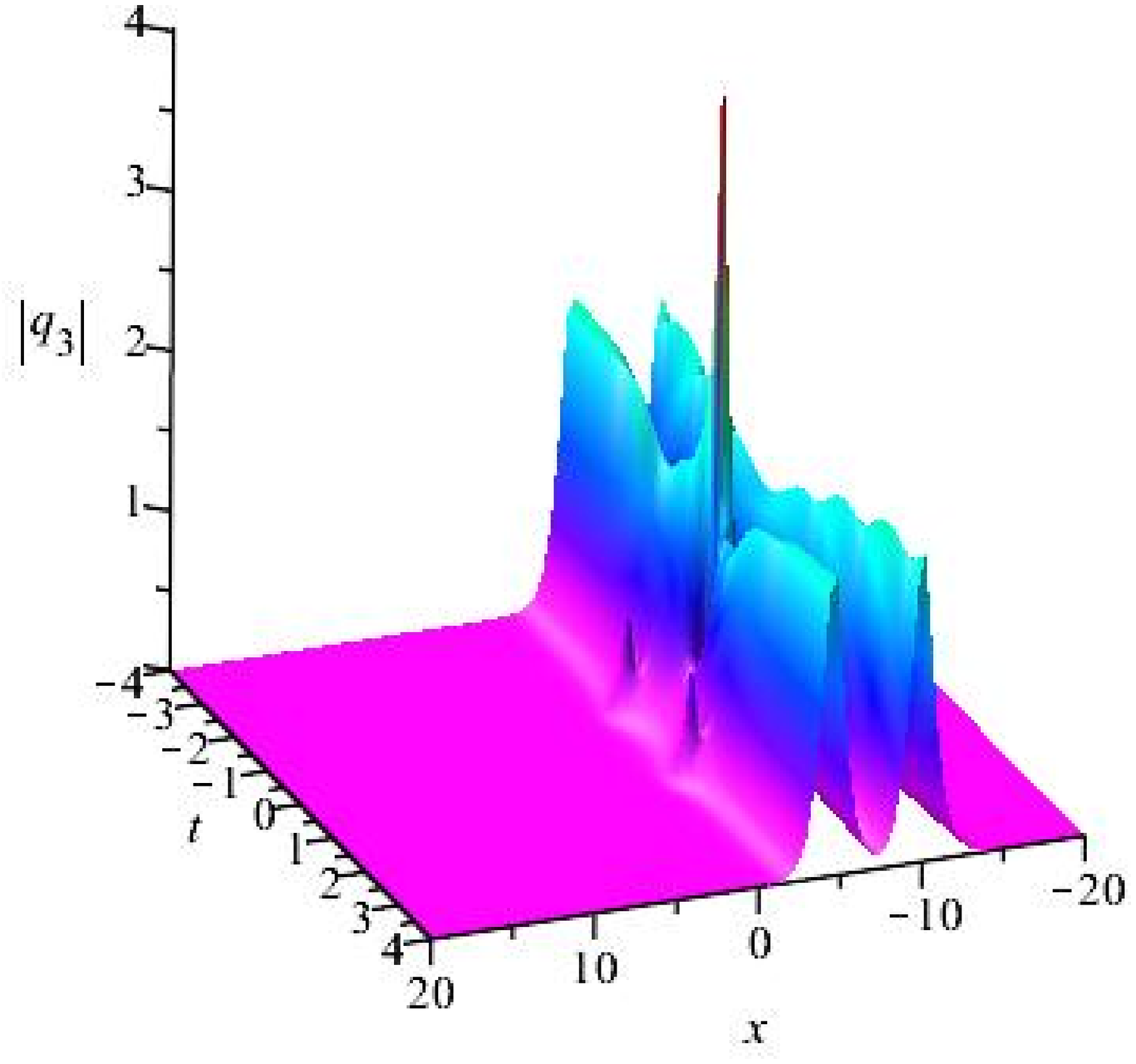}}
\centering
\caption{\small(Color online) Evolution plot of the interactional solution between the second-order RW of triangular pattern and two-amplitude-varying soliton or two-bright soliton with the parameters chosen by $d_1=1, d_2=-2, d_3=0,\alpha=0,\beta=\tfrac{1}{2000},m_1=100,n_1=-100$: (a) a second-order RW of triangular pattern and two amplitude-varying solitons split in $q_1$ component;  (b) a second-order RW of triangular pattern and two amplitude-varying solitons split in $q_2$ component;  (c) a second-order RW of triangular pattern and two bright solitons split in $q_3$ component. \label{xt-6f-15}}
\end{figure}

(\textrm{iii}) When $\alpha=0,\beta\neq0$, $d_1\neq0$, $d_2\neq0$ and $d_3=0$, one of the three disturbing terms is zero and one of three plane backgrounds is vanished. Choosing $m_1\neq0,n_1\neq0$, it can be found that the hybrid solution between a second-order RW of  triangular pattern and two amplitude-varying solitons  exists in $q_1$ and $q_2$ components, and the hybrid solution between a second-order RW of  triangular pattern exists in $q_3$ component from Fig. \ref{xt-6f-15}. The second-order RW of triangular pattern can merge with  two amplitude-varying solitons or two bright solitons by increasing the absolute values of $\alpha$ and $\beta$, here we omit these figures. In Fig. \ref{xt-6f-15}(c), we can find that two first-order RWs emerge at the bottom of the left bright soliton and a first-order RW appears at the top of the left bright soliton.  These three first-order RWs in Fig. \ref{xt-6f-15}(c) can be easily observed since they emerge at non-zero background.

\begin{figure}[H]
\renewcommand{\figurename}{{Fig.}}
\subfigure[]{\includegraphics[height=0.3\textwidth]{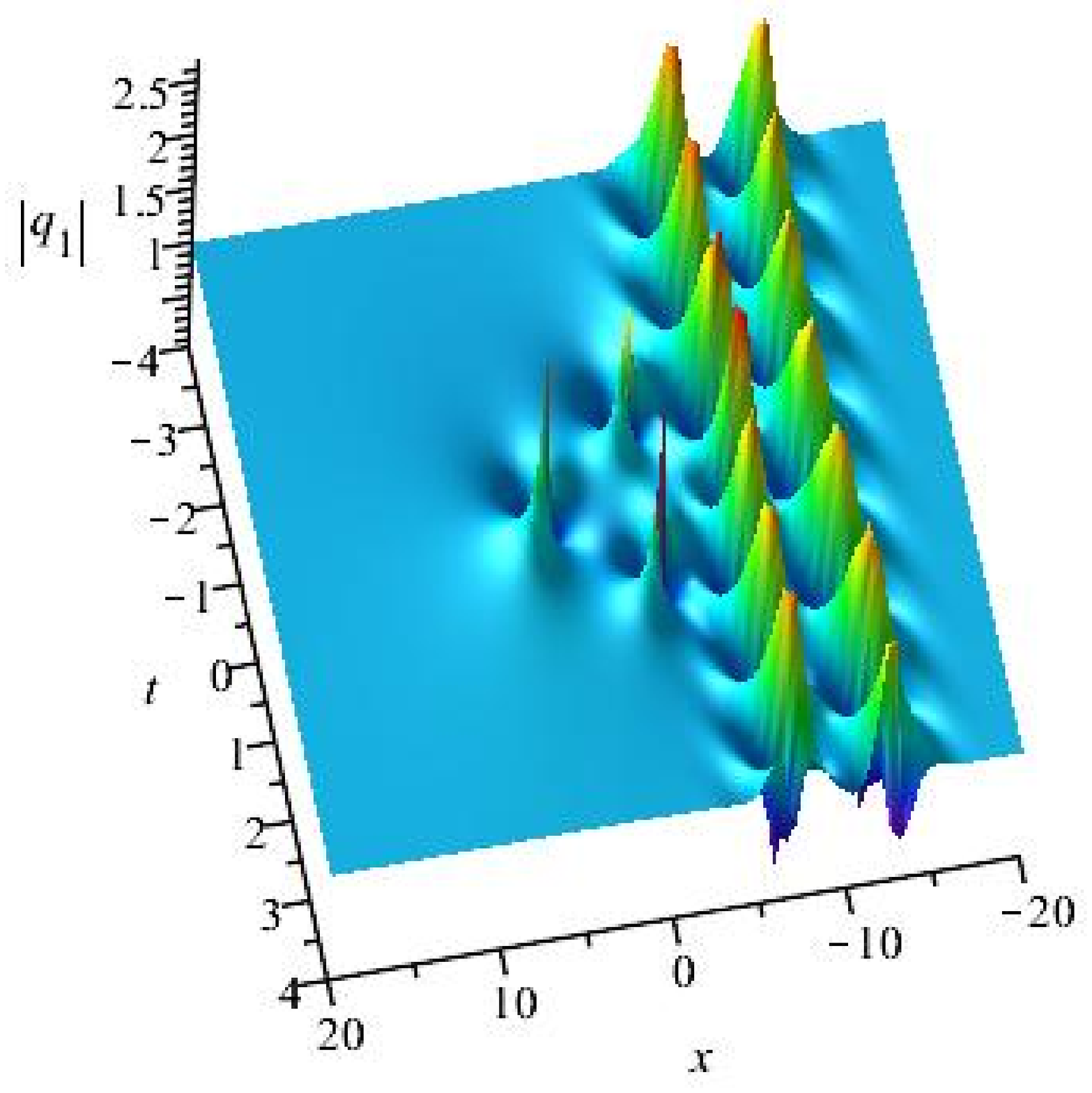}}
\centering
\subfigure[]{\includegraphics[height=0.3\textwidth]{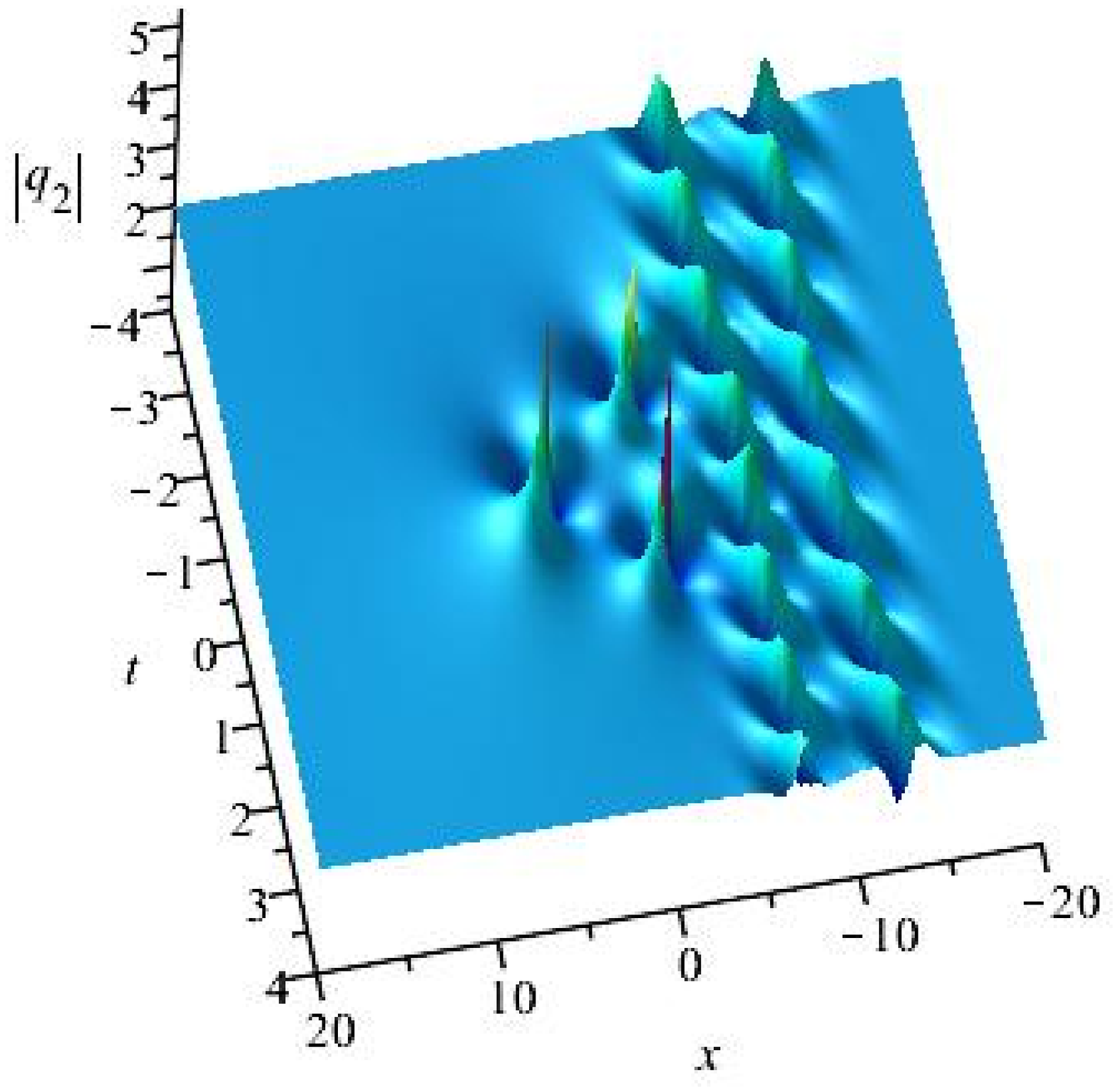}}
\centering
\subfigure[]{\includegraphics[height=0.3\textwidth]{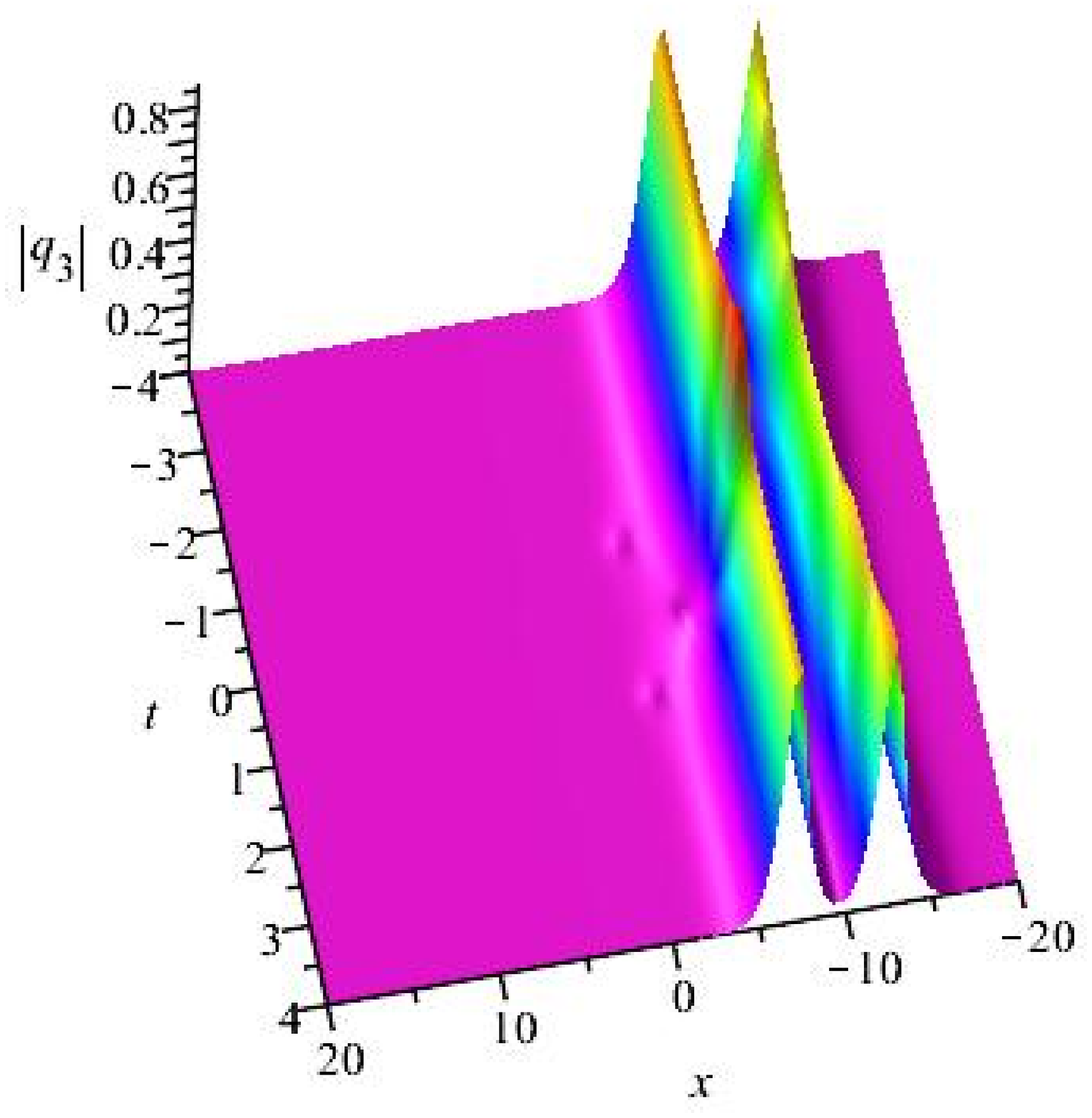}}
\centering
\caption{\small(Color online) Evolution plot of the interactional solution between the second-order RW of triangular pattern and two-breather or two-bright soliton with the parameters chosen by $d_1=1, d_2=-2, d_3=0,\alpha=\tfrac{1}{200000},\beta=-\tfrac{1}{200000},m_1=100,n_1=-100$: (a) a second-order RW of triangular pattern and two breathers split in $q_1$ component;  (b) a second-order RW of triangular pattern and two breathers split in $q_2$ component; (c) a second-order RW of triangular pattern and two bright solitons split in $q_3$ component. \label{xt-6f-16}}
\end{figure}

(\textrm{iv}) When $\alpha\neq0,\beta\neq0$, $d_1\neq0$, $d_2\neq0$ and $d_3=0$, all the three disturbing terms are non-zero and one of three plane backgrounds is vanished. In Fig. \ref{xt-6f-16}, it is shown that a second-order RW of  triangular pattern and two breathers split in $q_1$ and $q_2$ components with $m_1\neq0,n_1\neq0$, and a second-order RW of  triangular pattern and two bright solitons split in $q_3$ component. In the same way, the second-order RW of triangular pattern merges with two breathers or two bright solitons significantly by increasing the absolute values of $\alpha$ and $\beta$. Similar with Fig. \ref{xt-6f-15}(c),  two first-order RWs emerge at the bottom of the left bright soliton and a first-order RW appears at the top of the left bright soliton in Fig. \ref{xt-6f-16}(c).

\begin{figure}[H]
\renewcommand{\figurename}{{Fig.}}
\subfigure[]{\includegraphics[height=0.3\textwidth]{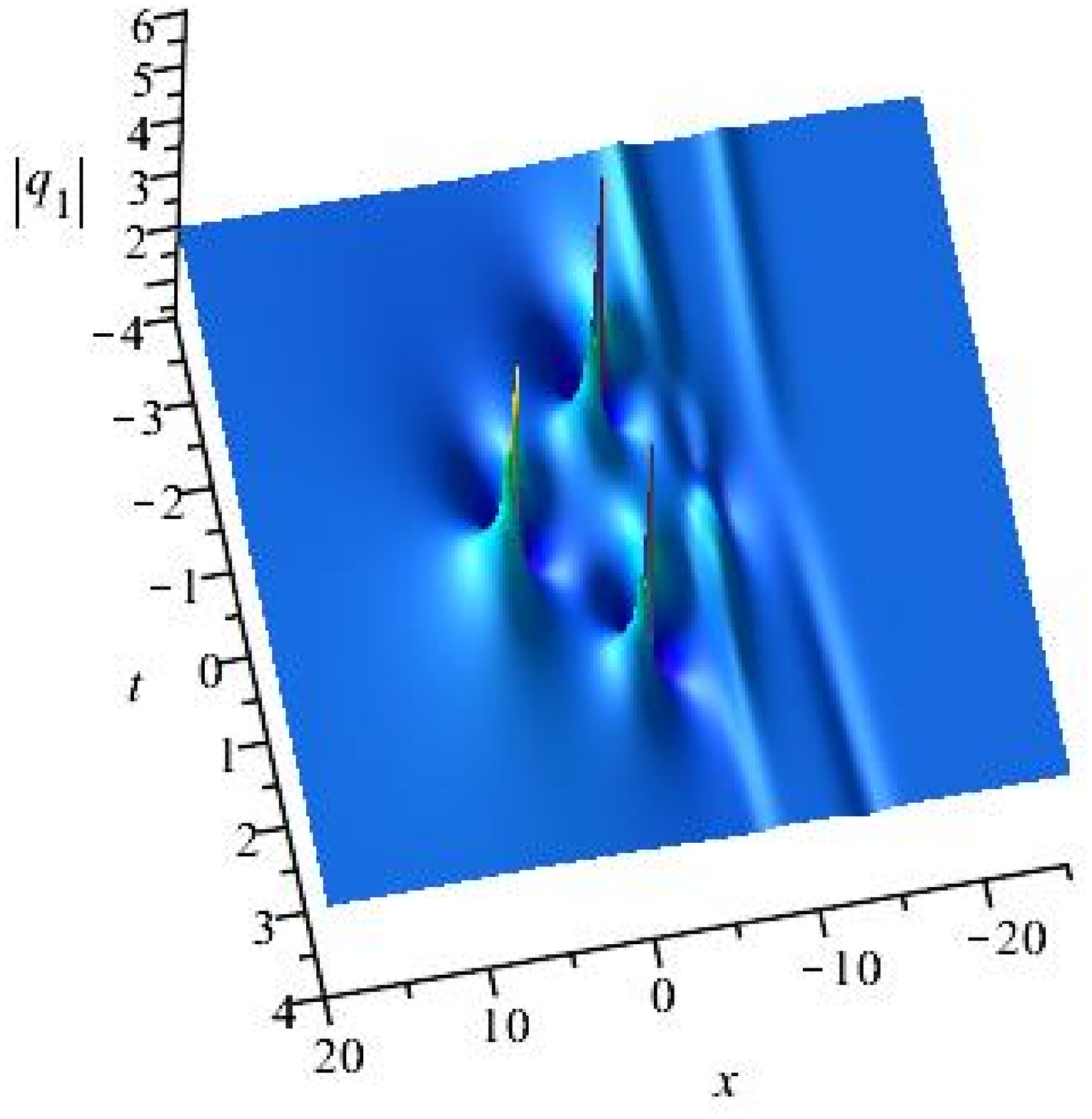}}
\centering
\subfigure[]{\includegraphics[height=0.3\textwidth]{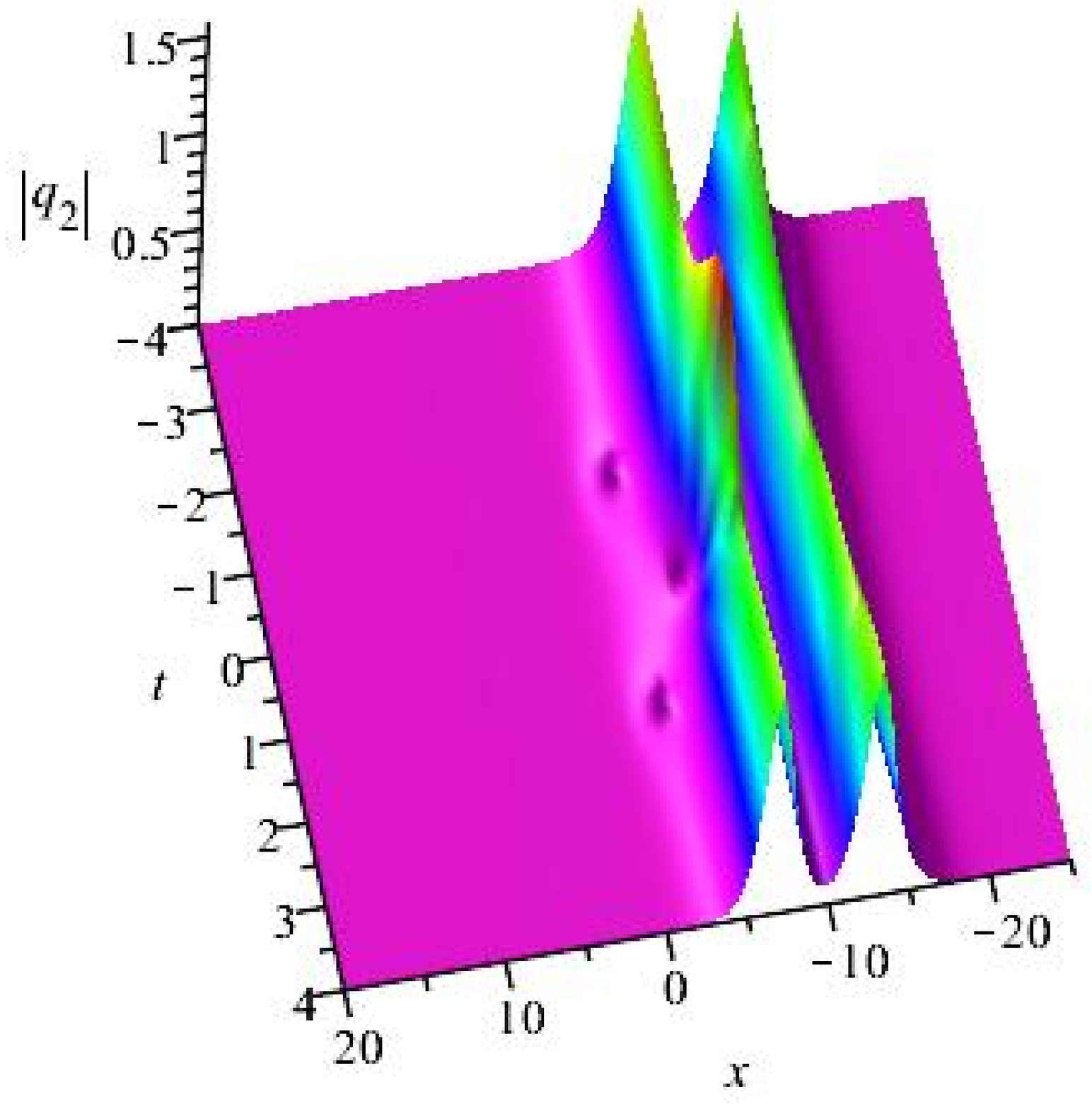}}
\centering
\subfigure[]{\includegraphics[height=0.3\textwidth]{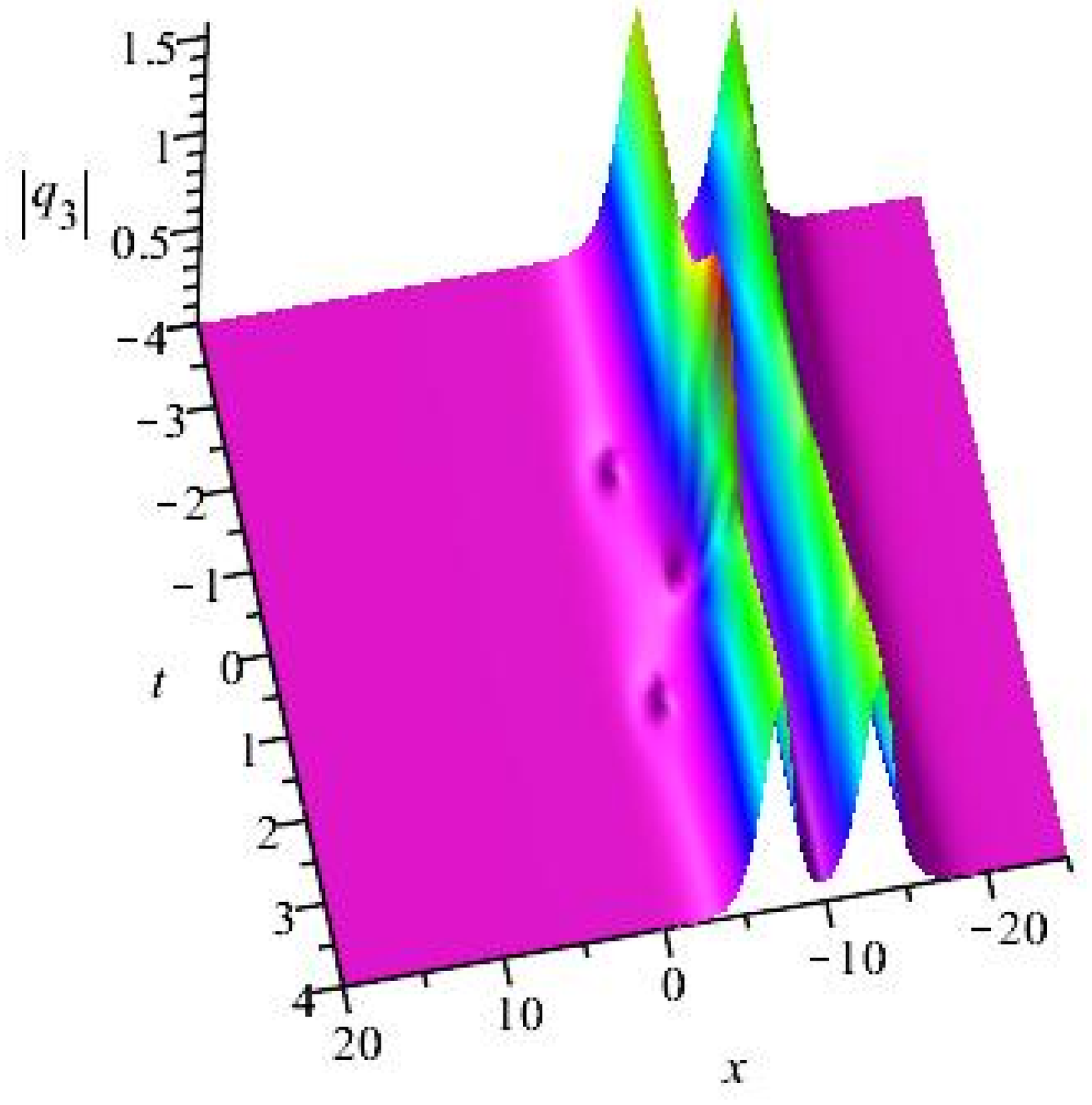}}
\centering
\caption{\small(Color online) Evolution plot of the interactional solution between the second-order RW of triangular pattern and two-amplitude-varying soliton or two-bright soliton in the three-component coupled DNLS equations with the parameters chosen by $d_1=2, d_2=0, d_3=0,\alpha=\tfrac{1}{20000},\beta=-\tfrac{1}{20000},m_1=100,n_1=-100$: (a) a second-order RW of triangular pattern and two  amplitude-varying solitons split in $q_1$ component; (b) a second-order RW of triangular pattern and two bright solitons split in $q_2$ component; (c) a second-order RW of triangular pattern and two bright solitons split in $q_3$ component. \label{xt-6f-17}}
\end{figure}

(\textrm{v}) When $\alpha\neq0,\beta\neq0$, $d_1\neq0$, $d_2=0$ and $d_3=0$, all the three disturbing terms are non-zero and two of three plane backgrounds are vanished. Setting $m_1\neq0,n_1\neq0$, it demonstrates that $q_2$ and $q_3$ components are the hybrid solution including a second-order RW of  triangular pattern and two bright solitons, and $q_1$ component is the hybrid solution including a second-order RW of  triangular pattern and two amplitude-varying solitons from Fig. \ref{xt-6f-17}. Increasing the absolute values of disturbing coefficients $\alpha$ and $\beta$, the second-order RW of  triangular pattern can merges with two  amplitude-varying solitons or two bright solitons significantly.

\begin{figure}[H]
\renewcommand{\figurename}{{Fig.}}
\subfigure[]{\includegraphics[height=0.3\textwidth]{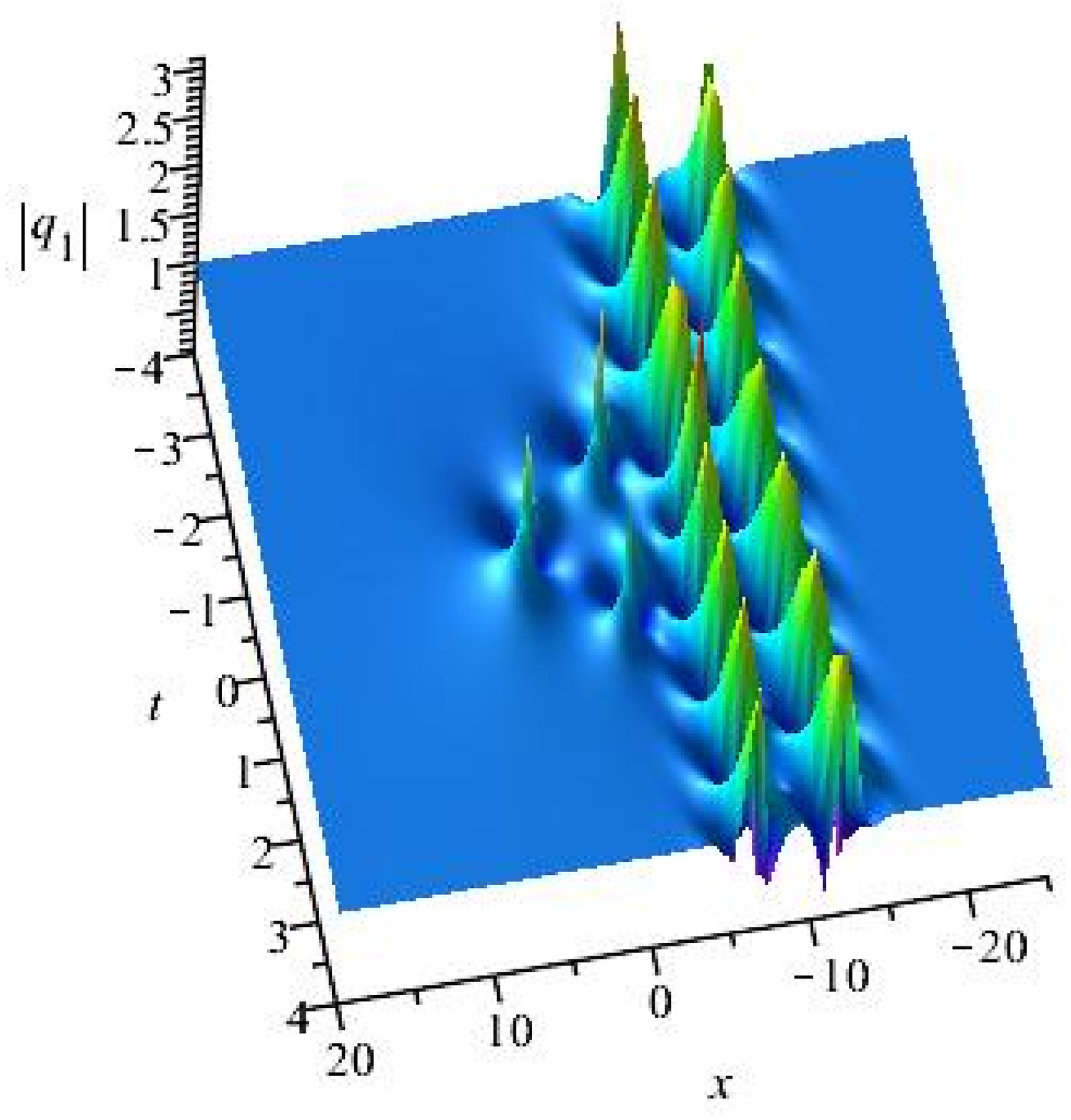}}
\centering
\subfigure[]{\includegraphics[height=0.3\textwidth]{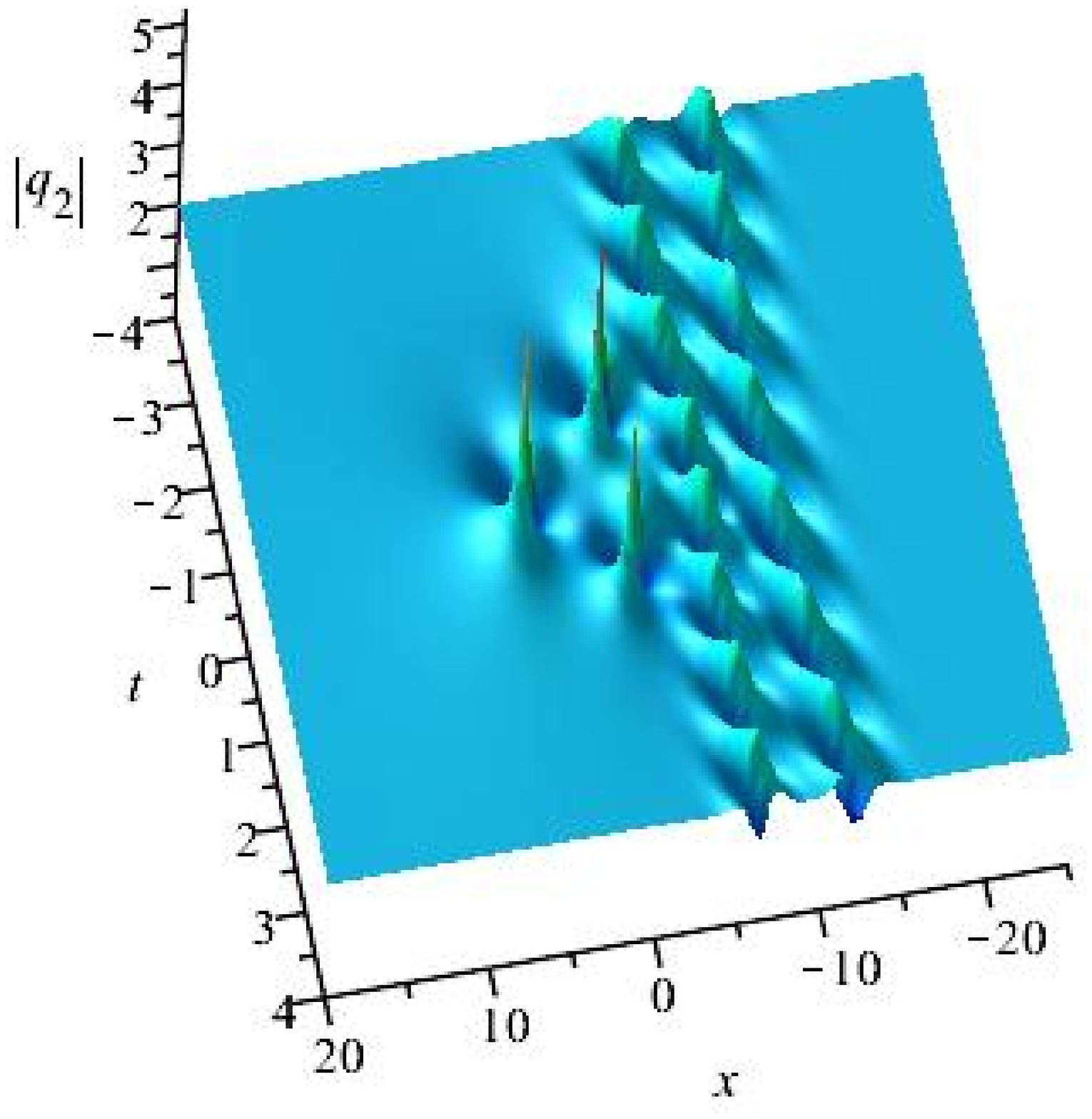}}
\centering
\subfigure[]{\includegraphics[height=0.3\textwidth]{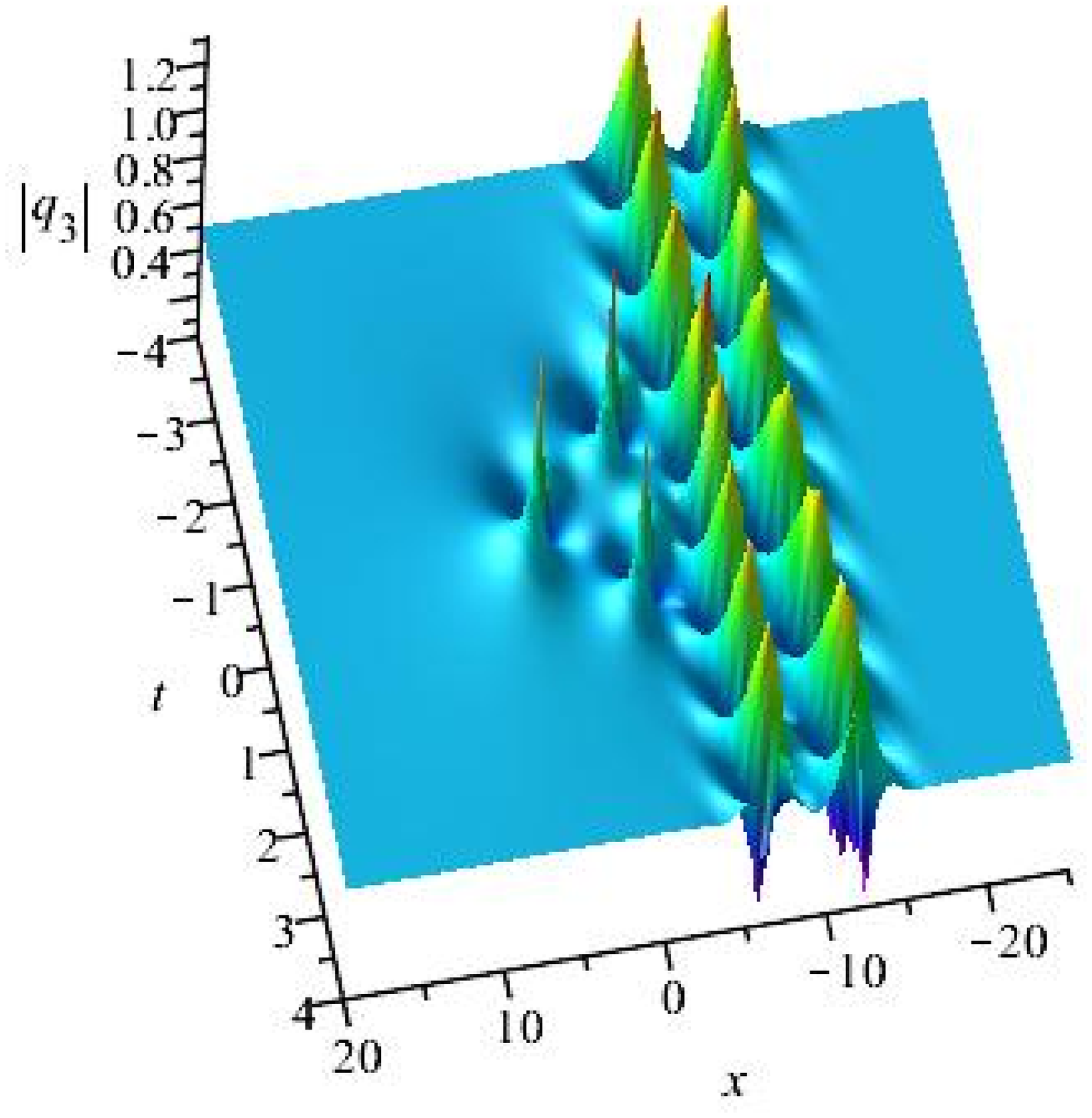}}
\centering
\caption{\small(Color online) Evolution plot of the interactional solution between the second-order RW of triangular pattern and two-breather in the three-component coupled DNLS equations with the parameters chosen by $d_1=2, d_2=-2, d_3=\tfrac{1}{2},\alpha=\tfrac{1}{200000},\beta=-\tfrac{1}{200000},m_1=100,n_1=-100$, three components are all a second-order RW of triangular pattern separate with two breathers: (a) $q_1$, (b) $q_2$, (c) $q_3$. \label{xt-6f-18}}
\end{figure}

(\textrm{vi}) When $\alpha\neq0,\beta\neq0$ and $d_j\neq0~(j=1,2,3)$, all the three disturbing terms are non-zero and the three backgrounds are all non-vanished. Choosing $m_1\neq0,n_1\neq0$, we can see that a second-order RW of triangular pattern and two breathers split in three components $q_1$, $q_2$ and $q_3$ from Fig. \ref{xt-6f-18}. Analogously, the second-order RW of triangular pattern can merge with the corresponding two breathers markedly by increasing the absolute values of $\alpha$ and $\beta$. Here, it demonstrates the interesting  phenomenon that the two breathers in $q_2$ component are greatly different from ones in $q_1$ and $q_3$ components.

In order to classify the higher-order interactional solutions of the coupled system (\ref{xt-6-1}), we define the same combination as the same type interactional solution. In the firs-order interactions of localized waves, the concrete expressions of the general interactional solutions can be given and  these solutions are classified into six types. For the second-order case, the  corresponding expressions  are very tedious and complicated and we only give their figures. The second-order interactional solutions are also discussed in six cases by reference to the first-order case.  Iterating the DT of the coupled system (\ref{xt-6-1}), we can get arbitrary higher-order interactional solutions, and the first- and second-order interactional solutions are discussed in detail.

 In these six types of interactional solutions, we can get four mixed interactions of localized waves among three components $q_1$, $q_2$ and $q_3$: (1) two components are the hybrid solutions including higher-order RWs and multi-breather (higher-order RWs$+$multi-breather), and one component is the interactional solution including higher-order RWs and multi-amplitude-varying soliton (higher-order RWs$+$multi-amplitude-varying soliton); (2) two components are higher-order RWs$+$multi-amplitude-varying soliton, and one component is higher-order RWs$+$multi-bright soliton; (3) two components are higher-order RWs$+$multi-breather, and one component is higher-order RWs$+$multi-bright soliton; (4) two components are higher-order RWs$+$multi-bright soliton, and one component is higher-order RWs$+$multi-amplitude-varying soliton.

\section{Conclusion}
Starting from the periodic seed solution, a special vector solution of the Lax pair (\ref{xt-6-2})-(\ref{xt-6-3}) is elaborately constructed. Combining a limiting process and the peculiar vector solution (\ref{xt-6-14}),  various novel and interesting higher-order interactional solutions in the three-component coupled DNLS equations are generated  through DT technique.  In the vector solution (\ref{xt-6-14}),  these  three disturbing terms $-(\alpha d_2+\beta d_3)e^{ix\lambda^2}$, $\alpha d_1e^{ix\lambda^2}$ and $\beta d_1e^{ix\lambda^2}$ are added in Eq. (\ref{xt-6-14}) to construct interactions of localized waves, but they are not necessary to only construct  RWs. We define the two free parameters $\alpha$ and $\beta$ as disturbing coefficients. These three free parameters $d_j~(j=1,2,3)$ determine the plane backgrounds on which different localized waves emerge. Here, when $d_j=0$, we call it vanished background; otherwise $d_j\neq0$, we name it non-vanished background. Considering both disturbing terms and the backgrounds, different interactional solutions are completely classified. The parameter $s_j=m_j+n_j~(j=1,2,\cdots,N)$ control the structures of high-order rogue waves in the hybrid solutions, for example, the fundamental second-order RW can split into three first-order ones with $s_1\neq0$.

Instead of considering various arrangements among the three components $q_1$, $q_2$ and $q_3$, we define the same combination as the same type interactional solution. The first- and second-order interactional solutions are discussed in six types in detail:  (1) the interactional solutions degenerate to rational ones and these three components are all rogue waves; (2) two components are hybrid solutions between rogue wave and breather (RW$+$breather), and one component is hybrid solutions between RW and amplitude-varying soliton (RW$+$amplitude-varying soliton); (3) two components are RW$+$amplitude-varying soliton, and one component is RW$+$bright soliton; (4) two components are RW$+$breather, and one component is RW$+$bright soliton; (5) two components are RW$+$bright soliton, and one component is RW$+$amplitude-varying soliton; (6) three components are all RW$+$breather. Among the above six types interactional solutions, there are four mixed interactions of localized waves in three different components, such as types (2), (3), (4) and (5).

In this work, we generalize Baronio's results \cite{6-37} to the higher-order case in the three-component DNLS equations (\ref{xt-6-1}) and obtain four mixed interactions of localized waves. The interactional solutions were constructed in some two-component coupled systems \cite{6-37,6-38,6-39}. Besides, we constructed mixed interactions of localized waves in three-component NLS equations \cite{6-40} and Hirota equations \cite{6-41} through generalized DT. However, these mixed interactions can not be obtained in single- and two-component systems by DT technique. Based on the above facts, a conclusion can be drawn that these kinds of mixed interactions of localized waves may only be obtained by DT in the nonlinear systems,  whose components are more than 3 with the corresponding Lax pair including the matrices larger than $3\times3$. In \cite{6-37},  Baronio gave the experimental conditions for observing the first-order interactional solutions among RW, one-bright (dark) soliton and one-breather in two-component coupled NLS equations. Here, we expect that these interactions of localized waves obtained in this article will be verified and observable in physical experiments in the future.

\section*{Acknowledgment}
We would like to express our sincere thanks to other members of our discussion group for their valuable
comments. The project is supported by the Global Change Research Program of China (No.2015CB953904),
National Natural Science Foundation of China (No.11675054, 11435005). and Shanghai Collaborative
Innovation Center of Trustworthy Software for Internet of Things (No.ZF1213).


\end{document}